\documentclass[a4paper, 12pt]{article}
\usepackage{jheppub}
\pdfoutput=1

\usepackage[latin1]{inputenc}

\usepackage{empheq}
\usepackage{epsfig,verbatim}
\usepackage{amsmath,amssymb,graphics}
\usepackage{color}
\usepackage{slashed}            
\usepackage{bbm}                
\usepackage{units}              
\usepackage{hyperref}           
\usepackage{xspace}             
\usepackage{enumerate}          
\usepackage{soul}
\usepackage{cancel}
\usepackage{amstext}    
\usepackage{array} 
\usepackage{pbox}

\usepackage{subfig}
\usepackage{graphicx}
   
\newcolumntype{L}[1]{>{\raggedright\let\newline\\\arraybackslash\hspace{0pt}}m{#1}}
\newcolumntype{C}[1]{>{\centering\let\newline\\\arraybackslash\hspace{0pt}}m{#1}}
\newcolumntype{R}[1]{>{\raggedleft\let\newline\\\arraybackslash\hspace{0pt}}m{#1}}

\renewcommand\[{\left[}
\renewcommand\]{\right]}
\newcommand{\la}{\langle}
\newcommand{\ra}{\rangle}

\renewcommand{\Re}{\operatorname{Re}}
\renewcommand{\Im}{\operatorname{Im}}

\newcommand{\rmi}{\mathrm{i}}
\newcommand{\rme}{\mathrm{e}}

\newcommand{\bea}{\begin{eqnarray}}
\newcommand{\eea}{\end{eqnarray}}
\newcommand{\be}{\begin{equation}}
\newcommand{\ee}{\end{equation}}

\newcommand{\cB}{\mathcal{B}}
\newcommand{\cC}{\mathcal{C}}

\newcommand{\cF}{\mathcal{F}}
\newcommand{\cG}{\mathcal{G}}
\newcommand{\cH}{\mathcal{H}}
\newcommand{\cI}{\mathcal{I}}
\newcommand{\cL}{\mathcal{L}}
\newcommand{\cM}{\mathcal{M}}
\newcommand{\cN}{\mathcal{N}}
\newcommand{\cO}{\mathcal{O}}

\newcommand{\cS}{\mathcal{S}}

\newcommand{\cV}{\mathcal{V}}

\newcommand{\SO}{\mathop{\rm SO}}
\newcommand{\Sp}{\mathop{\rm Sp}}

\newcommand{\unity}{\mathbbm{1}}
\newcommand{\del}{\nabla}
\newcommand{\pd}{\ensuremath{\partial}}

\usepackage[dvipsnames]{xcolor}

\title{\Large Towards a Complete Mass Spectrum of Type-IIB \\ Flux Vacua at Large Complex Structure}

\author[1,2]{Jose J. Blanco-Pillado,}
\author[3]{Kepa Sousa,}
\author[1]{Mikel A. Urkiola,}
\author[4]{Jeremy M. Wachter.}

\affiliation[1]{Department of Theoretical Physics and History of Science, University of the Basque Country UPV/EHU, 48080 Bilbao, Spain}
\affiliation[2]{IKERBASQUE, Basque Foundation for Science, 48011, Bilbao, Spain}
\affiliation[3]{Institute of Theoretical Physics, Faculty of Mathematics and Physics, Charles University in Prague, V Holes\v{o}vi\v{c}k\'ach 2, Prague, Czech Republic}
\affiliation[4]{Skidmore College Physics Department, 815 North Broadway Saratoga Springs, New York 12866}

\emailAdd{josejuan.blanco@ehu.eus}
\emailAdd{kepa.sousa@utf.mff.cuni.cz}
\emailAdd{mikel.alvarezu@ehu.eus}
\emailAdd{jwachter@skidmore.edu}

\abstract{The large number of moduli fields arising in a generic string theory 
compactification makes a complete computation of the low energy effective theory infeasible.
A common strategy to solve this problem is to consider Calabi-Yau manifolds with discrete symmetries, 
which effectively reduce the number of moduli and make the computation 
of the truncated Effective Field Theory possible. In this approach, however, the couplings 
(e.g., the masses) of the truncated  fields  are left undetermined. 
In the present paper we discuss the tree-level mass spectrum of type-IIB flux 
compactifications at Large Complex Structure, focusing on models with a reduced 
one-dimensional complex structure sector. We compute the  tree-level 
spectrum  for the dilaton and complex structure moduli, \emph{including the truncated fields},  
which can be expressed entirely in terms of the known couplings of the reduced theory. 
We show that the masses of this set of fields are naturally heavy at vacua consistent with the KKLT construction, 
 and we discuss other phenomenologically interesting
 scenarios where the spectrum involves fields much lighter than the gravitino.
We also derive the  probability distribution for the masses on the ensemble of flux vacua, 
and show that it exhibits universal features independent of the details of the compactification. 
We check our results on a large sample of flux vacua constructed in an orientifold of the 
Calabi-Yau $\mathbb{WP}^4_{[1,1,1,1,4]}$. Finally, we also discuss the conditions under which 
the spectrum derived here could arise in more general compactifications.}

\date{\today}

\begin{document}

\maketitle

\section{Introduction}

The need to compactify the 6 or 7 extra dimensions of supersymmetric string theories leads 
to significant technical problems, which make the study of the phenomenological 
and cosmological implications of the Landscape of $4d$ Effective Field Theories (EFTs) exceedingly difficult. One of these problems is the huge number 
of fields arising in these EFTs, \emph{the moduli}, which describe the geometry 
of the compact space. This makes the computation of the complete EFT prohibitively complex, 
and as a consequence it has only been obtained for simple compactifications. Another difficulty
is the vast number of possible ways to compactify the extra dimensions, which makes it infeasible 
to characterise every possible four dimensional vacuum of the theory.

In the last few decades, several complementary strategies have been followed to overcome these technical 
problems. On the one hand, many efforts have been dedicated to studying explicit models where most of the 
moduli can be truncated or integrated out, leaving only a few fields (up to ten) for which the EFT can be 
computed \cite{Kachru:2003aw,Balasubramanian:2005zx,Conlon:2005ki,Balasubramanian:2004uy,Westphal:2006tn,Giryavets:2003vd,Giryavets:2004zr,DeWolfe:2004ns,Denef:2004dm,Louis:2012nb,BlancoPillado:2012cb,Cicoli:2013cha}. In such models the observational implications 
can be studied in detail, and these explicit computations have been used as lampposts to guide the analysis 
of more complex scenarios. On the other hand, instead of attempting an exhaustive examination of all possible 
string compactifications, one can take a statistical perspective. In this approach one regards this set of solutions 
as a statistical ensemble and characterises the probability distributions of the relevant observables in such a 
Landscape \cite{Bousso:2000xa,Douglas:2003um,Denef:2004ze,Denef:2004cf,Douglas:2004kp}. It is expected 
(or hoped) that some of these quantities will exhibit universal properties, i.e., independent of the specific details 
of each particular EFT, which would partially alleviate the need to compute the effective theories. Following this
approach one could try to model the complicated low-energy effective potential as a random function with
some particular statistics; for example, as a multidimensional Gaussian random field. This procedure has
been recently developed in the literature in relation to different aspects of the distribution of vacua
of this potential as well as its applications to 
cosmology \cite{Tegmark:2004qd,Easther:2005zr,Battefeld:2012qx,Marsh:2013qca,Masoumi:2016eag,Wang:2016kzp,Pedro:2016sli,Freivogel:2016kxc, Bjorkmo:2017nzd,Dias:2017gva,Masoumi:2017gmh,Masoumi:2017xbe,Blanco-Pillado:2017nin,Paban:2018ole,Bjorkmo:2018txh,Blanco-Pillado:2019mnq,Low:2020kzy}. 

Finally, the more recent \emph{Swampland program} is directed to find universal constraints that should 
be satisfied by any EFT arising in a consistent theory of quantum gravity \cite{Garg:2018reu,Ooguri:2018wrx,ArkaniHamed:2006dz} 
(see also \cite{Palti:2019pca,Brennan:2017rbf} and references therein). These constraints determine 
conditions under which the EFTs are under computational control, and serve to identify which low-energy
solutions can be regarded as plausible string theory vacua. In this respect, the Swampland program 
sets an outermost limit for the boundaries of the Landscape.

In the present paper we will take a conservative approach and discuss one of the best studied domains of the 
Landscape:  the tree-level flux vacua on Calabi-Yau compactifications of type-IIB superstrings at Large Complex Structure (LCS). 
The construction of the EFTs describing this corner of the Landscape, and the applicability of these theories, 
has been widely discussed in the literature \cite{Candelas:1990pi,Candelas:1990rm,Hosono:1994ax,Gukov:1999ya,Giddings:2001yu} (see \cite{Klemm:2005tw} for a review). 
Among this class of models, phenomenologically interesting compactifications generally involve a large 
number of complex structure moduli and only a few K\"ahler moduli (see, e.g., \cite{Cicoli:2013cha}). However, as we mentioned above, the detailed construction 
of the complete Effective Field Theory is prohibitive in general. Consequently, explicit constructions of flux 
vacua are often based on Calabi-Yau manifolds invariant under large groups of discrete symmetries which 
allow a consistent supersymmetric truncation of a large fraction of the complex structure moduli 
\cite{Giryavets:2003vd,Giryavets:2004zr,DeWolfe:2004ns,Denef:2004dm,Louis:2012nb,Cicoli:2013cha}. 
These groups of symmetries arise naturally when compactifying on hypersurfaces of complex projective spaces 
and toric varieties, and on Complete Intersection Calabi-Yaus (see \cite{Braun:2010vc,Braun:2017juz} and 
references therein). Particular and prominent examples are the discrete symmetry groups which 
allow the Greene-Plesser construction of the mirror Calabi-Yau pairs \cite{Greene:1990ud}. Provided 
only fluxes invariant under these symmetries are turned on, it is possible to \emph{freeze} a large set 
of complex structure moduli at a critical point of the resulting flux scalar potential, leaving a reduced 
theory for a few surviving fields \cite{Giryavets:2003vd,Giryavets:2004zr,DeWolfe:2004ns,Denef:2004dm}. The 
phenomenological and cosmological predictions of these models are then computed after including the relevant 
quantum corrections and supersymmetry breaking effects \emph{in the reduced theory}. However, the fate 
of the truncated fields is rarely discussed in detail \cite{Louis:2012nb,Cicoli:2013cha}. 

The main objective of this work is to take a first step towards a more precise understanding of the truncated 
moduli sector in this class of models. Note that, in the approach we just described, the truncated moduli are 
not \emph{integrated out}; instead, the resulting EFT is a \emph{consistent supersymmetric truncation} of the 
complete low energy theory at tree-level, and thus there is not necessarily a mass gap between the frozen 
moduli and those in the reduced theory \cite{Gallego:2011jm,Achucarro:2008fk,Sousa:2014qza}. Actually, 
although the truncated sector is guaranteed to be at a stable configuration at tree-level, the spectrum might 
contain arbitrarily light fields. Thus, in principle the quantum corrections and the breaking of supersymmetry 
could render some of these light fields tachyonic. Alternatively, the fixed point of the discrete symmetry group 
could cease to be a critical point of the corrected scalar potential. 

The perturbative stability of the complete complex structure and axio-dilaton sector has been proven using 
scaling arguments for generic KKLT constructions \cite{Kachru:2003aw}, and for Large Volume Scenarios (LVS) 
with an exponentially large compactification volume \cite{Balasubramanian:2005zx,Conlon:2005ki,Balasubramanian:2004uy}.  The case of K\"ahler uplifted vacua \cite{Westphal:2006tn} and LVS scenarios with moderately large 
volume \cite{Rummel:2014raa,Rummel:2013yta,Maharana:2015saa,Cicoli:2015wja} is more subtle. 
In \cite{Achucarro:2015kja}, it was argued, using statistical techniques, that these classes of 
vacua may contain a sizeable fraction of tachyonic fields in the truncated sector. It is important to emphasize 
that, despite of all of these efforts, the complete mass spectrum in the truncated sector has never been explicitly 
computed.

Here we will focus on the simplest possible class of these models, those where the reduced theory contains a 
single complex structure modulus. Although this is a rather restrictive type of compactification, it contains 
plenty of examples (see, e.g., \cite{Klemm:1992tx,Doran:2007jw,Candelas:2017ive,Braun:2011hd,Batyrev:2008rp}). Furthermore, the moduli space geometry is well characterised for many 
of them \cite{Doran:2005gu,Braun:2015jdy,Candelas:2019llw,Joshi:2019nzi}, including the well known family 
of quintic hypersurfaces $\mathbb{WP}^4_{[1,1,1,1,1]}$ and its generalisations \cite{Candelas:1990rm,Doran:2007jw,Font:1992uk,Klemm:1992tx}. 
We will prove, using only symmetry arguments and properties of the effective theory on type-IIB compactifications 
at LCS, that it is possible to compute the tree-level mass spectrum 
for the axio-dilaton and the complete set of $h^{2,1}$ complex structure moduli fields, \emph{including the truncated ones}. 

Interestingly, the resulting set of masses can be expressed entirely in terms of the known couplings of the reduced 
effective theory, and exhibits universal features independent on the details of the compactification. More specifically, to leading order in $\alpha'$ and the string coupling $g_s$,
we find that  the $2 h^{2,1}-2$ real scalar modes on the truncated sector have  squared masses $\mu^2_{\pm\lambda}$ given by
\be
\boxed{\mu^2_{\pm\lambda} =\left(m_{3/2} \pm m_{\text{susy}} \frac{1+\xi}{\sqrt{3(1-2\xi)}}\right)^2, \qquad \text{with}\qquad \lambda = 2,\ldots, h^{2,1}.}\nonumber
\ee

This expression involves only two scales which can be computed in the reduced theory: the gravitino mass,
$m_{3/2}$, and the scale of supersymmetric masses induced by the fluxes, $m_{\text{susy}}$. 
The real parameter $\xi\in [0,1/2]$ depends on the configuration of the complex structure of the reduced theory, 
and in particular it takes the value $\xi=0$ at the LCS point. Analogous universal properties of the spectrum of matrices arising in the effective theory were previously 
reported in \cite{Brodie:2015kza,Marsh:2015zoa}, where the authors considered generic points of the moduli 
space, i.e., not necessarily flux vacua. The computations of the mass spectra rely on the perturbative description 
of the moduli space geometry at large complex structure, and thus are valid provided the exponentially 
suppressed instanton corrections can be neglected. 

In order to illustrate our results, we have compared our analytic formulae with a numerical scan of flux vacua 
of type-IIB compactified on an orientifold of $\mathbb{WP}^4_{[1,1,1,1,4]}$ \cite{Font:1992uk,Klemm:1992tx}. This 
family of hypersurfaces has a $h^{2,1}=149$ dimensional complex structure moduli space, which can be 
consistently reduced to a single field at the fixed locus of a $ \mathbb{Z}_8^2 \times \mathbb{Z}_2$ symmetry. Using 
the known reduced effective theory, we construct a large ensemble of flux vacua and verify the validity of the formulae 
we derived for masses of the axio-dilaton and the complex structure field on the reduced theory. It is important to stress 
that, at each of these vacua, our results allow us to infer the masses of all of the truncated $148$ complex structure 
fields, \emph{without the need to compute the complete EFT}. 

For generic vacua, the mass spectrum has a dependence on the fluxes and thus, to have a characterisation of the 
perturbative stability independent of the flux choice, we resort to statistical methods. 
More specifically, we use the techniques derived in the seminal papers \cite{Denef:2004ze,Denef:2004cf}, whose 
only assumption is the continuous flux approximation. With this at hand, we are able to analytically compute the probability 
distribution for the complete set of masses in the ensemble of flux vacua, and show that the statistical properties 
of the spectrum are independent of the compactification. Then, we compare the ``empirical'' mass distributions from the 
ensemble of vacua in the $\mathbb{WP}^4_{[1,1,1,1,4]}$ model with the predicted probability distributions, and show they 
are in good agreement within the regime of validity of both the EFT and the continuous flux approximation. 

Regarding the validity of the statistical methods, our numerical scan shows a deficit in the number of generic no-scale 
vacua with respect to the theoretical statistical distributions in a small neighbourhood of the LCS point. This is in perfect 
agreement with previous analyses which predict a breakdown of the continuous flux approximation in this 
limit \cite{Denef:2004ze,Eguchi:2005eh,Torroba:2006kt,Marsh:2015zoa}. The failure of this approximation 
leads to an absence of generic no-scale vacua in the LCS limit \cite{Magda2} of one-parameter models, other than the vacuum
sitting on the LCS point itself \cite{Danielsson:2006xw} (see also \cite{Grimm:2019ixq}). 

For completeness we have also compared our analysis with an alternative method aimed at describing the statistical 
properties of the complex structure sector in the flux ensemble: the Random Matrix Theory (RMT) approach. These models were 
originally proposed in \cite{Denef:2004cf}, and further developed in \cite{Marsh:2011aa,Bachlechner:2012at,Sousa:2014qza,Achucarro:2015kja}. Here 
we show that this method, whose validity relies on the complexity of the couplings in generic string compactifications, 
does not correctly characterise the obtained mass spectra of the models we study here. This can easily be explained by noting 
that the large group of symmetries present in these models severely constrains the allowed values of couplings of the 
effective theory, and this resulting simplicity violates the premise on which Random Matrix Theory method is based. The 
failure of the RMT approach to describe the mass spectra arising in the LCS regime of more general type-IIB flux 
compactifications was also discussed in \cite{Brodie:2015kza}.

To conclude, we will argue that the universal spectra and mass distributions found here may also arise in more generic 
type-IIB compactifications. In particular we expect that our results may also apply to compactifications where no symmetry group is 
present, or when the reduced complex structure moduli space involves more than one field. Actually, as we shall show, provided we restrict to the LCS regime   this is certainly the case if we neglect flux quantization. This motivates future works on the search for this class of vacua 
with universal spectrum in compactifications with large values of the $D3$ tadpole, where the continuous flux approximation 
is expected be more accurate.\\

The paper is organized as follows. In section \ref{typeIIB} we review the effective theory for the axio-dilaton and complex 
structure sector on type-IIB compactifications, and collect the relevant formulae for the computation of the tree-level mass 
spectrum. In section \ref{sec:ESV} we revisit the effective reduction of the complex structure moduli space on a Calabi-Yau 
admitting a discrete group of symmetries. We also derive the restrictions that these symmetries impose on the structure 
of the Hessian and the fermion mass matrix. Section \ref{sec:totalSpectrum} contains the main results of the paper where 
we analytically derive the tree-level mass spectrum for the class of models we consider. In section \ref{sec:example} we present the EFT for the compactification of type-IIB in the  $\mathbb{WP}^4_{[1,1,1,1,4]}$ Calabi-Yau manifold. In section \ref{sec:statisticalAnalysis} 
we analyze the statistical properties of the computed spectra in the ensemble of flux vacua, and verify our conclusions 
by performing a numerical scan on the $\mathbb{WP}^4_{[1,1,1,1,4]}$ model. In section \ref{sec:otherModels} we discuss 
briefly how to extend our results to more general compactifications. We present our conclusions in section \ref{sec:conclusions}.

\section{Flux vacua on type-IIB compactifications}
\label{typeIIB}

In the next subsection we will summarize the relevant formulae for compactifications of type-IIB superstrings on the 
orientifold $\tilde M_3$ of a Calabi-Yau manifold $M_3$ (see \cite{Grimm:2004uq,Klemm:2005tw} for a review). We will work in 
units of the reduced Planck mass, $M_p^{-2} = 8 \pi G=1$.

\subsection{Effective theory for type-IIB flux compactifications}

The low-energy spectrum of type-IIB string theory compactified on a Calabi-Yau orientifold $\tilde M_3$
includes the axio-dilaton $\tau$, the complex structure moduli $z^i$, where $i=1,\ldots,h^{2,1}$, and 
the K\"ahler moduli $T^{\rho}$, where\footnote{On the orientifold $\tilde M_3$ a fraction of the deformations of $M_3$ are projected out, $(h^{1,1},h^{2,1}) \to (h^{1,1}_+,h^{2,1}_-)$, but we will omit the subscripts on the Hodge numbers to keep the notation simple.   We will also ignore
further degrees of freedom, such as possible $h^{1,1}_-$ axion multiplets, $D3$- and $D7$-brane moduli, or matter fields.} $\rho=1,\ldots, h^{1,1}$. 

To leading order in $\alpha'$ and $g_s$, the K\"ahler potential $K$ of the corresponding 4-dimensional effective supergravity theory reads
\be
K = - 2 \log \cV - \log(-\rmi (\tau-\bar \tau)) - \log \left( \rmi \int_{M_3} \Omega \wedge \bar \Omega\right)\,. 
\label{eq:KahlerPotential}
\ee
Here $\cV(T^\rho,\bar T^\rho)$ denotes the K\"ahler moduli-dependent volume of $\tilde M_3$, measured in the 
Einstein frame and in units of the string length $\ell_s= 2\pi \sqrt{\alpha'}$. The holomorphic three-form of the 
Calabi-Yau is denoted by $\Omega(z^i)$, and it encodes the dependence of the K\"ahler potential on the 
complex structure moduli.  For this K\"ahler potential to provide a good description of the moduli space geometry, and in particular for the $\alpha'$ corrections to remain under control, we will restrict ourselves to the large volume regime, $\cV\to\infty$.

The couplings of the theory are conveniently expressed by specifying a symplectic basis of three cycles of the 
Calabi-Yau $\{A^I,B_I\}$, with $I=0, \ldots, h^{2,1}$, and a dual basis of three-forms $\alpha^I$ and $\beta_I$ 
such that
\be
\int_{A^I} \alpha_J = \delta_J^I \qquad \int_{B_I} \beta^J = - \delta_I^J, \qquad \int_{M_3} \alpha_I \wedge \beta^J = \delta_I^J, \qquad \int_{A^I} \beta^J = \int_{B_I} \alpha_J =0\,.
\label{eq:symplecticBasis}
\ee
When $\Omega$ is expressed in this basis, it reads 
\be
\Omega = X^I \alpha_I - \cF_I \beta^I, \qquad \text{with}\qquad X^I = \int_{A^I} \Omega, \qquad \cF_I = \int_{B_I} \Omega\,.
\ee

The $X^I$ are projective coordinates in the complex structure moduli space, and the corresponding moduli fields can 
be defined to be $z^i \equiv -\rmi X^i/X^0$, $i = 1, \ldots, h^{2,1}$. In order to find a more convenient expression for 
the K\"ahler potential, the quantities $X^I$ and $\cF_I$ are grouped in a symplectic \emph{period vector} $\Pi^T = (X^I, \cF_I)$. Then, 
it is possible to write the K\"ahler potential of the complex structure moduli space $K_{cs}$ as
\be
\rme^{- \, K_{cs }} =\rmi \int_{M_3} \Omega \wedge \bar \Omega = -\rmi (X^I \bar \cF_I -\bar X^I \cF_I ) =\rmi \, \Pi^\dag \cdot \Sigma\cdot \Pi\,,
\ee
where $\Sigma$ is the symplectic matrix
\be
\Sigma = \begin{pmatrix}
0 & \unity\\
-\unity & 0
\end{pmatrix}.
\label{symInvMat}
\ee
The previous expression is invariant under transformations $\Sp(2h^{2,1} +2,\mathbb{Z})$ 
associated with different choices for the symplectic basis \eqref{eq:symplecticBasis}. These symplectic 
transformations act on the period vector as follows 
\be
\Pi \longrightarrow \cS \cdot \Pi \qquad \text{where} \qquad \cS^T \cdot \Sigma \cdot \cS = \Sigma.
\label{eq:symplectitTrans}
\ee
The quantities $\cF_I$ can be expressed as the derivatives of a holomorphic 
function of the $X^I$, \emph{the prepotential}, so that
\be
\cF_I (X) = \pd_I \cF(X).
\ee
The prepotential is a homogeneous function of degree 2, i.e., $\cF(\lambda X) = \lambda^2 \cF(X)$, and therefore it satisfies
\be
X^I \cF_I = 2 \cF(X).
\ee
Setting the gauge $X^0 =1$, and using the homogeneity of the prepotential, the period vector can be written as 
\be
\Pi(z^i) = 
\left(
\begin{array}{c}
1 \\
\rmi z^i \\
2 \cF- z^j \cF_j \\
-\rmi \cF_i
 \end{array}
\right) \, .
\label{eq:Period}
\ee
In the present paper we will consider compactifications in the LCS regime, where the prepotential $\cF(z^i)$ admits the expansion 
\be
\cF = \frac{\rmi}{6} \kappa_{ijk} z^i z^j z^k + \frac{1}{2} \kappa_{ij} z^i z^j +\rmi \kappa_i z^i + \frac{1}{2} \kappa_0 + \cF_{\text{inst}} \, .
\label{eq:F}
\ee
The terms $\kappa_{ijk}$, $\kappa_{ij}$ and $\kappa_i$ are numerical constants which can be computed from the 
topological data of the mirror manifold to $M_3$. In particular, for historical reasons the coefficients $\kappa_{ijk}$ are 
often referred to as the classical \emph{Yukawa couplings}. The constant contribution $\kappa_0$ originates from 
radiative $\alpha'$ corrections, and is determined by the Euler number $\chi(M_3)= 2(h^{1,1}-h^{2,1})$ of the Calabi-Yau:
\be
\kappa_0 = \rmi \frac{\zeta(3)}{(2 \pi)^3} \chi(M_3)\,,
\label{eq:kappa0}
\ee
where $\zeta$ is the Riemann zeta function. Finally, $\cF_{\text{inst}}$ denotes exponentially suppressed string worldsheet 
instanton contributions, which can be expressed as 
\be
\cF_{\text{inst}} = - \frac{\rmi}{(2 \pi)^3} \sum_{\vec d} n_{\vec d} \; \mathrm{Li_3}[\rme^{-2 \pi d_i z^i}]\,.
\ee

Here the integers $n_{\vec d}$ are the genus zero Gopakumar-Vafa invariants, which are labeled by the vector 
$d^i\in \mathbb{Z}^+$, and the function $\mathrm{Li}_3(q)$ is the polylogarithm $\mathrm{Li}_p(q) = \sum_{k>0} \frac{q^k}{k^p}$ \cite{Cicoli:2013cha}. In the 
LCS regime, the contribution to $\cF$ from instantons is subleading, and in the following calculations we will neglect it entirely. 

When the K\"ahler potential is written in terms of the prepotential, provided we discard the instanton contribution, it takes the simple form
\be
K_{cs} = - \log\left( \frac{1}{6} \kappa_{ijk} (z + \bar z)^i (z + \bar z)^j (z +\bar z)^k - 2 \Im (\kappa_0) \right)\,. 
\label{eq:CSkahler}
\ee

It is straightforward to check that the field space metric
derived from the K\"ahler potential $K$ in \eqref{eq:KahlerPotential} is real and block-diagonal in the 
axio-dilaton and complex structure sectors, namely,
\bea
K_{\tau \bar \tau} &=& \frac{1}{2 (\Im \tau)^2}\,,\nonumber \\
K_{i\bar j} &=& - \mathring{\kappa}_{ijk} (z + \bar z)^k + \frac14 \mathring \kappa_{ilm} \mathring \kappa_{jnp} (z +\bar z)^l (z + \bar z)^m (z + \bar z)^n (z+\bar z)^p\,,
\label{eq:KahlerMetric}
\eea
where subscripts denote the derivatives of the K\"ahler functions, i.e., $K_{\tau \bar \tau} \equiv \pd_{\tau} \pd_{\bar \tau}K$ and $K_{i \bar j} \equiv\pd_i \pd_{\bar j}K$, and the quantities $\mathring \kappa_{ijk}\equiv \rme^{K_{cs}} \kappa_{ijk}$ are usually called the \emph{rescaled Yukawa couplings} \cite{Candelas:1990rm}.

\subsection{No-scale flux vacua}

The presence of three-form fluxes induces the following superpotential for the dilaton and complex structure moduli \cite{Gukov:1999ya}:
\be
W= \frac{1}{ \ell_s^2\, \sqrt{4 \pi}}\int_{M_3} G_{(3)} \wedge \Omega\,,
\ee
where $G_{(3)} = F_{(3)} - \tau H_{(3)}$, denoting by $F_{(3)}$ and $H_{(3)}$ the RR and NS-NS 3-form field 
strengths respectively. These fluxes satisfy the quantization conditions
\bea
\frac{1}{\ell_s^2} \int_{A^I} F_{(3)} = - f_A^I\in \mathbb{Z}\,, &\qquad& \frac{1}{\ell_s^2} \int_{B_I} F_{(3)} = - f_I^B \in \mathbb{Z}\,,\nonumber \\
\frac{1}{\ell_s^2} \int_{A^I} H_{(3)} = - h_A^I\in \mathbb{Z}\,, &\qquad& \frac{1}{\ell_s^2} \int_{B_I} H_{(3)} = - h_I^B\in \mathbb{Z}\,.
\label{eq:flux_defs}
\eea
Here the minus signs in all expressions have been introduced for convenience. Then, these fluxes can be 
decomposed in the symplectic basis as
\be
 F_{(3)} = - \ell_s^2 \, (f_{A}^I \alpha_I - f_I^B \beta^I),\qquad \qquad 
 H_{(3)} = - \ell_s^2\, (h_{A}^I \alpha_I - h_I^B \beta^I). 
\ee
If we define the symplectic flux vectors $f^T =(f^I_A,f_I^B)$, $h^T =(h^I_A,h_I^B)$, and $N=f - \tau\, h$, we 
can write the flux superpotential in a compact way as 
\be
W= \frac{1}{ \sqrt{4 \pi}} \left[(f_A^I- \tau h_A^I) \cF_I - (f_I^B - \tau h_I^B ) X^I\right] 
= \frac{1}{ \sqrt{4 \pi}}\, N^T \cdot \Sigma\cdot \Pi\,.
\label{fluxW}
\ee
At tree-level, the K\"ahler sector satisfies the no-scale property $K^{\rho \bar \sigma}K_\rho K_{\bar \sigma}=3$, and 
therefore the scalar potential of the effective supergravity action reads\footnote{We denote by $K^{\rho \bar \sigma}= (K_{\rho \bar \sigma})^{-1}$, $K^{\tau \bar \tau}=(K_{\tau \bar \tau})^{-1}$ and $K^{i\bar j}=(K_{i\bar j})^{-1}$ the inverses of the field space metrics on the K\"ahler, axio-dilaton and complex structure sectors respectively. }
\be
V_{\text{tree}} = \rme^{K} \, [K^{i \bar j} D_i W D_{\bar j}\bar W+ K^{\tau \bar \tau} D_\tau W D_{\bar \tau}\bar W ]\ge0\,,
\label{eq:treeLevelV}
\ee
where $D_{\tau} W= (\pd_\tau + K_\tau) W$ and $D_i W = (\pd_i + K_i)W$ are K\"ahler covariant derivatives of the superpotential.
In this work we will only consider critical points, denoted by $\{\tau_c,z_c^i\}$, of the no-scale potential 
where the axio-dilaton/complex structure sector configuration 
preserves supersymmetry, namely those satisfying 
\be
D_\tau W|_{\tau_c,z_c^i}= 0,\qquad \text{and} \qquad D_i W|_{\tau_c,z_c^i}= 0 \quad \text{for all} \quad i= 1 \ldots, h^{2,1}\,. 
\label{eq:susyeqs}
\ee
Note, however, that in general supersymmetry is still broken by the K\"ahler sector, since $D_\rho W = K_\rho W \neq0$ unless the expectation value of the flux superpotential vanishes, $W|_{\tau_c,z_c^i}=0$. In what follows, field configurations satisfying \eqref{eq:susyeqs} will be referred to as \emph{no-scale vacua}.

The allowed values of fluxes are subject to the tadpole cancellation condition which requires that the 
$D3$-brane charge induced by the fluxes, together with the contribution from $D3$-branes, cancels 
the negative charge from  $D7-$branes and  orientifold planes. The charge induced by the fluxes is given by (see \cite{Douglas:2006es,Dimofte:2008jg})
\be
N_{\text{flux}} \equiv \frac{1}{\ell_s^4}\int_{M_3} F_{(3)} \wedge H_{(3)} = (f^{B}_I h_{A}^I - h^B_I f_{A}^I)= h^T \cdot \Sigma \cdot f = \frac{N^\dag\cdot \Sigma\cdot N}{\tau- \bar \tau}\,,
\label{eq:tadpole_def}
\ee
and then, denoting by $L$ the negative contribution from the $D7$'s and the orientifolds,  we have the bound
 \be
N_{\text{flux}} \le N_{\text{flux}} + N_{D3} = L\,, 
\label{eq:tadpole1}
\ee
where $N_{D3}\ge0$ is the number of $D3$-branes.

Note that the expressions for the flux superpotential \eqref{fluxW} and the previous one for the $D3$-charge 
are both manifestly invariant under the action of the symplectic group $\Sp(2h^{2,1} +2,\mathbb{Z})$, provided the flux vector also transforms as
\be
N \longrightarrow \cS \cdot N, \qquad \cS \in \Sp(2h^{2,1} +2,\mathbb{Z}).
\label{eq:fluxTransform}
\ee
Actually, the combined actions \eqref{eq:symplectitTrans} and \eqref{eq:fluxTransform} represent redundancies 
of the supergravity description, and therefore no-scale solutions related by these transformations should be 
regarded as equivalent. In addition, the previous characterisation of flux vacua is also invariant under $\mathrm{SL}(2,\mathbb{Z})$ 
transformations acting simultaneously on the axio-dilaton $\tau$ and the fluxes as
\be
\begin{array}{c}
\tau \rightarrow \dfrac{a \tau + b}{c \tau + d}\,,\qquad \qquad 
\begin{pmatrix}
F_3 \\ H_3
\end{pmatrix}
\rightarrow
\begin{pmatrix}
a & b \\ c & d
\end{pmatrix}
\cdot
\begin{pmatrix}
F_3 \\ H_3
\end{pmatrix}
\end{array},
\label{eq:SL2Z}
\ee
with $a,b,c,d \in \mathbb{Z}$ and $ad - bc = 1$. As in the case of symplectic transformations, these actions should 
also be regarded as redundancies, thus, different no-scale vacua connected by them represent the same physical state.\\

As a final remark, it is important to emphasize that the no-scale structure leading to the potential \eqref{eq:treeLevelV} is broken by $\alpha'$  and  non-perturbative effects \cite{Becker:2002nn,Balasubramanian:2004uy,Balasubramanian:2005zx,Kachru:2003aw,Conlon:2005ki,Anguelova:2010ed}. However, provided these corrections remain under control, they will only induce subleading contributions to the mass spectra on the axio-dilaton/complex structure sector that we compute below\footnote{Here, following \cite{Kachru:2003aw,Balasubramanian:2005zx,Conlon:2005ki,Balasubramanian:2004uy} we assume that no-scale configurations \eqref{eq:susyeqs}  represent a good classical background for the computation of quantum corrections in string theory, including the case when $W|_{\tau_c,z_c^i}\neq0$. For a criticism  of this approach see \cite{Sethi:2017phn} (see also \cite{Kachru:2018aqn}).}.  Regarding the K\"ahler moduli, the leading  $\alpha'$ corrections  generically induce a run-away direction for the compactification volume, i.e. when $W|_{\tau_c,z_c^i}\neq0$. Therefore, to have a fully stabilised vacuum would require including further ingredients, such as non-perturbative corrections. However,  the stabilisation of the K\"ahler moduli is out of the scope of the present work.

\subsection{Mass spectrum at tree-level vacua}
\label{sec:spectra}

The main focus of the present work is the study of the mass spectrum at no-scale vacua, 
$\{\tau_c, z^i_c\}$, satisfying \eqref{eq:susyeqs}. In this subsection we will enumerate the relevant 
properties of the Hessian of the potential \eqref{eq:treeLevelV} at these points and its spectrum of 
eigenvalues. This information will in turn determine the tree-level masses of the moduli fields. 

At no-scale vacua the scalar potential vanishes identically, regardless of the configuration of the K\"ahler moduli, 
as a consequence, the K\"ahler moduli remain flat directions of $V_{\text{tree}}$. This means that to study
the spectrum of excitations of these configurations, it is sufficient to focus on the axio-dilaton/complex structure sector, 
since all the K\"ahler moduli are massless.
Additionally, in order to simplify the computations, we will make use of the freedom to perform a field redefinition 
to bring the field space metric to a canonical form at the vacuum $\{\tau_c, z_c^i\}$. To be more specific, since 
the K\"ahler metric \eqref{eq:KahlerMetric} is real and block-diagonal in the axio-dilaton and complex structure 
sectors, we can redefine the complex structure fields as 
$z^a = e^a_i z^i$ with 
$e_i^a \in \mathrm{GL}(h^{2,1},\mathbb{R})$, 
so that
\be
(e^{-1})^i_a (e^{-1})^j_b \, K_{i\bar j}|_{\tau_c z_c^i} = \delta_{ab}
\label{eq:canonicalFields}
\ee
with $a, b = 1,\ldots, h^{2,1}$. Then, the matrices $e^i_a \equiv (e^{-1})^i_a$ can 
be identified with a \emph{real} vielbein basis for the metric $K_{i\bar j}$ at the point $\{\tau_c, z_c^i\}$. Note that 
this does not completely fix the freedom to choose a matrix $e^i_a$, as we are still allowed to make field 
redefinitions $z^a \to \Lambda^a_b z^b$ (equivalently $e^a_i\to \Lambda^a_b e^b_i$) preserving the 
canonical form of the metric, that is with $\Lambda\in \SO(h^{2,1})$. Similarly, we can use the real 
vielbein $e^\tau_0 = \rmi (\tau-\bar \tau)$ to obtain the canonical normalisation of the axio-dilaton at the 
vacuum $\{\tau_c, z_c^i\}$. For convenience we will also use the index $A = 0,\ldots, h^{2,1}$ to collectively label 
the canonically normalized axio-dilaton and the complex structure fields, so that the full K\"ahler metric in 
the axio-dilaton/complex structure sector takes the form $K_{A\bar B} = \delta_{A\bar B}$ at the no-scale vacuum.

After bringing the field-space metric to a canonical form, it is straightforward to check that the Hessian of the scalar 
potential \eqref{eq:treeLevelV} at no-scale vacua $\{\tau_c, z_c^i\}$ has the following structure\footnote{Indices are here 
raised and lowered with the canonical form of the metric $\delta^{A \bar B}$ and $\delta_{A \bar B}$.} 
\be
\cH \equiv \begin{pmatrix}
\del_A \del_{\bar B} V &\del_A \del_{ B} V\\
 \del_{\bar A} \del_{\bar B} V & \del_{\bar A} \del_{B} V
\end{pmatrix} = \begin{pmatrix}
Z_{AC} \bar Z^C{}_{\bar B} + \delta_{A \bar B} \, m_{3/2}^2& 2 m_{3/2} \, Z_{AB}\, \rme^{-\rmi \alpha_W}\\
2 m_{3/2} \, \bar Z_{\bar A\bar B}\, \rme^{\rmi \alpha_W} &\bar Z_{\bar A \bar C} Z^{\bar C}{}_{B} + \delta_{\bar A B}\, m_{3/2}^2
\end{pmatrix},
\label{eq:Hessian}
\ee
where $m_{3/2}\equiv\rme^{K/2}|W|$ is the gravitino mass, $\alpha_W = \arg(W)$ is the phase of the flux 
superpotential and $Z_{AB} \equiv \rme^{K/2} D_A D_B W ~$. Equivalently, we can rewrite the Hessian as
\be
\cH= \Big( m_{3/2}\, \unity + \cM\Big)^2 \qquad \text{with} \qquad \cM \equiv \begin{pmatrix}
0& Z_{AB}\, \rme^{-\rmi \alpha_W}\\
\bar Z_{\bar A \bar B}\, \rme^{\rmi \alpha_W}& 0
\end{pmatrix}.
\label{eq:FMM}
\ee
Since the field space metric is already in a canonical form, the eigenvalues of the matrix $\cH$ can be identified 
with the squared masses of the $2(h^{2,1}+1)$ real scalar fields in the axio-dilaton/complex structure sector 
at $\{\tau_c, z^i_c\}$. Therefore to find the spectrum of $\cH$ it suffices to diagonalize the matrix $\cM$, which 
can be identified with the fermion mass matrix (see, e.g., \cite{freedman2012supergravity}). Moreover, note that the eigenvalues 
of $\cM$
come in pairs of opposite signs $\pm m_\lambda$, and therefore the mass spectrum of the scalar sector at tree-level is 
simply \cite{Sousa:2014qza}
\be
\mu_{\pm \lambda}^2 =( m_{3/2} \pm m_\lambda)^2\ge0\,,\label{susyMAsses}
\ee
where $\lambda =0,\ldots, h^{2,1}$. The positivity of the masses squared $\mu_{\pm\lambda}^2$ ensures that all 
no-scale vacua are perturbatively stable, which could have been anticipated by noting that the tree-level 
potential \eqref{eq:treeLevelV} is always non-negative, and vanishes at no-scale vacua.
 
In practice, the simplest way to find the fermion masses $m_\lambda$, and thus also the scalar mass spectrum, is to consider 
the $(h^{2,1} + 1) \times(h^{2,1} + 1)$ hermitian matrix $(Z Z^\dag )_{AB} \equiv Z_{AC} \bar Z^C{}_{\bar B}$, whose $h^{2,1}+1$ 
eigenvalues $m_\lambda^2$ coincide with those of
\be
\cM^2 = \begin{pmatrix}
Z_{AC} \bar Z^C{}_{\bar B}& 0\\
0& \bar Z_{\bar A}{}^C Z_{CB}
\end{pmatrix}.
\label{eq:sqFMM}
\ee

Regarding the structure of the matrix $Z_{AB}$, it is straightforward to prove that, at no-scale vacua, we always have 
$Z_{0 0}= \rme^{K/2} (e^\tau_0)^2\, D_\tau D_\tau W=0$. Moreover, when our model is defined in terms of 
a prepotential as in \eqref{eq:F}, we can simplify the computations with the identity \cite{Candelas:1990pi,Denef:2004ze}
\be
Z_{ij} = - (\tau - \bar \tau) \rme^{K_{cs}} \, \kappa_{ijk}\, K^{k \bar l} \, \bar Z_{\bar \tau \bar l},
\label{Zidentity}
\ee
which we have written in a form invariant under redefinitions of the $z^i$ fields to ease comparison with previous 
works. If we instead use canonically normalised fields, 
plus the definition of the rescaled Yukawa couplings $\mathring \kappa_{abc} = \rme^{ K_{cs}} \kappa_{abc}$, the
previous identity takes the simpler form
\be
 Z_{ab} = \rmi \, \mathring \kappa_{abc} \, \bar Z_{ 0 c}\,.
\label{Zidentity2}
\ee 
For later reference we will also collect here the following form of the tadpole constraint \eqref{eq:tadpole1} which, at no-scale vacua, 
can be expressed in terms of the expectation value of the gravitino mass and the quantities $Z_{0a}$ as (see appendix \ref{app:HodgeDecomp})
\be
0 \le 4 \pi \cV^2 \, \left(m_{3/2}^2 + |Z_{0a}|^2 \right) = N_{\text{flux}}\le L.
\label{tadpoleBound}
\ee

To summarise, the scalar mass spectrum $\mu^2_{\pm \lambda}$ at no-scale vacua \eqref{susyMAsses} can be computed 
from the gravitino mass $m_{3/2}$, the quantities $Z_{0a}$, and the canonically normalised and rescaled Yukawas $\mathring \kappa_{abc}$, 
using the formulae \eqref{Zidentity2} and diagonalising the matrix $Z Z^\dag$. In the next section we will discuss compactifications on Calabi-Yau 
manifolds invariant under a group of discrete symmetries. As we shall see, at no-scale vacua preserving those symmetries, the structure 
of both the Yukawa couplings and $Z_{0a}$ is severely constrained.

\section{Flux vacua with enhanced symmetries}
\label{sec:ESV}

In this section we will consider the special case where the Calabi-Yau geometry 
is invariant under a global group of discrete isometries. As discussed in \cite{Giryavets:2003vd}, provided that only fluxes which 
are invariant under these symmetries are turned on, the low energy action is consistent with the \emph{supersymmetric truncation} 
of a subset of the complex structure fields. Indeed, in this setting the spacetime isometries act non-trivially on the complex structure fields, 
while leaving the low energy supergravity action invariant. Then, the consistent truncation of the theory is defined by 
restricting the complex structure moduli space to the fixed locus of this symmetry, in other words, a subset of the fields 
is \emph{frozen} at the fixed locus. The consistency of the truncation ensures that any solution of the reduced theory 
obtained after freezing a subset of the fields is also a solution of the complete theory. In particular, critical points of the 
reduced scalar potential are also critical points in the full effective theory. Moreover, 
if the fields surviving the truncation are stabilized at a supersymmetric critical point, 
the full complex structure sector also preserves supersymmetry \cite{Giryavets:2003vd} (see also discussion in \cite{Louis:2012nb}). 
 
In the next paragraphs we will review how the presence of discrete symmetries in the Calabi-Yau geometry can be used to truncate 
a sector of the complex structure fields. We will also discuss the restrictions that these symmetries impose on the 
couplings of the resulting reduced theory.

\subsection{Invariant fluxes and low energy symmetries}
\label{sub5a}

As we mentioned in the introduction, in many interesting compactifications the Calabi-Yau geometry 
is invariant under the action of a discrete group of transformations, $\cG$. 
These transformations act on the complex structure fields, $z^i\to \hat z^i$, and thus also induce a change on the period 
vector $\Pi(z^i)$. Since the Calabi-Yau geometry is left invariant under these symmetries, these transformations 
must also leave the geometry on its moduli space invariant. Therefore, the action of a transformation $g \in \cG$
on the period vector must be of the form
\be
\Pi (z^i) \longrightarrow \Pi (\hat z^i) = \rme^{\Lambda_g(z)}\, \cS_g \cdot \Pi(z^i)\,,
\ee
with $\Lambda_g(z^a)$ a holomorphic function of the complex structure fields and $\cS_g$ a constant symplectic 
matrix in $\Sp(2h^{2,1}+2, \mathbb{Z})$, both determined by the group element $g$. 
In addition, when the three-form fluxes are turned on, the invariance of the effective action under the group $\cG$ requires that the flux vector $N=f-\tau h$ transforms as 
in \eqref{eq:fluxTransform}.
Then, it is easy to check that under a transformation $g \in\cG$, the K\"ahler potential $K_{cs}$ and 
the superpotential $W$ experience a $g$-dependent K\"ahler transformation
\be
K_{cs}(\hat z^i{}, \hat {\bar z}^i{}) = K_{cs}(z^i, \bar z^i) + \Lambda_g(z^i) + \bar \Lambda_g(\bar z^i)\,, \qquad W_{\hat f,\hat h}(\hat z^i) = \rme^{-\Lambda_g(z)} \, W_{f,h}(z^i)\,.
\ee
Here  we have explicitly indicated for clarity the dependence of the superpotential on the flux vectors $(f,h)$ and their transformed values   $(\hat f,\hat h)$ under \eqref{eq:fluxTransform}. However, the symmetry groups $\cG$ that we are considering are discrete and of finite order, and
thus it is always possible to choose a K\"ahler gauge so that $K_{cs}$ and $W$ transform as scalars 
under\footnote{The invariant K\"ahler gauge $K^{\text{inv}}_{cs} = K_{cs} + \Lambda^{\text{inv}}(z) + \bar \Lambda^{\text{inv}}(\bar z)$ 
can be found noting that under a transformation $g: z^i \to \hat z^i$ we must have $\Lambda^{\text{inv}}(\hat z) =\Lambda^{\text{inv}}(z) - \Lambda_g(z)$. It is 
easy to check that this condition is solved by $\Lambda^{\text{inv}}(z) = \frac{1}{[\cG]}\sum_{g\in \cG} \Lambda_g(z)$, where $[\cG]$ is the 
order of the group $\cG$. } $\cG$ (see \cite{Gates:1983nr}), that is,
\be
K_{cs}(\hat z^i, \hat{ \bar z}^i{}) = K_{cs}(z^i, \bar z^i)\,,\quad W_{\hat f,\hat h}(\hat z^i{}) = W_{f,h}(z^i)\,.
\label{scalarKW}
\ee

It is important to note that despite of the behaviour \eqref{scalarKW} of the K\"ahler potential and the flux 
superpotential under the group of transformations $\cG$, generically they do not constitute 
a proper symmetry of the low energy effective theory for the moduli fields \cite{DeWolfe:2004ns}.
Indeed, each choice of fluxes defines an effective theory for the moduli, where the flux integers $(f,h)$ appear as coupling 
constants (see \cite{D'Auria:2005yg}). Therefore, since the group $\cG$ generally acts non-trivially on the fluxes, i.e., the 
couplings of the EFT, in general it will not correspond to a low energy symmetry for the moduli effective action. On the contrary, if we 
restrict the flux configuration $N= f-\tau h$ to be invariant under the transformations \eqref{eq:fluxTransform}, then $\cG$ 
will be a symmetry of the low-energy action defined by this choice of fluxes. Indeed, from \eqref{scalarKW} we have that for an invariant set of fluxes
\be
K_{cs}(\hat z^i, \hat {\bar z}^i) = K_{cs}(z^i, \bar z^i)\,, \quad W_{f,h}(\hat z^i) = W_{f,h}(z^i)\,,
\label{eq:KWscalars}
\ee
so the low energy supergravity theory of the moduli is properly invariant under the action of $\cG$.

In the following we will assume that the fluxes are invariant under the action of $\cG$, and we will again omit
the subscripts $(f,h)$ in the superpotential in order to simplify the notation.

\subsection{Consistent truncation of the moduli space}
\label{sub5b}

We will now discuss how the symmetry group $\cG$ aids in the task of finding solutions to the 
no-scale equations \eqref{eq:susyeqs}. In general, for a given group $\cG$ we can always split the complex structure 
fields into two sets, $z^i =\{ z^{\alpha}, w^{\alpha'}\}$: those invariant under the action of the symmetry group, $z^{\alpha}$ with $\alpha=1,\ldots,h_{\text{red}}^{2,1}$, 
and those fields which transform non trivially, $w^{\alpha'} \to \hat w^{\alpha'}$, where $\alpha' = h_{\text{red}}^{2,1}+1,\ldots, h^{2,1}$. 

Then, if the symmetry group $\cG$ admits a fixed locus \emph{on the moduli space}, i.e., a configuration of 
the fields $w_*^{\alpha'}$ satisfying $\hat w^{\alpha'}_*= w^{\alpha'}_*$, the derivatives of the scalar potential 
$V$ and the K\"ahler potential along the non-invariant fields $w^{\alpha'}$ must vanish there
\be
\pd_{w^{\alpha'}} V =0\,, \quad K_{w^{\alpha'}} =0\quad 
\text{at} \quad z^i=(z^{\alpha}, w^{\alpha'}_{*}) \quad \text{for all} \quad \tau, z^{\alpha}\,.
\label{truncatedVK}
\ee
To prove this it is sufficient to note that, for an invariant choice of fluxes, both the no-scale potential
$V$ and the K\"ahler potential transform as scalar fields under the action of $\cG$, and thus $\pd_{z^i} V$ and $K_{i}$ will transform as tensors. Then, in the case of the scalar potential, we have that a generic point of the moduli space satisfies
\be
\pd_{w^{\alpha'}} V(z^{\alpha},w^{\alpha'}) =\frac{\pd \hat w^{\beta'}}{\pd w^{\alpha'}}\, \pd_{\hat w^{\beta'}} V(z^{\alpha}, \hat w^{\beta'})\,.
\ee 
At the fixed locus, where $\hat w^{\alpha'}_*= w^{\alpha'}_*$, the previous expression can be seen as a system of 
equations for $\pd_{w^{\alpha'}} V(z^{\alpha}, w^{\alpha'}_*)$. But this system only admits the trivial solution 
\eqref{truncatedVK} because, by assumption, all the fields $w^{\alpha'}$ transform non-trivially away from the 
fixed point implying that all equations are independent. Moreover, in the previous discussion the expectation 
values of the dilaton or the $\cG$-invariant fields are irrelevant, and therefore the fixed point will always be a 
stationary point of the superpotential regardless of the field configuration $(\tau,z^{\alpha})$. Although our 
argument has been derived in the particular K\"ahler gauge where $K_{cs}$ and $W$ transform 
as scalars \eqref{scalarKW}, our conclusion is a K\"ahler invariant statement. Different derivations can be 
found in \cite{Giryavets:2003vd,Louis:2012nb,Cicoli:2013cha}. 
 
The left condition in \eqref{truncatedVK} implies that the fixed locus of the symmetry group $\cG$ is always a 
critical point of the scalar potential, while the right one leads to a consistency condition on 
the geometry of the moduli space. In particular, this geometric condition implies that the moduli space metric 
on the fixed locus is block-diagonal in the truncated and surviving sectors $K_{z^{\alpha} \bar w^{\beta'}}=0$. Moreover, 
the reduced moduli space defined by the fixed locus $w^{\alpha'} = w^{\alpha'}_*$ must be a \emph{totally geodesic submanifold} 
of the full moduli space (see \cite{Sousa:2014qza}). In other words, any geodesic on the moduli space manifold 
with at least one point located at the fixed locus of $\cG$, and which is locally tangent to it, should be entirely contained in the reduced moduli space.
 
These are very strong requirements which ensure the consistency of freezing the moduli $w^{\alpha'}$ 
at the level of the EFT Lagrangian $\cL_{EFT}(\tau,z^i)$, thus defining a reduced theory involving the surviving fields alone:
\be
\cL^\text{red}_{EFT} (\tau, z^{\alpha}) \equiv \cL_{EFT} (\tau, z^{\alpha}, w^{\alpha'} =w^{\alpha'}_*)\,.
\ee
Indeed, the conditions \eqref{truncatedVK} guarantee that any solution of the reduced theory given by $\cL^\text{red}_{EFT}$ is also a solution of the complete EFT. Moreover, using \eqref{eq:KWscalars} and a similar argument to the one given above, it is possible to prove that the flux potential is also extremized at the fixed locus of $\cG$
\be
D_{w^{\alpha'}} W|_{w = w_*} = \pd_{w^{\alpha'}} W|_{w = w_*} =0\qquad \text{for all} \qquad \tau, z^{\alpha}\,,
\label{eq:truncatedW}
\ee 
which means that the truncated fields $w^{\alpha'}$ preserve supersymmetry there. If supersymmetry 
is preserved in the reduced theory, it is also unbroken in the original EFT. Then, the process of freezing the 
non-invariant fields $w^{\alpha'}$ constitutes a consistent supersymmetric truncation of the 
theory\footnote{It is important to emphasize that by this procedure the non-invariant deformations are \emph{not projected out}, as it happens when orientifolding a Calabi-Yau or in the Green-Plesser construction of  mirror Calabi-Yau duals. Here the non-invariant fields are still degrees of freedom of the EFT. } (see \cite{Achucarro:2008fk,Achucarro:2008sy,Sousa:2014qza}).  

From \eqref{eq:truncatedW} it follows that compactifications admitting a discrete symmetry 
group are particularly convenient for the search of no-scale vacua since at the fixed locus of $\cG$, the non-invariant fields 
automatically satisfy the no-scale equations \eqref{eq:susyeqs}. No-scale vacua located at the fixed locus of the 
symmetry group $\cG$ are often called \emph{enhanced symmetry vacua}. Moreover, provided we are interested 
only in this class of vacua, the consistency of the truncation ensures that it is sufficient to calculate the 
couplings, i.e., the period vector, of the reduced action 
(see, e.g., \cite{Candelas:1990rm,Candelas:1993dm,Berglund:1993ax,Candelas:1994hw,Candelas:2000fq,Cicoli:2013cha}), 
which renders the computation of the EFT tractable.

\subsection{Mass matrix structure at enhanced symmetry vacua}
\label{sub5c}

The high degree of symmetry present in low energy theories with $\cG$-invariant fluxes provides 
valuable information regarding the structure of the fermion mass matrix and the Hessian at enhanced
symmetry vacua. First, as we saw above, at the fixed locus of $\cG$ the moduli space metric is block 
diagonal in the truncated and surviving sectors, and from \eqref{truncatedVK} and \eqref{eq:truncatedW} it also follows that 
\be
D_\tau D_{w^{\alpha'}} W|_{w_*} = D_{z^{\alpha}} D_{w^{\beta'}} W|_{w_*} = 0\,, \qquad \del_{z^{\alpha}} \del_{w^{\beta'}} V|_{ w_*} = \del_{z^{\alpha}} \del_{\bar w^{\beta'}} V|_{w_*} =0\,,
\label{eq:truncatedMH}
\ee
regardless of the configuration of the reduced moduli $(\tau,z^{\alpha})$. This, in turn, implies that the fermion 
mass matrix $\cM = \cM_{\{\tau, z^{\alpha}\}} \otimes \cM_{\{w^{\alpha'}\}}$ and the Hessian of the 
potential $\cH = \cH_{\{\tau,z^{\alpha}\}} \otimes \cH_{\{w^{\alpha'}\}}$ are block-diagonal in the two sectors at no-scale 
vacua, which means that it is consistent to study the perturbative stability of the fields $(\tau, z^{\alpha})$ 
and $w^{\alpha'}$ separately. Moreover, the particular structure of the fermion mass matrix on the EFTs 
we are considering, i.e., the identity \eqref{Zidentity}, leads to an additional simplification. From \eqref{eq:truncatedMH} it is 
easy to see that the only non-vanishing quantities $Z_{\tau z^i}$ are those with components on the surviving sector, $Z_{\tau z^{\alpha}}$.

Collecting all the previous results, and 
using \eqref{Zidentity2}, we can see that the components of the canonically normalised matrix $Z_{AB}$ which appears in $\cM$ satisfy the relations
\be
Z_{0 w^{a'}} = Z_{z^{\tilde a} w^{b'}}=0\,,\qquad Z_{z^{\tilde a} z^{\tilde b}} =\rmi \, \kappa_{\tilde a \tilde b \tilde c} \, \bar Z_{\bar 0 \bar z^{\tilde c}}\,,\qquad 
Z_{w^{a'} w^{b'}} = \rmi \, \kappa_{a' b' \tilde  c}\, \bar Z_{\bar 0 \bar z^{\tilde c}}\,,
\label{fmmBlocks-1}
\ee
where fields $z^{\tilde a}$ correspond to the canonically normalised fields of the reduced theory, $\tilde a, \tilde b = 1, \ldots,h^{2,1}_\text{red}$ (see eq. \eqref{eq:canonicalFields}), and $w^{a'}$ to those of the truncated sector, $a',b' = h^{2,1}_{\text{red}}+1,\ldots,h^{2,1}$. 

Thus, the main result of this section can be summarized as follows: at enhanced symmetry vacua the canonically 
normalized Hessian of the no-scale potential can be entirely expressed in terms of the derivatives of the flux 
superpotential \emph{of the reduced theory}, $Z_{0 z^{\tilde a}}$, plus 
 $\mathring{k}_{\tilde  a\tilde  b\tilde  c}$ and $\mathring{k}_{a' b' \tilde  c}$ of the canonically normalized invariant 
 Yukawa couplings. Furthermore, in the class of models we are interested in, the sector surviving the truncation is 
 one dimensional (see the sample in Table \ref{table:ModelSelection}), and thus the indices $\tilde a$, $\tilde b$ and $\tilde c$ in \eqref{fmmBlocks-1} can only take one 
 value, which we choose to be ``1" without loss of generality. The non-vanishing components of the matrix $Z_{AB}$ then read
\be
Z_{11} = \rmi \, \mathring \kappa_{111}\, \bar Z_{\bar 0\bar 1},\qquad 
Z_{ a' b'} = \rmi \, \mathring \kappa_{a' b' 1}\, \bar Z_{\bar 0 \bar 1}.
\label{fmmBlocks}
\ee
As we shall see in section \ref{sec:totalSpectrum}, for the class of models we discuss here, the 
quantities $\mathring \kappa_{111}$ and $\mathring \kappa_{a' b'1}$ appearing in these 
expressions can also be completely expressed in terms of the field expectation values and the 
known couplings of the reduced theory. 

\begin{table*}[t]
\setlength\extrarowheight{2pt}

\centering
\begin{tabular}{|>{$}C{0.5cm}<{$}|>{$}C{3.5cm}<{$}|>{$}C{2.1cm}<{$}|>{$}C{2.3cm}<{$}
|C{2.2cm}|}
\hline
\#&\text{Manifold} & (h^{1,1},h^{2,1})& 
\cG
 & \text{Reference} \\
\hline
\hline
 1&\mathbb{WP}_{[1,1,1,1,1]}^4& (1,101) &  \begin{minipage}{20mm}$ \mathbb{Z}_5^3$, \nolinebreak$\mathbb{Z}_{41}$, \nolinebreak $\mathbb{Z}_{51}$, \\  $\mathbb{Z}_{5}\times \mathbb{Z}_{13}$   
\end{minipage}  & 
\cite{Candelas:1990rm,Klemm:1992tx,Doran:2007jw}\\[2pt]
\hline
2&\mathbb{WP}_{[1,1,1,1,2]}^4& (1,103) & \mathbb{Z}_3\times \mathbb{Z}_6^2&
 \cite{Font:1992uk,Klemm:1992tx} \\[2pt]
\hline
3&\mathbb{WP}_{[1,1,1,1,4]}^4& (1,149) & \mathbb{Z}_8^2 \times \mathbb{Z}_2& 
 \cite{Font:1992uk,Klemm:1992tx}\\[2pt]
\hline
 4&\mathbb{WP}_{[1,1,1,2,5]}^4& (1,145) & \mathbb{Z}_{10}^2& 
 \cite{Font:1992uk,Klemm:1992tx} \\[2pt]
\hline
5&\mathbb{WP}_{[1,1,1,1,1]}^4/\mathbb{Z}_5\times \mathbb{Z}_5& (1,5) & \text{Dic}_3,\,  \text{Dic}_5&  
\cite{Candelas:2017ive}\\[2pt]
\hline
\vspace{-.35cm}
 6&\vspace{-.35cm}X^{20,20} & \vspace{-.35cm}(20,20) & \begin{minipage}{15mm}$\mathrm{SL}(2,3)$, \\  $\mathbb{Z}_3\times\mathbb{Z}_8$, \\ $\mathbb{Z}_3 \times \mathbb{Q}_8$\end{minipage} 
&
\vspace{-.35cm} \hspace{-.2cm}\cite{Braun:2011hd,Braun:2015jdy}\\[2pt]
\hline
7&X^{20,20}/\mathbb{Z}_3 & (3,3) & \mathbb{Q}_8
&
 \cite{Braun:2011hd,Braun:2015jdy}\\[2pt]
\hline
\end{tabular}
\caption{Selection of  Calabi-Yau geometries which admit a consistent supersymmetric truncation of the moduli space. The truncation is defined by the fixed locus (on the moduli space) of  the  symmetry group $\cG$, and  effectively  reduces  the number of moduli to  $(h^{1,1},h^{2,1})\to (h^{1,1}_\text{red}=1,h^{2,1}_\text{red}=1)$. In the cases $1$, $5$ and $6$, the fixed locus of each of the groups $\cG$ leads to one or more  distinct Calabi-Yau families. The manifold $X^{20,20}$ is the``24-cell" Calabi-Yau threefold with hodge numbers $h^{1,1}=h^{2,1}=20$ \cite{Braun:2011hd}, and  the symbols  $\mathbb{Q}_8$ and $\text{Dic}_n$ stand for the quaternion and dicyclic groups respectively.  }  
\label{table:ModelSelection}
\end{table*}

\section{Complete tree-level mass spectrum}
\label{sec:totalSpectrum}

We begin the present section by deriving certain universal properties of the type-IIB couplings which are 
valid in a generic Calabi-Yau compactification at LCS. We will then restrict ourselves to 
Calabi-Yau manifolds admitting a symmetry group which enables a consistent reduction of the complex structure 
moduli space to a single surviving field. Using these results together with the ones in the previous sections we will 
show how to compute the tree-level spectrum for the complete axio-dilaton/complex structure 
sector at no-scale vacua.

\subsection{Universal features of the type-IIB effective field theory}

In this subsection we will obtain general properties satisfied by the canonically normalised Yukawa couplings 
in the large complex structure regime. More specifically, we will show that a subset of the rescaled Yukawas 
$\mathring \kappa_{abc}$ can be expressed in terms of a single parameter $\xi\in [0,1/2]$, which can 
be defined in terms of known quantities appearing in the reduced theory as
\be
\xi \equiv \frac{-2 \rme^{K_{cs}} \Im \kappa_0}{1+2 \rme^{K_{cs}} \Im \kappa_0}\,.
\label{eq:LCS_param}
\ee
This quantity can be understood as a coordinate parametrising the complex structure moduli space, with the 
LCS point located at $\xi=0$. For the models we are interested in, with a few K\"ahler moduli and a large complex 
structure sector $h^{1,1}\ll h^{2,1}$, we have from \eqref{eq:kappa0} that $\Im \kappa_0<0$. Combined with the 
definition \eqref{eq:LCS_param}, the latter condition also implies that physical configurations satisfy $\xi\ge0$.
Then, it is easy to check that field configurations with $\xi=1/2$ are those at the boundary of the moduli space, that is, 
for $\xi>1/2$ the K\"ahler metric has a negative eigenvalue, leading to unphysical solutions. 

The argument below will proceed along the lines of \cite{Cremmer:1984hj,Farquet:2012cs,Marsh:2015zoa}, where analogous properties for the Yukawas  where found strictly at the LCS point ($\xi=0$). But here we will only assume that the exponentially suppressed instanton 
contributions to the prepotential \eqref{eq:F} can be entirely neglected. Therefore, the results presented below 
generalize those of \cite{Cremmer:1984hj,Farquet:2012cs,Marsh:2015zoa}, as the regime of validity of our analysis can be extended to the 
entire region of the moduli space where the polynomial approximation of the prepotential \eqref{eq:F} is under control. 

The starting point of the derivation is the K\"ahler metric \eqref{eq:KahlerMetric} on the complex structure moduli space.
Following \cite{Marsh:2015zoa} we introduce the following real vector of unit norm
\be
e_1^i \equiv \frac{1}{x} (z + \bar z)^i,\qquad K_{i\bar j}\, e^i_1 \bar e^j_1 =1,
\label{eq:vielbein}
\ee 
where the parameter $x$ is a normalisation constant which has yet to be determined. Without loss of generality, and 
making use of the residual $\SO(h^{2,1})$ freedom to define the canonically normalised fields, we rotate the vielbein basis $e^i_a$ 
so that the first vector coincides with $e_1^i$.
Since the K\"ahler metric \eqref{eq:KahlerMetric} has the canonical form $\delta_{ab}$ when expressed in the 
basis $e^i_a$, we find that the rescaled and canonically normalised Yukawa couplings should satisfy
\be 
\delta_{ab} = -\mathring \kappa_{ab1} x+ \frac14 \mathring \kappa_{a11} \mathring \kappa_{b11} x^4\,.
\ee
Note also that from the definition of the Yukawa couplings, $\mathring \kappa_{abc} =\rme^{K_{cs}} \kappa_{abc}$, and 
the expression for the K\"alher potential \eqref{eq:CSkahler}, we have
\be
\rme^{-K_{cs}} = \frac16 \kappa_{111} x^3 (1 + \xi) \qquad \Longrightarrow \qquad \frac16 \mathring \kappa_{111} x^3 (1 + \xi) = 1\,.
\ee
Solving the two previous conditions for the Yukawas of the form $\mathring \kappa_{ab1}$, it is straightforward to obtain
\be
\mathring \kappa_{111} = \frac{2(1+\xi)^2}{\sqrt{3(1 - 2\xi)^3}}\,, \qquad \mathring \kappa_{a' 11} = 0 \quad \text{and} \quad \mathring \kappa_{a' b' 1} = - \frac{1+\xi}{\sqrt{3 (1 - 2 \xi)} } \delta_{a ' b'}\,. 
\label{eq:Yukawas}
\ee
with $a',b'= 2, \ldots h^{2,1}$. The rest of the rescaled Yukawa couplings $\kappa_{a'  b' c'}$ 
are not constrained by the conditions above, and therefore a priori they can be generic. The normalisation constant $x$ 
of the vielbein $e^i_1$ is found to be
\be
x^2 = \frac{3 (1- 2 \xi)}{(1+\xi)^2}\,.
\ee
To the best of our knowledge these relations have never been presented before in the literature.
 
The direction specified by the vielbein $e_1^i$ has a concrete geometrical significance. It corresponds 
to \emph{the no-scale direction} of the complex structure moduli space \cite{Covi:2008ea,Covi:2008zu}
\be
K_a = -\frac12 \mathring \kappa_{a11} x^2 = -\sqrt{3/(1- 2 \xi)} \delta_a^1\,. 
\label{eq:noscaleDirection}
\ee
The previous relation also implies the following generalised no-scale property 
\be
K_i K_{\bar j} K^{i\bar j} = 3/(1- 2 \xi)\ge3~,
\ee
which is satisfied by any type-IIB compactification with $h^{1,1}\le h^{2,1}$ at LCS (see appendix A in \cite{Covi:2008ea}). 

Note that in the models we are interested in, where only one field survives the truncation, the v.e.v. of the complex 
structure field $z^a$ is necessarily aligned with the vector $Z_{0a}$, since both of them point along the unique 
direction of the reduced complex structure moduli space. Therefore, the Yukawa couplings $\mathring \kappa_{a' b'1}$ 
computed above are precisely those also appearing in the expression \eqref{fmmBlocks}, and thus we already have 
all the necessary ingredients we require to compute the tree-level spectrum at a generic no-scale vacuum.

\subsection{Fermion and scalar mass spectra at no-scale vacua}

\begin{figure}
    \centering
    \includegraphics[width=0.6\textwidth]{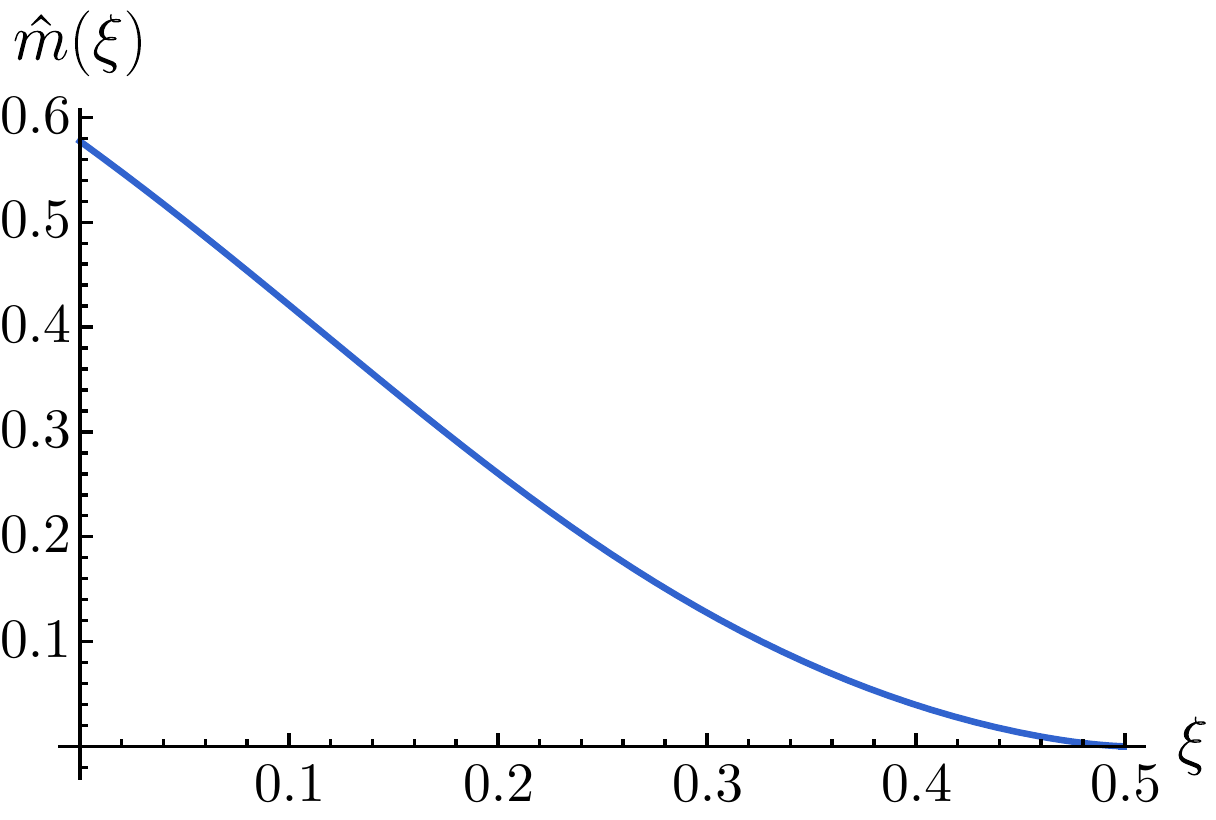}
    \caption{Plot of $\hat{m}(\xi)$, as defined in \eqref{eq:mhat}.}
    \label{fig:mhat}
\end{figure}

We begin by computing the fermion mass spectrum as described in section \eqref{sec:spectra}, that is,
diagonalising the hermitian matrix $Z Z^\dag$, and using the formula \eqref{susyMAsses} to 
obtain the masses of the scalar fields. First, since the vector $Z_{0a} = \delta_{a1} Z_{01}$ is necessarily 
aligned with the no scale direction, from the expressions \eqref{eq:Yukawas} for the rescaled Yukawa couplings, 
and the relations \eqref{fmmBlocks}, we find that matrix $Z_{AB}$ has the following structure
\be
Z_{AB} = \begin{pmatrix}
0 &Z_{01}&0\\
Z_{01} & \rmi \mathring {\kappa}(\xi) \bar Z_{01} & 0\\
0 & 0 & -\frac{1+\xi}{\sqrt{3(1-2\xi)}} \delta_{a' b'} \bar Z_{01}
\end{pmatrix}\,, 
\ee
where we used the shorthand $\mathring \kappa(\xi) \equiv \mathring \kappa_{111}(\xi)$. Then, after factorising 
an overall scale $m_{\text{susy}} \equiv |Z_{01}|=|e^{K/2} D_0 D_1 W|$, and computing 
the spectrum of eigenvalues $m_\lambda^2$ of $Z Z^\dag$, we obtain 
\be
\boxed{m_{\lambda}/m_{\text{susy}} =
\left\{
\begin{array}{l c l }
\hat m(\xi)&~~& \lambda=0\\
1/\hat m(\xi)&~~& \lambda=1\\
\frac{1+\xi}{\sqrt{3(1-2\xi)}} &~~& \lambda=2,\ldots, h^{2,1}
\end{array}
\right.}\,,
\label{eq:gralFermionSpectrum}
\ee
where we defined
\be
\hat m(\xi) \equiv \frac1{\sqrt{2}}\left(2 + \mathring \kappa(\xi)^2 - \mathring \kappa(\xi) \sqrt{4 + \mathring \kappa(\xi)^2}\right)^{1/2}\,,
\label{eq:mhat}
\ee
which is shown in figure \ref{fig:mhat}.

Interestingly, it can be seen that all the fermions on the truncated sector have the same mass. In particular, at the LCS 
point ($\xi=0$) the fermion mass spectrum reads simply 
\be
m_0/m_{\text{susy}} =\frac{1}{\sqrt{3}}, \qquad m_1/m_{\text{susy}} =\sqrt{3}, \qquad \text{and} \qquad m_{\lambda'} /m_{\text{susy}}= \frac{1}{\sqrt{3}}.
\label{eq:fermionSpectrumLCS}
\ee
with $\lambda' =2, \ldots,h^{2,1}$.

The mass spectrum of the scalar fields can be immediately obtained from equation \eqref{susyMAsses}. To write it down, it is 
convenient to introduce the angular parameter $\theta_W$ (dependent on the choice of flux) as
\be
\cos \theta_W \equiv  \frac{\cV\, m_{3/2}}{\sqrt{N_{\text{flux}}/4 \pi}}\,, \qquad \text{with} \qquad \theta_W \in [0,\pi/2]\,,
\label{eq:thetaDef}
\ee
where the range of values of $\theta_W$ follows from the tadpole constraint \eqref{tadpoleBound}. Despite appearances, the 
parameter $\theta_W$ has no dependence on the Calabi-Yau volume or the K\"aher moduli, since the 
combination $\cV m_{3/2} =\frac1{\sqrt{2}} \rme^{K_{cs}/2} \Im(\tau)^{-1/2} |W|$, often denoted by $W_0$ in 
the literature, depends solely on the axio-dilaton and complex structure fields.
Furthermore, recalling that 
$Z_{0 a'}=0$, we find from \eqref{tadpoleBound} that the total $D3$-charge induced by fluxes is simply 
\be
N_{\text{flux}} =4 \pi \cV^2 \, \left(m_{3/2}^2 + m_{\text{susy}}^2\right)\qquad \Longrightarrow\qquad \tan\theta_W =m_{\text{susy}}/m_{3/2}\,,
\label{eq:simpleTadpole}
\ee
and then it is straightforward to check that the complete set of scalar masses at tree-level in the axio-dilaton/complex structure sector is given by
\be
\boxed{\mu_{\pm\lambda}^2/m_{3/2}^2 =
\left\{
\begin{array}{l c l }
\left( 1 \pm \tan \theta_W \, \hat m(\xi)\right)^2&~~& \lambda=0\\
\left(1 \pm \frac{\tan \theta_W}{\hat m(\xi)}\right)^2&~~& \lambda=1\\
\left(1 \pm \frac{(1+\xi) \tan \theta_W }{\sqrt{3(1-2\xi)}} \right)^2 &~~& \lambda=2,\ldots, h^{2,1}
\end{array}
\right.}\,.
\label{eq:gralSpectrum}
\ee
This mass spectrum is the main result of this paper. All the parameters appearing in the previous expression 
can easily be computed in the reduced theory, as  $\xi$ is determined by the configuration of the complex 
structure fields surviving the truncation, and $\theta_W$ depends only on the expectation value of the flux 
superpotential $W_0$ and the total $D3$-charge induced by fluxes $N_{\text{flux}}$.

It is worth noticing that this result is independent of both the specific details of the compactification and the number of 
moduli fields. Moreover, these masses only depend on the choice of flux via an overall scale given by the gravitino 
mass $m_{3/2} = W_0/\cV$ and the angular parameter $\theta_W$. We have chosen to present the masses normalised 
by the gravitino mass in order to eliminate their dependence on the Calabi-Yau volume $\cV$, which appears as an 
overall multiplicative factor. 
 
An interesting case to mention is that of the KKLT scenario \cite{Kachru:2003aw}, where the consistency of the EFT 
requires that the value of $W_0\ll1$ is very close to zero, or equivalently $\theta_W \sim \pi/2$. 
In this limit the scalar spectrum simplifies to 
\be
\text{\bf KKLT scenario:} \qquad \mu_{\pm\lambda}^2/m_{3/2}^2 \approx
\left\{
\begin{array}{l c l }
\tan^2 \theta_W \, \hat m(\xi)^2&~~& \lambda=0\\
\frac{\tan^2 \theta_W}{\hat m(\xi)^2}&~~& \lambda=1\\
\frac{(1+\xi)^2 \tan^2 \theta_W }{3(1-2\xi)} &~~& \lambda=2,\ldots, h^{2,1}
\end{array}
\right.\,.
\label{eq:KKLTspectrum}
\ee
implying that all the masses in the spectrum are very large compared to the gravitino mass, $\mu^2_{\pm \lambda} \gg m_{3/2}^2$. As 
discussed in detail in \cite{Abe:2006xi,Gallego:2008qi,Gallego:2009px}, in generic situations this guarantees the consistency of neglecting the complete 
axio-dilaton/complex structure sector in KKLT constructions, even after including quantum corrections and supersymmetry 
breaking effects in the theory. However, very light modes might still appear in the spectrum when considering no-scale solutions at special points of the moduli space  \cite{Bena:2018fqc,Demirtas:2019sip}.

In the following two subsections we will discuss another two special cases where the value of the parameter $\theta_W$ is 
fixed, and thus we can write the entire (normalised) mass spectrum as a function of the parameter $\xi$ alone.

\begin{figure}[t]   
    \centering
    \subfloat[]{
        \includegraphics[width=0.47\textwidth]{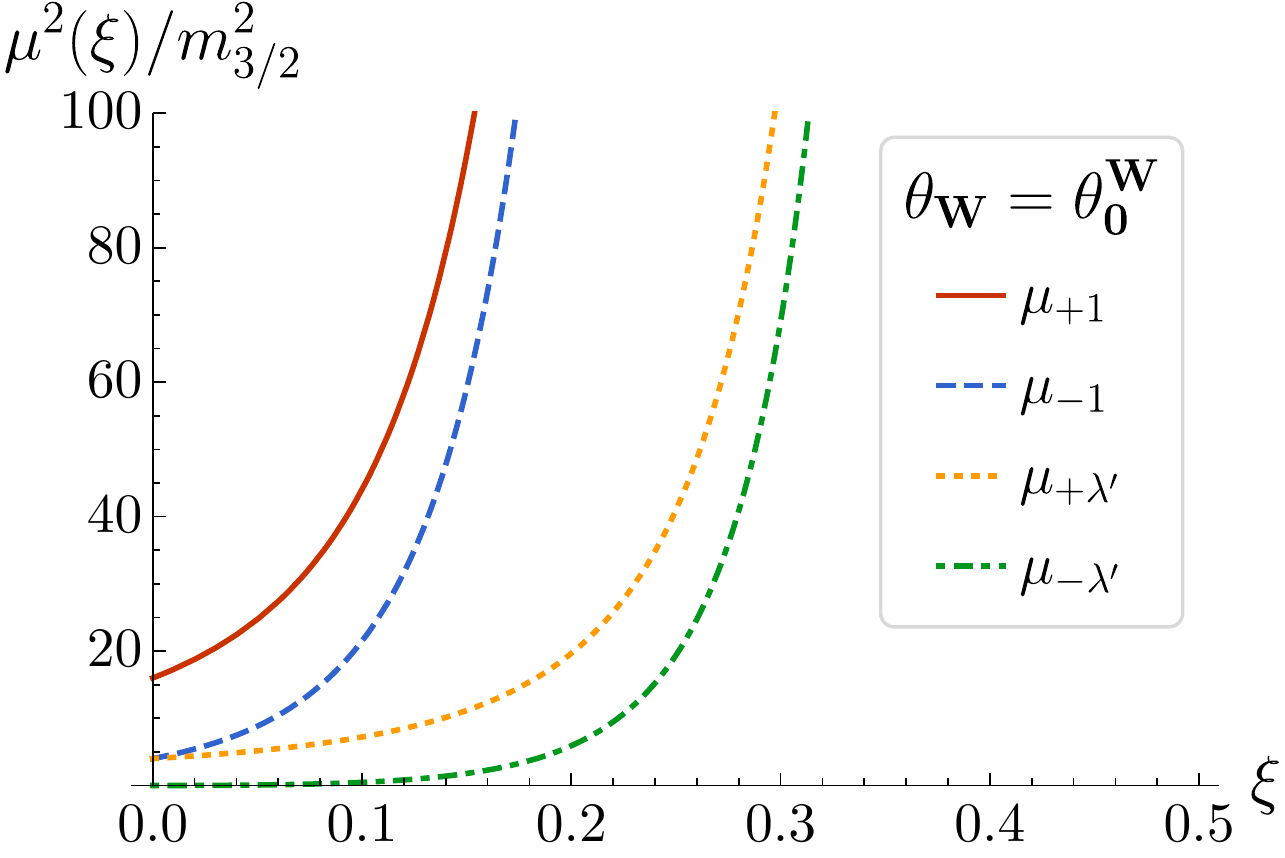}
    }
    \hfill  
    \subfloat[]{
        \includegraphics[width=0.47\textwidth]{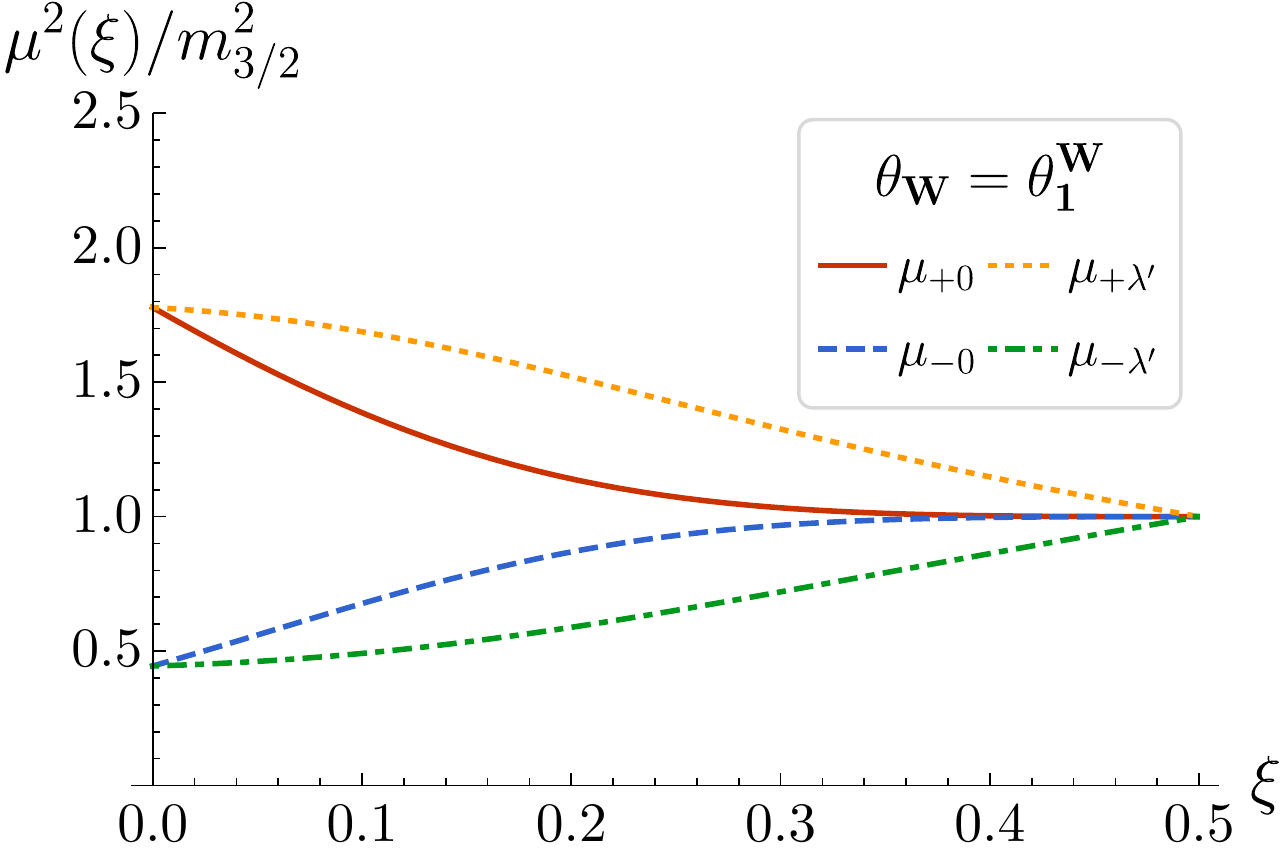}
    }
    \caption{Scalar mass spectra associated with the critical values for $\theta_W$ \eqref{eq:criticalTheta} where the mass spectrum contains at least a zero mode. (a) Spectrum for $\theta_W = \theta^W_0$ where $\mu_{-0}^2=0$ and $\mu_{+0}^2=4m_{3/2}^2$. (b) Spectrum for $\theta_W = \theta^W_1$ where $\mu_{-1}^2=0$ and $\mu_{+1}^2=4m_{3/2}^2$. }
    \label{fig:zermodeSpectra}
\end{figure} 

\subsection{Flux vacua with massless scalars}
\label{sec:masslessFields}

An important consequence of \eqref{eq:gralSpectrum} is that at no-scale vacua we might encounter spectra with very light or 
even massless scalar fields, since $\tan \theta_W \in [0,\infty)$. In general the presence of those light fields is not convenient for phenomenological applications, as such vacua might become tachyonic after including quantum corrections or due to supersymmetry  breaking  effects. However, no-scale solutions with light (or massless) modes
are still of interest for certain constructions of dS 
vacua \cite{Marsh:2014nla,Gallego:2017dvd}, and for implementing inflation. Thus, we will now 
briefly discuss the properties of their mass spectra.

\begin{figure}[t]   
\centering
        \includegraphics[width=0.7\textwidth]{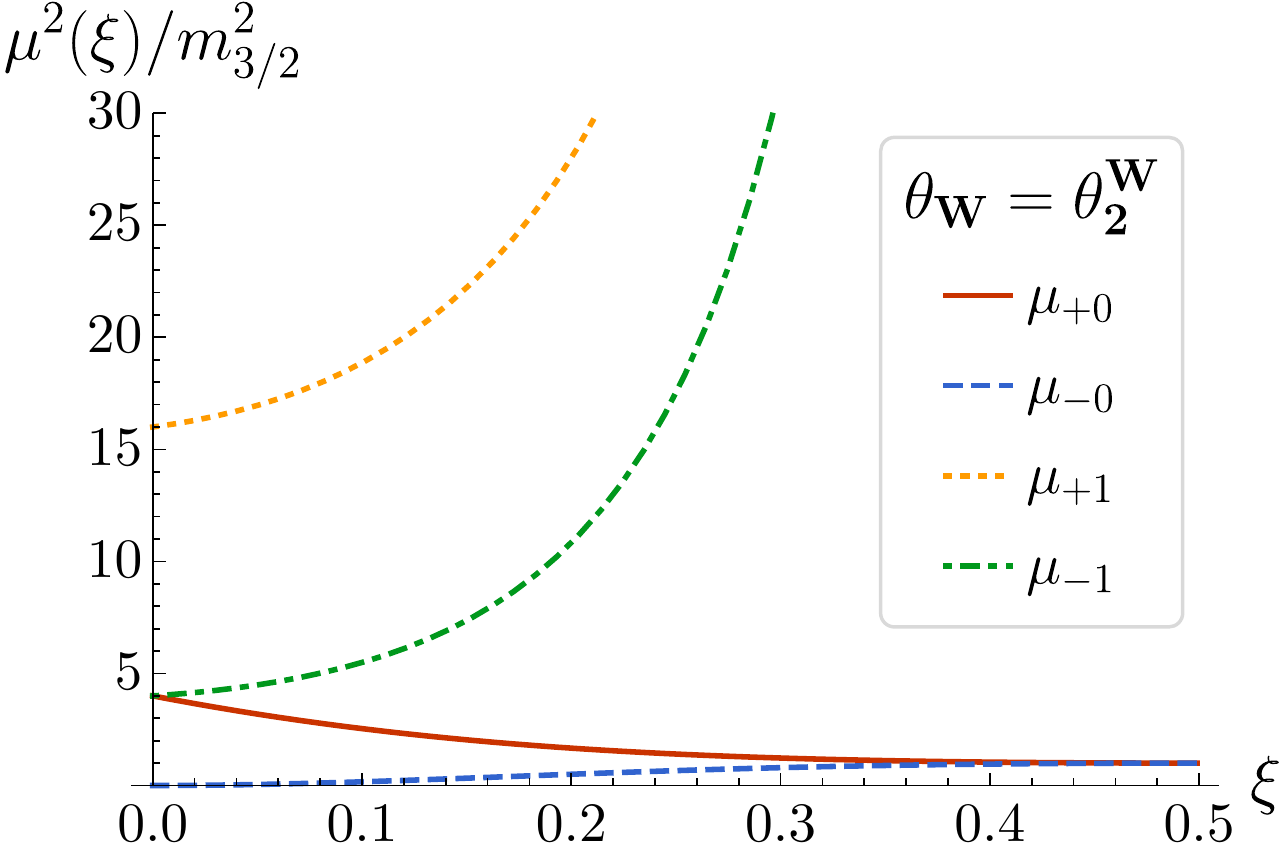} 
    \caption{Scalar mass spectrum for $\theta_W = \theta^W_2$ where there are $h^{2,1}-1$ massless modes in the truncated sector $\mu_{-\lambda'}^2=0$ and $\mu_{+\lambda'}^2=4m_{3/2}^2$.}
    \label{fig:zermodeSpectra2}
\end{figure}

Note that, for any given value of the parameter $\xi$, there are three values of $\theta_W$ such 
that the spectrum contains one, or several massless modes. They are given by
\be
\tan \theta_0^W= \hat m(\xi )^{-1}\,, \qquad 
\tan \theta_1^W= \hat m(\xi)\,, \qquad \text{and} \qquad
\tan \theta_2^W= \frac{\sqrt{3(1-2\xi)}}{1+\xi }\,,
\label{eq:criticalTheta}
\ee 
and for each of these values the massless field(s) correspond(s) to $\mu^2_{-0}$, $\mu^2_{-1}$, and $\mu^2_{-\lambda'}$, respectively.
The corresponding spectra associated to these branches of vacua are displayed in figures 
\ref{fig:zermodeSpectra} and \ref{fig:zermodeSpectra2}. In the case of the critical 
values $\theta_W = \theta_0^W$ and $\theta_W=\theta^W_1$, away from the LCS point ($\xi>0$) the 
spectrum contains exactly one vanishing mass, corresponding to fields in the reduced 
theory: $\mu_{-0}^2=0$ and $\mu_{-1}^2=0$, respectively. 
In those two cases all the other fields have masses of at least the order of the gravitino mass. These 
classes of vacua might be particulary interesting to realise the construction of dS vacua 
of \cite{Marsh:2014nla,Gallego:2017dvd}, which required a massless field in the complex 
structure sector at tree-level. Regarding the last branch, $\theta_W = \theta_2^W$, away from the 
LCS point the spectrum contains $h^{2,1}-1$ massless modes $\mu_{-\lambda'}^2=0$, that 
is \emph{half of the scalar modes} in the truncated sector.

In section \ref{sec:statisticalAnalysis} we will discuss the statistics of this mass spectrum in the 
ensemble of no-scale flux vacua. This will help us to estimate how generic these classes of vacua are 
in the Landscape.

\subsection{No-scale vacua with $N_A^0=0$}
\label{fluxAlinged}

In the present subsection we consider the second class of no-scale vacua for which the parameter $\theta_W$ is 
fixed in terms of $\xi$, namely flux vacua where the flux vector satisfies the constraint $N_A^0=0$. The flux 
$N_A^0$ is associated to the period \eqref{eq:Period} that grows without bound in the LCS limit. The main consequence 
of setting this flux to zero is that the terms of the superpotential which are cubic in $z^i$ are also identically zero.

The main motivation to study this class vacua is the analyses done in \cite{Brodie:2015kza,Marsh:2015zoa}. On the one hand, 
in \cite{Brodie:2015kza} it was argued (via a numerical analysis) that for generic choices of the fluxes, and at 
points of the moduli space near the LCS point the cubic terms of the superpotential typically become 
dominant.\footnote{Actually this is true even for points in field space not associated with a vacuum.} On the other hand, as 
proven in \cite{Brodie:2015kza,Marsh:2015zoa}, when the cubic terms of $W$ dominate no vacua can exist in the 
region of the moduli space where $\xi\approx0$. As a consequence no-scale vacua with $N_A^0\neq0$ are 
expected to be very scarce, or even non-existent, in a small neighbourhood of the LCS point. On the contrary, the 
conclusions in \cite{Brodie:2015kza,Marsh:2015zoa} cannot be applied to the class of vacua where the flux 
$N_A^0$ is set to zero, since the cubic terms of the superpotential are identically zero, and therefore can never 
become dominant. Thus, it is expected that the constrained class of vacua with $N_A^0=0$ may still be present, 
and even become the dominant type of vacua in a small neighbourhood of the LCS point.

To give further support to this conclusion, in appendix \ref{sec:bounds} we have estimated the minimal values 
of $\xi$ for which it is possible to find no-scale solutions with both non-vanishing $N_A^0$ and when subject 
to the constraint $N_A^0=0$. We find
\be
\xi_\text{min}|_{N_A^0\neq0} \gtrsim \frac{|\Im \kappa_0|}{4 \sqrt{N_\text{flux}}}
\label{eq:xiMinTextBound}
\ee 
for $N_A^0\neq 0$, while in the case $N_A^0=0$ the parameter $\xi$ remains unbounded below.
In agreement with the analyses in \cite{Brodie:2015kza,Marsh:2015zoa}, we can see that
vacua with $N_A^0=0$ are expected to be dominant in a small neighbourhood of the LCS point.
As we shall show in section \ref{sec:statisticalAnalysis}, the numerical scan of no-scale vacua in the $\mathbb{WP}^4_{[1,1,1,1,4]}$ model confirms this expectation, and matches perfectly with the conclusions of \cite{Brodie:2015kza,Marsh:2015zoa}.

\begin{figure}[t]
    \centering
    \subfloat[]{
        \includegraphics[width=0.47\textwidth]{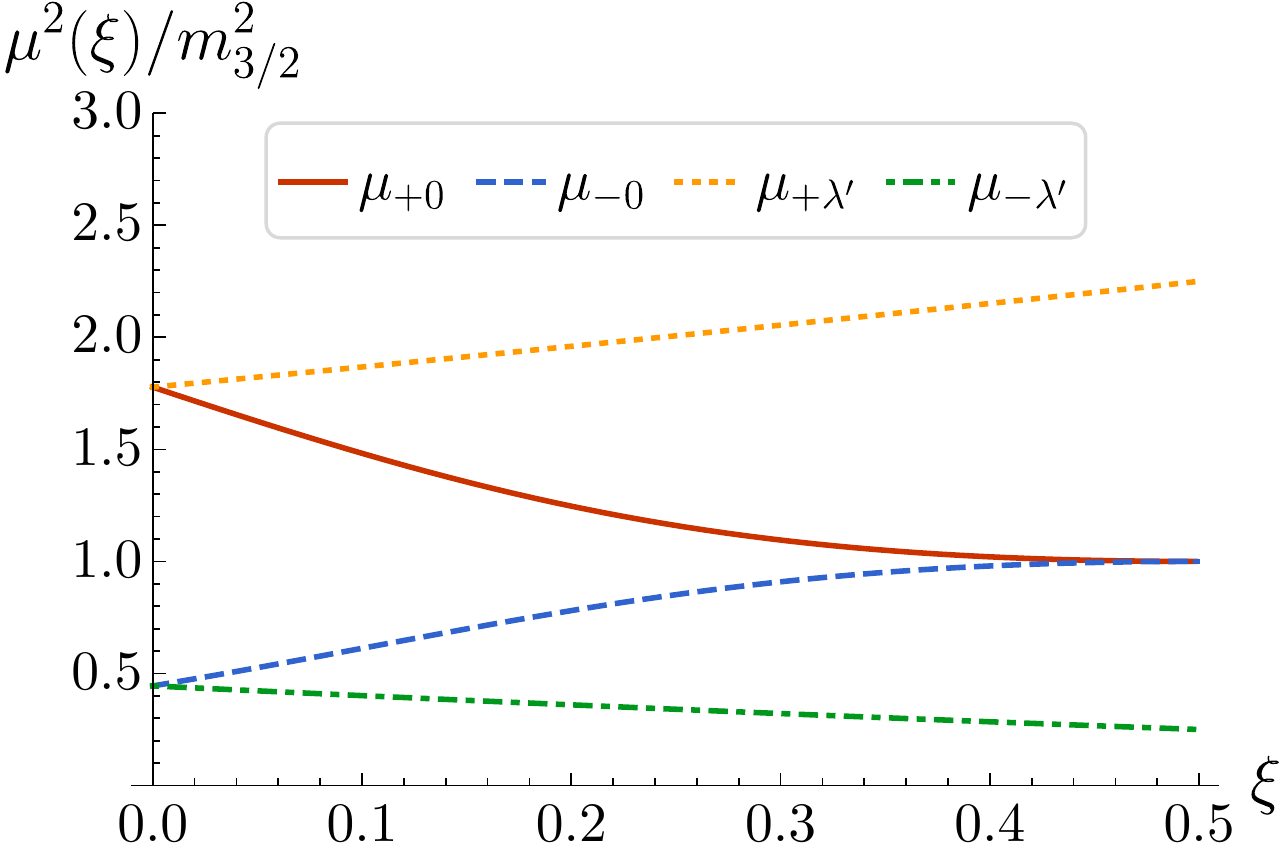}
    }
    \hfill  
    \subfloat[]{
        \includegraphics[width=0.47\textwidth]{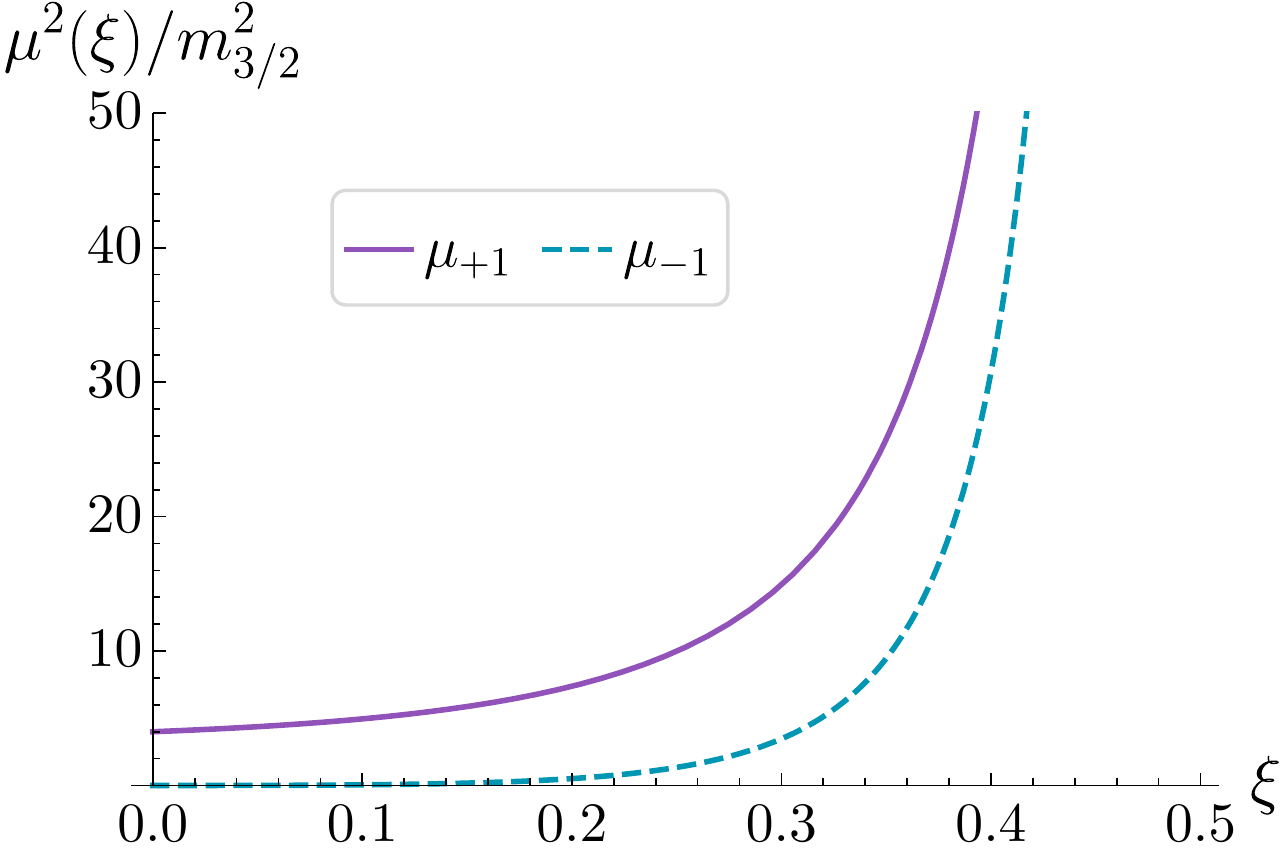}
    }
    \caption{Spectrum of scalar masses for vacua with the restriction $N_A^0=0$ on the flux configuration. The masses are plotted as a function of the LCS parameter $\xi\in[0,1/2]$ and are normalised by the gravitino mass $m_{3/2}$. (a) Branches corresponding to the masses $\{\mu_{\pm 0}^2, \mu_{\pm\lambda'}^2\}$, where $\mu_{\pm\lambda'}^2$ are the $2(h^{2,1}-1)$ masses in the truncated sector. (b) Scalar mass branches $\{\mu_{+1}^2, \mu_{-1}^2\}$. We can see that in these branches of vacua there are no light truncated fields $\mu^2_{\pm1}\ll m_{3/2}^2$ in the entire LCS regime.}
\label{fig:scalarMasses1}
\end{figure} 

To prove that in this type of vacua the angular parameter $\theta_W$ is determined by the value 
of $\xi$, it is convenient to make use of the Hodge decomposition of the flux vector $N$ \cite{Denef:2004ze}. As 
we review in appendix \ref{app:HodgeDecomp}, at any given no-scale vacuum $\{\tau_c,z^i_c\}$ the flux 
vector can be written in terms of the period vector $\Pi$ and its K\"ahler covariant derivatives $D_a \Pi = (\pd_a +K_a) \Pi$ as
\be
N =\sqrt{4 \pi} \rme^{K_{cs}} \left(\rmi W \, \bar \Pi + D_{\bar 0} D_{\bar a} \bar W \, D_a \Pi\right)\,.
\label{eq:hodgeN1}
\ee
Setting $N_A^0=0$ in this expression, we find
\be
W = i D_{\bar 0} D_{\bar a} \bar W \, K_a\,, 
\ee
where we have used that in the gauge \eqref{eq:Period} the period vector satisfies $\Pi_A^0=1$. Finally, taking into account the result \eqref{eq:noscaleDirection}
and the definition of the angular parameter $\theta_W$ \eqref{eq:thetaDef} together with \eqref{tadpoleBound}, we arrive at the constraint 
\be
\tan \theta_W = \sqrt{(1-2 \xi)/3} \qquad \Longrightarrow \qquad  \theta_W\in \[0,\frac{\pi}{6}\]\,.
\label{correlation}
\ee
Alternatively, this relation can be expressed in the following useful way.
\be
W_0^2 = \cV^2 m_{3/2}^2 = \frac{3N_{\text{flux}}}{8\pi( 2 - \xi )}\ge \frac{N_{\text{flux}}}{8 \pi}\sim \cO(10-10^3),
\label{eq:NAWNfrel}
\ee
which relates the flux parameter $W_0$ and the total $D3$-charge $N_{\text{flux}}$.
From here we can see immediately that these solutions are not compatible with the KKLT construction 
of dS vacua, since that scenario requires $W_0\ll1$. On the contrary, this class of no-scale vacua 
is suitable for the construction of LVS vacua, where $W_0\sim \cO(1-10)$.\\

In order to find the scalar spectrum at these no-scale solutions, we just need to substitute the relation 
\eqref{correlation} into our main result \eqref{eq:gralSpectrum}, which leads to 
\be
\hspace{.3cm}\mathbf{N_A^0=0:}\qquad \mu_{\pm\lambda}^2/m_{3/2}^2 =
\left\{
 \begin{array}{l c l }
\left( 1 \pm \sqrt{(1-2 \xi)/3}\; \hat m(\xi) \right)^2&~~& \lambda=0\\
\left( 1 \pm \frac{\sqrt{(1-2 \xi)}}{\sqrt{3}\hat m(\xi)} \right)^2&~~& \lambda=1\\
\left(1 \pm \frac{1+\xi}{3}\right)^2 &~~& \lambda=2,\ldots, h^{2,1}
\end{array}
\right.\,.
\label{eq:N0Spectrum}
\ee
We have displayed the dependence of these masses on the parameter $\xi$ in figure \ref{fig:scalarMasses1}.
Note that the previous spectrum 
is independent of the details of the Calabi-Yau compactification. With the aid of \eqref{eq:NAWNfrel}, it can 
be computed entirely from the total $D3$-charge $N_\text{flux}$, the LCS parameter $\xi$, and the Calabi-Yau 
volume $\cV$. 

Finally, as we approach the LCS point, $\xi\to0$, the mass spectrum \eqref{eq:N0Spectrum} takes 
the universal form
\be
\hspace{-.9cm}\mathbf{N_A^0=0, \, \xi=0:} \qquad \mu_{\pm\lambda}^2/m_{3/2}^2 =
\left\{
\begin{array}{l c l }
\left( 1 \pm\frac{1}{3} \right)^2&~~& \lambda=0,2,\ldots, h^{2,1}\\
\left( 1 \pm 1 \right)^2&~~& \lambda=1
\label{eq:na0_mu_LCS}
\end{array}\,.
\right.
\ee
This result is reminiscent of the deterministic spectra found in \cite{Brodie:2015kza,Marsh:2015zoa} at generic moduli configurations (not necessarily vacua) near the LCS point.

\section{Example: the $\mathbb{WP}_{[1,1,1,1,4]}^4$ model}
\label{sec:example}

In order to illustrate our results we have analysed a large sample of no-scale vacua of an 
orientifold of the Calabi-Yau hypersurface $\mathbb{WP}^4_{[1,1,1,1,4]}$ (the \emph{octic}). We will 
now briefly review the effective field theory for the compactification of type-IIB superstrings in this 
Calabi-Yau, and we will discuss the statistical properties of the resulting ensemble in 
section \ref{sec:statisticalAnalysis}. For a more detailed description of this compactification we 
refer the reader to \cite{Giryavets:2004zr,Klemm:1992tx}.

\subsection{Effective theory}

The Calabi-Yau geometries that we will consider can be defined in terms of the following family of hypersurfaces
\be
4 x_0^2 + x_1^8 + x_2^8+x_3^8 + x_4^8 - 8 \psi x_0 x_1 x_2 x_3 x_4 =0
\label{eq:definingPol}
\ee
in the complex projective space $x^i \in \mathbb{WP}^4_{[1,1,1,1,4]}$ (Model 3 of Table \ref{table:ModelSelection}). 

This family of hypersurfaces is characterised by a single complex deformation parameter 
$\psi$, with $\arg \psi \in[0,\frac{\pi}{4}]$. However, this Calabi-Yau three-fold has $h^{1,1}=1$ 
K\"ahler moduli and $h^{2,1} = 149$ complex structure fields, and thus there are many other 
deformations that one could consider. 
The Calabi-Yau geometries described by \eqref{eq:definingPol} are all invariant under a large 
group of discrete symmetries, namely $\cG =  \mathbb{Z}_8^2 \times \mathbb{Z}_2$ 
with order $[\cG]=128$, and all the deformations that we have not included in \eqref{eq:definingPol} are 
those transforming non-trivially under this group \cite{Giryavets:2004zr,Giryavets:2003vd}. Thus, by 
retaining only the deformation parametrised by $\psi$, we are realizing a consistent truncation of the 
complex structure moduli space, just as we discussed in section \ref{sec:ESV}.

In the neighbourhood of the large complex structure point, $\psi \to \infty$, the truncated action is 
characterised by the prepotential \cite{Font:1992uk,Klemm:1992tx}

\begin{figure}[t]
    \centering 
\vspace{-1cm}   \hspace{-1cm}\includegraphics[width=0.8\textwidth]{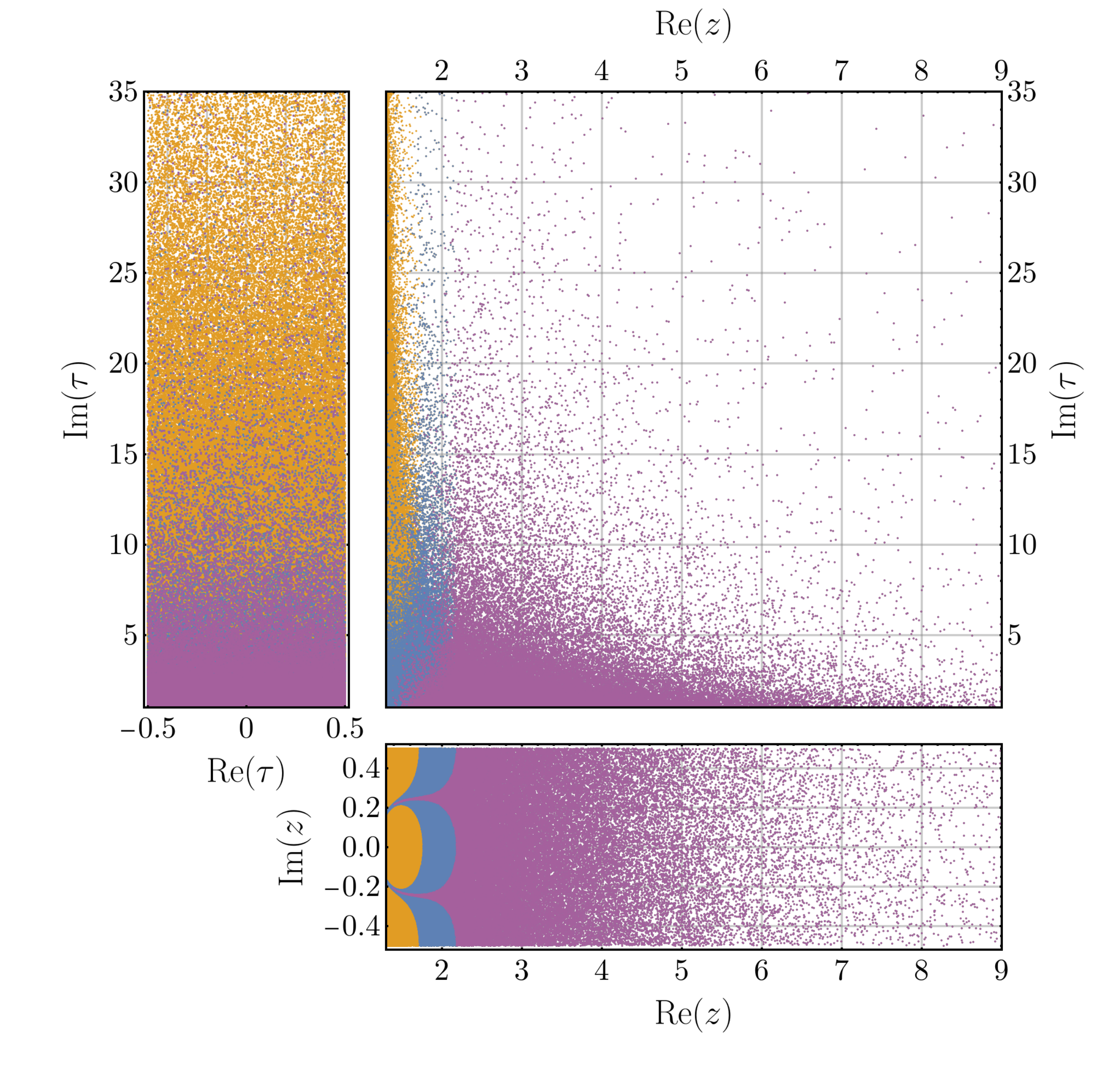} 
    \caption{Distribution of the numerically generated set of generic no-scale solutions on the $(\tau,z)$ field space.
 We have represented  in orange vacua with large instanton corrections  $>20\%$ (leading term in \eqref{eq:instCorrectDom}),  in blue when  corrections are in the range $1-20\%$, and in purple when corrections are $<1\%$. The $(\Re z, \Im \tau)$ plane exhibits nicely delineated regions, which are nevertheless likely to be blurred by higher order contributions to \eqref{eq:instCorrectDom}. The generic ensemble of vacua analysed in the text  is comprised of those solutions with small instanton corrections $<20\%$, and small string coupling $g_s = (\Im \tau)^{-1} <1$ (blue and purple, 119,139 solutions).} 
    \label{fig:scatterPlots}
\end{figure} 

\be
\cF(z) = \frac{\rmi}{3} z^3 + \frac{3}{2} z^2 + \rmi \frac{11}{6} z - \rmi \frac{37}{2\pi^3} \zeta(3) + \cF_{\text{inst}}\,, 
\label{eq:octic_prep}
\ee
where $z \approx \frac{4}{\pi} \log (4 \psi)$ and $\Im z \in [-1/2,1/2)$.
Here, $\cF_{\text{inst}}$ represents exponentially suppressed instanton corrections to the prepotential. Its leading term is of the form 
\be
\cF_{\text{inst}} \approx - \frac{\rmi n_1}{(2 \pi)^3} \rme^{-2 \pi z} + \ldots, \qquad \text{with} \qquad n_1 =29504~.
\label{eq:instCorrectDom}
\ee
The expansion for the prepotential \eqref{eq:octic_prep} around the LCS point converges in the region 
$|\psi|>1$, or equivalently $\xi \lesssim \xi_{\text{cnf}}\equiv 0.39$, away from the conifold singularity at $\psi=1$ \cite{Klemm:1992tx,Font:1992uk}.
However, here we will require in addition that the instanton corrections cause small 
variations on the moduli space geometry and the relevant physical quantities (e.g., the Yukawa couplings 
$\mathring \kappa_{abc}$, the vielbeins $e^a_i$, and $m_{3/2}$). As we discuss in appendix \ref{sec:instantonEstimate}, the 
most restrictive bound is found when imposing that the relative corrections to the moduli space vielbein are
small. Although this is checked for each particular vacuum, a simple estimate shows that the corrections 
remain moderately small ($<20\%$) as long as the LCS parameter satisfies
\be
\xi \lesssim \xi_\text{max} = 0.185 < \xi_{\text{cnf}}\,,
\label{eq:xiBoundInst}
\ee
which is a more conservative bound than just requiring the convergence of \eqref{eq:octic_prep}.

Following \cite{Denef:2004ze,Giryavets:2004zr,Giryavets:2003vd}, we will regard this compactification 
as an orientifold limit of a compactification of $F$-theory on the  fourfold $M_4=\mathbb{WP}^5_{[1,1,1,1,8,12]}$, 
where the orientifold action is defined by the transformations\footnote{As shown in \cite{Giryavets:2003vd}, it is possible to turn on the three-forms $F_{(3)}$ and $H_{(3)}$ on the four periods of the reduced theory consistently with the orientifold action.}  $x_0\to -x_0$ and $\psi \to -\psi$ \cite{Sen:1997gv,Klemm:1996ts}. 
The advantage of considering the embedding in $F$-theory is that compactifications on a fourfold allow 
a great deal of freedom in the choice of fluxes, which is particularly appropriate for performing a 
statistical analysis \cite{Denef:2004ze}. Indeed, the tadpole constraint $L$ is 
\be
L = \frac{\chi(M_4)}{24}\,,
\ee
where $\chi(M_4)$ is the Euler number of the fourfold, which typically greatly exceeds the one of the associated Calabi-Yau orientifold $\tilde M_3$. In the case at hand, 
the Euler number of the fourfold $\mathbb{WP}^5_{[1,1,1,1,1,8,12]}$ is $\chi(M_4) =23,328$, and thus the 
upper bound on the $D3$-brane charge induced by the fluxes is $N_{\text{flux}}\le L = 972$. 

As a final remark, note that the F-theory embedding also requires including in the theory  additional $D7$-brane moduli  (see \cite{Denef:2008wq}). The problem of the stabilisation of those moduli is however beyond the scope of the present paper, and we refer the reader to    \cite{Lust:2005bd,Collinucci:2008pf,Alim:2009bx,Honma:2017uzn,Honma:2019gzp} and the references therein  for works on the subject.

\subsection{Numerical search for flux vacua}

In order to perform a numerical exploration of the flux landscape of the octic, we have 
used \texttt{Paramotopy} \cite{bates2018paramotopy}. This software uses a numerical technique 
known as the \emph{Polynomial Homotopy Continuation} (PHC) method \cite{SW96,NSSP}, which 
efficiently finds all roots of non-linear polynomial systems, such as the no-scale 
equations \eqref{eq:susyeqs} (see appendix \ref{app:paramotopy}). 
Therefore, given a flux ensemble satisfying the tadpole condition \eqref{eq:tadpole1}, the 
PHC method allows for an exhaustive search of all the solutions to the no-scale equations 
\eqref{eq:susyeqs} \cite{MartinezPedrera:2012rs,He:2013yk}. 

As described in detail in appendix \ref{app:paramotopy}, we have constructed two separate 
ensembles of no-scale vacua: one with generic fluxes satisfying the tadpole constraint, and
one where fluxes additionally satisfy the condition $N_A^0=0$, as considered in section \ref{fluxAlinged}. We shall refer to them 
as the \emph{generic} and \emph{constrained} ensembles, respectively. The starting point 
for the construction of each of the ensembles is a collection of fluxes $f$ and $h$ randomly 
selected from a uniform distribution with support $[-50,50]$. This starting set consists of $10^7$ 
choices of flux for the generic ensemble, and $10^6$ choices for the constrained one.

For each choice of flux, the corresponding set of no-scale vacua were found using the PHC method. We then 
selected all solutions which have a small string coupling constant $g_s=(\Im \tau)^{-1}<1$ 
and small instanton corrections, i.e., which satisfy \eqref{eq:xiBoundInst}. In addition, when constructing the 
ensemble we checked that there was no double-counting of vacua related by either an $\mathrm{SL}(2,\mathbb{Z})$ 
action \eqref{eq:SL2Z}, or the symplectic transformations \eqref{eq:symplectitTrans} and 
\eqref{eq:fluxTransform}. Regarding the symplectic transformations, as proposed in \cite{DeWolfe:2004ns}, 
all no-scale solutions have been mapped to the fundamental domain of the axio-dilaton, where the 
redundant copies have been identified and discarded.\footnote{We avoided imposing conditions 
on the fluxes to eliminate the redundancies, as done, e.g., in \cite{MartinezPedrera:2012rs,He:2013yk}. In 
particular, our analysis showed that the constraints on the fluxes proposed in \cite{MartinezPedrera:2012rs} 
to deal with the $\mathrm{SL}(2,\mathbb{Z})$ symmetry lead to spurious correlations arising in the statistical 
analysis, and which are incompatible with the predictions derived from the continuous flux approximation 
\cite{Denef:2004ze}.} As for symplectic transformations, there is the monodromy around the LCS point 
\cite{Font:1992uk,Klemm:1992tx} which we have treated similarly, by mapping all solutions to a fundamental 
domain of the complex structure modulus $z$ and eliminating duplicate solutions. 

The ensemble of vacua with unconstrained fluxes that we obtained with this method contains 
$119,139$ solutions, while the constrained ensemble has $57,487$.
The results of this procedure for the generic ensemble are displayed in figure \ref{fig:scatterPlots}, where 
we show the distribution of no-scale vacua in the fundamental domain of the axio-dilaton $\tau$, in
the complex structure field $z$, and in the $(\Re z, \Im \tau)$. For completeness, let us mention that 
more conservative constraints could be imposed on the vacua, e.g., $g_s<0.1$ and 
instanton corrections below $<1\%$, leading to a considerably smaller ensemble with $427$ vacua. However, 
in order to have a sufficiently large sample to perform the statistical analysis, in the following 
we will consider all vacua in the weak coupling regime $g_s<1$ and with moderately small 
instantons corrections $<20\%$.\\

\begin{figure}[t]
    \centering
    \includegraphics[width=0.75\textwidth]{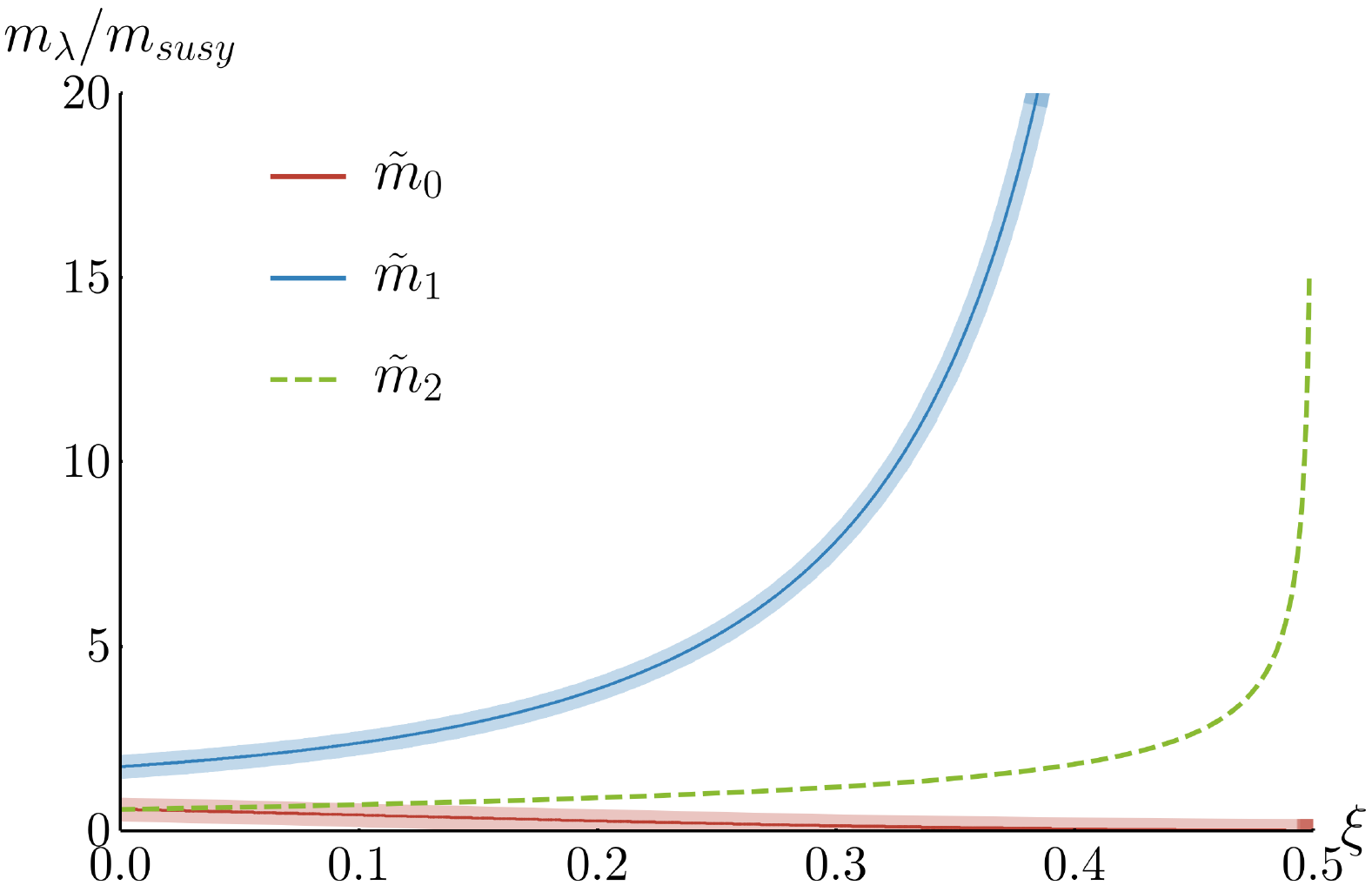} 
    \caption{Fermion masses in the $\mathbb{WP}^4_{[1,1,1,1,4]}$ compactification versus the LCS parameter $\xi$ at no-scale vacua. The fermion masses are normalised by the supersymmetric mass scale, $\tilde{m}_{\lambda}\equiv m_\lambda /m_{\text{susy}}$. The uppermost and lowermost thick curves represent the analytic result in \eqref{eq:gralFermionSpectrum} for the two fermions in the reduced theory. The thin solid curves, composed of (indistinguishable) data points, 
show masses obtained diagonalising numerically the fermion mass matrix \eqref{eq:sqFMM} at each vacuum of the ensemble.  The middle dashed curve represents the mass of the 148 fermions in the truncated sector (with a priori unknown EFT couplings), which was 
    computed via the third equation in \eqref{eq:gralFermionSpectrum}.}
    \label{fig:massFormulaCheck}
\end{figure}

In order to check the validity of our main result, \eqref{eq:gralFermionSpectrum}, at each of the 
no-scale solutions, we computed the eigenvalues of the fermion mass matrix \eqref{eq:FMM} for the 
reduced theory (involving only $\tau$ and $z$) using two different methods: first via the direct diagonalisation 
of the mass matrices obtained numerically, and then using the analytic formula \eqref{eq:gralFermionSpectrum}. We 
display the outcome of these computations in figure \ref{fig:massFormulaCheck}, which demonstrates 
the perfect the agreement of both methods. Regarding the 148 truncated complex structure moduli, although 
the EFT given above has no specific information about them, the expressions \eqref{eq:gralFermionSpectrum} 
allowed us to determine the fermionic masses corresponding to this sector at each no-scale solution. Finally, the 
scalar mass spectra in the whole axio-dilaton/complex structure sector for the ensemble of no-scale solutions 
were computed via \eqref{susyMAsses}. We checked that these masses coincide with those obtained by diagonalizing the Hessian \eqref{eq:Hessian} at each no-scale vacuum.
The statistical properties of these spectra will be analysed in the next section.

\section{Statistics of vacua}
\label{sec:statisticalAnalysis}

As we discussed in section \ref{sec:totalSpectrum}, the no-scale mass spectrum will depend in general 
on the flux configuration. To determine the properties of the spectra that may arise in the ensemble of flux 
vacua, we will adopt the statistical approach of \cite{Denef:2004ze}, and derive the 
probability distributions for the masses and other quantities of interest. We begin by presenting the 
relevant formulae for general compactifications before using them to study the particular models 
we consider.

Our starting point for this analysis will be the formula for the density of flux vacua derived in \cite{Denef:2004ze} 
using the continuous flux approximation. This approximation is based on the assumption that for large tadpoles, $L \gg1$, 
flux quantisation can be neglected, and thus it is possible to replace the sums over flux configurations by integrals.
\be
\sum_{N^I_A,N^B_I} \longrightarrow \int d^{2n} N_A d^{2n} N^B\,,
\ee
where $n= h^{2,1}+1$ and each component of $N = f - \tau h$ is a complex number parametrized 
by the two tuples of integers $f$ and $h$. Furthermore, as was proven in \cite{Denef:2004ze} 
and reviewed in appendix \ref{app:HodgeDecomp}, by using the Hodge decomposition of the flux vector, it is possible to 
establish a one-to-one correspondence between the $2n$ \emph{continuous}  flux complex variables $\{N_A^I,N^B_I\}$ 
and the $2n$ complex quantities $\{Z_A, F_A\}$ ($A= 0,\ldots,h^{2,1}$) given by 
\be
Z_0 \equiv \cV \rme^{K/2} W\,, \qquad Z_a\equiv \cV Z_{0a}\,, \qquad F_0 \equiv \cV \rme^{K/2}D_0 W\,, \qquad F_a \equiv \cV \rme^{K/2}D_a W\,.
\label{eq:fluxModulusVariables}
\ee
These variables are particularly convenient choices for describing the flux ensemble, as no-scale vacua can be equivalently 
characterised as flux configurations satisfying the conditions $F_A=0$. Thus, assuming a flat probability distribution 
on the fluxes, and using a generalisation of the Kac-Rice formula \cite{kac1943,rice1944mathematical} 
(see \cite{adler2009random} for a review), the \emph{density function} for no-scale vacua follows as~\cite{Denef:2004ze}
\be
d\mu_{vac} (Z_A, u^A) =\cN\cdot |\det \cH|^{1/2} |\det g| \rme^{-|Z|^2} \cdot d^{2n} Z \cdot d^{2n} u\,, 
\label{eq:vacuaDensity0}
\ee
where we denote the fields collectively by $u^A = \{\tau, z^a\}$, $g$ is the moduli space metric, and $\cH$ is 
the canonically normalised Hessian of the no-scale potential given by \eqref{eq:Hessian} and $\eqref{Zidentity2}$ 
(see appendix \ref{app:DDdistribution}). Here, and throughout the text, $\cN$ indicates some normalization constant
which must be computed for each particular distribution. The previous formula should not be confused with the \emph{index density} 
of flux vacua\footnote{The index density obtained  in \cite{Ashok:2003gk}  counts vacua weighted with the sign of $\det(\unity + \cM)$. }, first derived in \cite{Ashok:2003gk} and subsequently verified numerically in \cite{Giryavets:2004zr} and \cite{Conlon:2004ds}. 

In the class of models we are interested in, the number of complex structure moduli can be arbitrarily high but only 
one survives the truncation. In addition, as explained in section \ref{sec:ESV}, the truncation requires that only the 
components $I=0,1$ of the flux vector $N=\{N_A^I,N^B_I\}$ are turned on (8 flux integers).
Therefore, the statistics of these models can be described by \eqref{eq:vacuaDensity0}, setting 
$h^{2,1}=1$ ($n=2$). In this case, the determinant of the Hessian $\cH$ takes the particularly simple form
\be
 |\det \cH|^{1/2} = ||Z_0|^4 +|Z_1|^4 -(2+ \mathring \kappa^2) |Z_0|^2 |Z_1|^2|\,,
\label{eq:detHessian}
\ee
which will considerably simplify the computation of the mass distributions. 

\subsection{Moduli space distribution of generic no-scale vacua.}

\begin{figure}[t]
    \centering
    \subfloat[]{
        \includegraphics[width=0.47\textwidth]{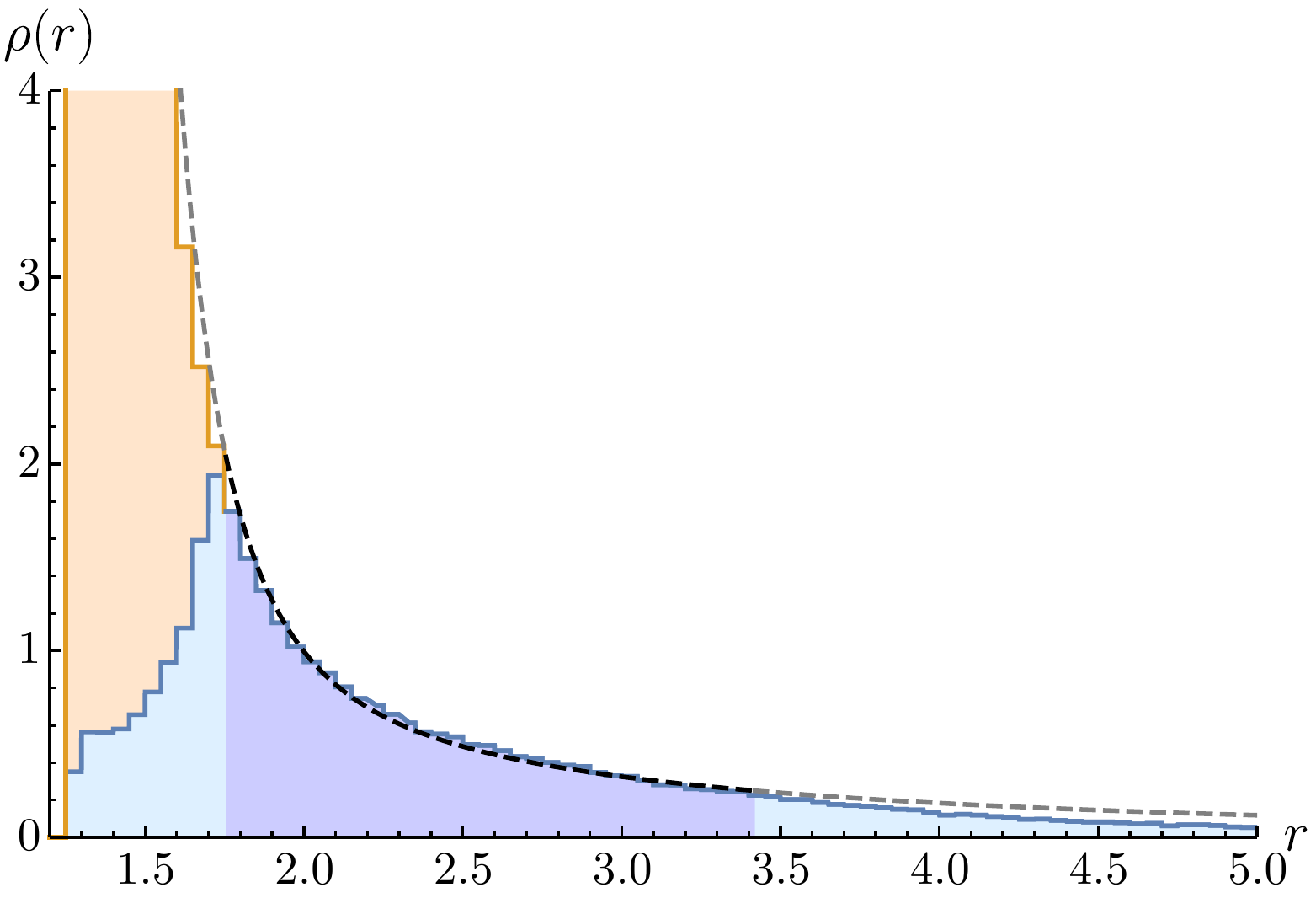}
    }
    \hfill  
    \subfloat[]{
        \includegraphics[width=0.47\textwidth]{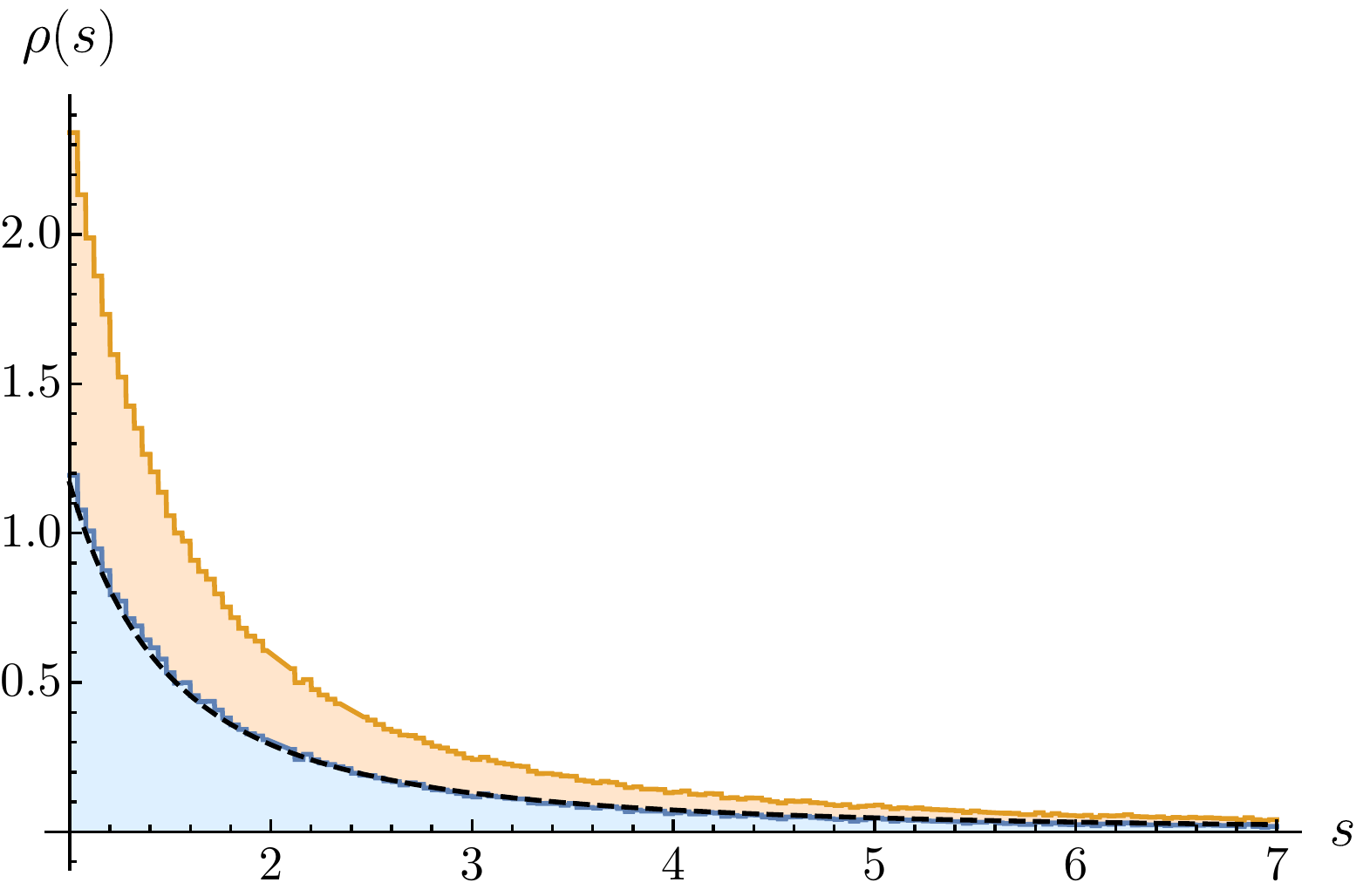}
    }
    \caption{(a) Marginalised density of no-scale vacua on the complex structure sector \eqref{eq:densityFields} (dashed line), and numerically obtained histogram of generic no-scale solutions in the $\mathbb{WP}^4_{[1,1,1,1,4]}$ flux ensemble. The quantity $r \equiv(2 \kappa_{zzz}/3 |\Im \kappa_0|)^{1/3} \, \Re z$ represents the complex structure field at the vacua, with the boundaries of the moduli space located at $r=2^{1/3}$ and $r\to \infty$ (the LCS point). The orange area represents excluded solutions with large instanton corrections ($>20\%$). In dark blue we indicate the subset of the remaining vacua well described by \eqref{eq:densityFields} (normalised in $r \in [1.75,3.42]$).   
    (b) Marginalised distribution \eqref{eq:densityFields} for the imaginary part of the axio-dilaton, $s\equiv\Im \tau$ (dashed line), and histogram of solutions in the generic ensemble of no-scale vacua. 
    }
\label{fig:fieldSpaceDist}
\end{figure} 

\begin{figure}[t]
    \centering\hspace{-.0cm}
    \includegraphics[width=0.75\textwidth]{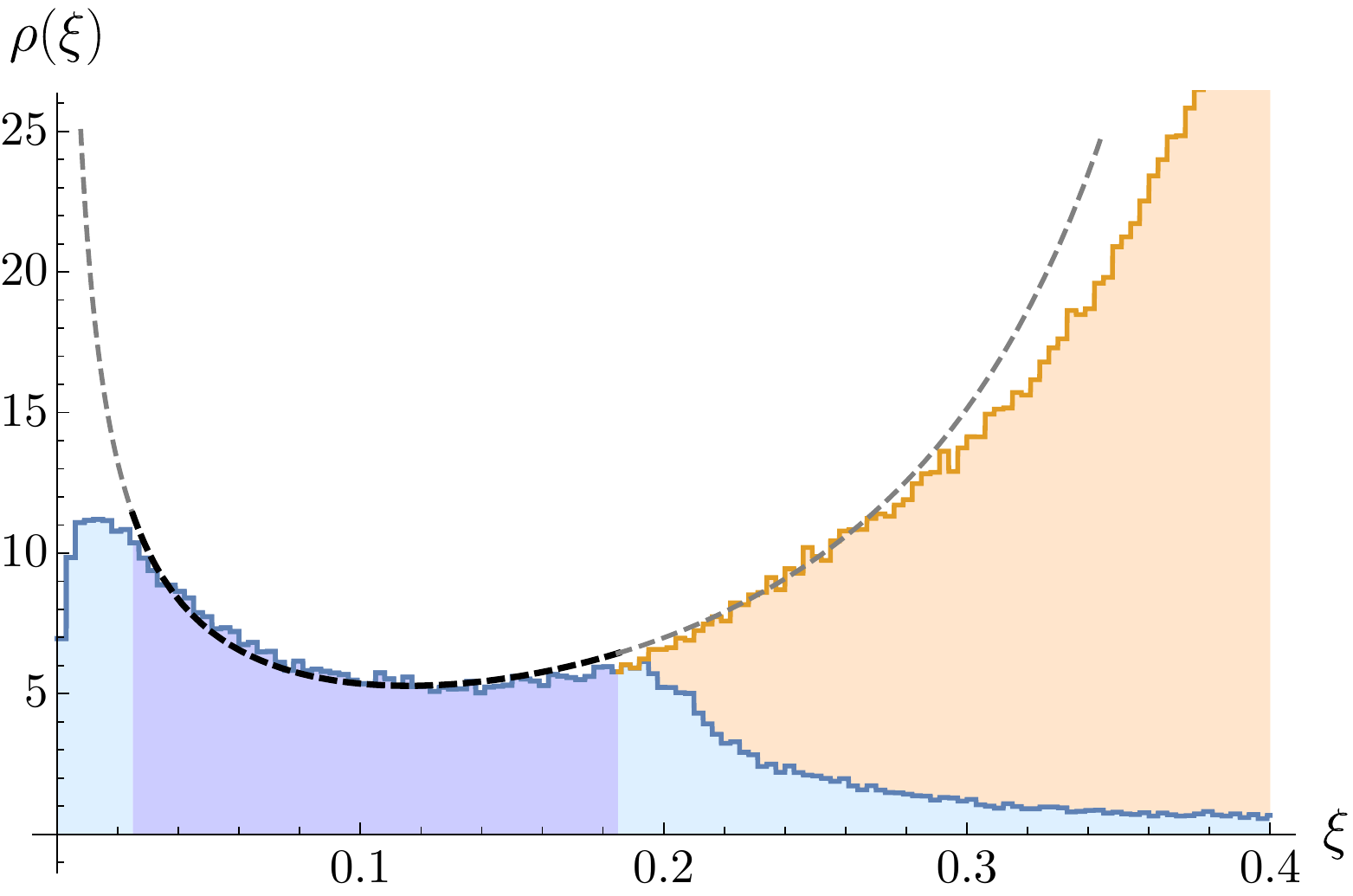}
    \caption{Distribution for the LCS parameter $\xi$ \eqref{eq:distributionXi} in the flux ensemble with unconstrained $N_A^0$ (dashed line), and histogram of solutions obtained from the numerical scan in the $\mathbb{WP}^4_{[1,1,1,1,4]}$ model. The colours are the same as in figure \ref{fig:fieldSpaceDist}.} 
    \label{fig:genericXi}
\end{figure}

Integrating \eqref{eq:vacuaDensity0} over the flux parameters $Z_A$ (with $h^{2,1}=1$) one obtains the following 
density distribution of no-scale vacua \cite{Denef:2004ze}:
\be
d\mu(z,\tau) = \cN \cdot |\det g| \cdot \left(2 - \mathring \kappa^2 +\frac{2 \mathring \kappa^3}{\sqrt{4+\mathring \kappa^2}}\right) d^2 \tau d^2 z\,,
\label{eq:densityFields}
\ee
where
\be
|\det g|=\frac{3}{16} \left(\frac{2 \kappa_{zzz}}{3 |\Im \kappa_0|}\right)^{2/3} \frac{(r^3- 2)r}{(r^3+1)^2\, s^2}
\ee
is the determinant of the moduli space metric, with $\kappa_{zzz}$ denoting Yukawa coupling for the (non-canonically normalised) field $z$.
We have also introduced the shorthands 
\be
s \equiv \Im \tau \qquad \text{and}\qquad r \equiv 1/\xi^{1/3} =(2 \kappa_{zzz}/3 |\Im \kappa_0|)^{1/3} \, \Re z \,.
\label{eq:srDefs}
\ee
Thus the quantity $\mathring \kappa$, defined in \eqref{eq:Yukawas} in terms of $\xi$, should be understood as a 
function of $\Re z$ in the expression \eqref{eq:densityFields} for the no-scale vacua density function. The corresponding 
marginal probability distributions for $\Re z$ and $\Im \tau$ are displayed in figure \ref{fig:fieldSpaceDist}. The plots show a 
remarkable agreement with the histograms obtained from the numerical scan of the octic model. Combining \eqref{eq:densityFields} and \eqref{eq:srDefs} it is also straightforward to find the probability distribution function for the LCS parameter $\xi$,
\be
d \mu(\xi) = \cN \cdot \frac{(1- 2 \xi)}{(1+\xi)^2\, \xi^{2/3}} \left(2 - \mathring \kappa(\xi)^2 +\frac{2 \mathring \kappa(\xi)^3}{\sqrt{4+\mathring \kappa(\xi)^2}}\right) d\xi\,,
\label{eq:distributionXi}
\ee
which we have displayed in figure \ref{fig:genericXi}, together with the histogram obtained from the direct computation 
from the octic flux ensemble data.

It is interesting to note that the density of vacua grows without bound as we move towards small values of $r$ (i.e.,  $\xi\to 1/2$), where 
the conifold point is located, $r_{\text{cnf}}\approx1.37$ ($\xi_{\text{cnf}}=0.39$). This is consistent with the expectation that the density of 
vacua is enhanced in regions of large moduli space curvature \cite{Ashok:2003gk,Giryavets:2004zr,Conlon:2004ds}. Actually, 
the marginalised density functions for $\Re z$  obtained from \eqref{eq:densityFields} are not normalisable when we 
define its support to be the entire range\footnote{Recall that we obtained the condition $\xi<1/2$, satisfied  away from the  moduli space boundaries,
  neglecting completely the instanton contributions, and thus it gives no information about the position of the conifold 
point at $\xi_{\text{cnf}} = 0.39$.} $r \in [2^{1/3},\infty)$. This property of the ensemble has an observable effect: as the underlying distribution from which we are extracting the vacua is not 
normalisable, regardless of the size of the sample, the histograms will always exhibit a deficit of vacua in with respect to the 
probability distribution (see region $\xi\gtrsim0.3$ in 
figure \ref{fig:genericXi}). Nevertheless, the complications due to the distribution being non-normalisable can be easily 
avoided: recall that the EFT for the octic can only be trusted in the region where the instanton corrections can be safely 
neglected, that is, in the region given by the bound \eqref{eq:xiBoundInst}, or equivalently with $r\in [1.75,\infty)$. Thus, as 
long as the support of \eqref{eq:densityFields} is taken to be the region of validity of the EFT, the probability distribution 
will be finite and normalizable, and thus well defined.

 In figure \ref{fig:genericXi}, it can also be seen that the density of generic no-scale vacua 
near the LCS point ($\xi \approx 0$) is considerably lower that the statistical prediction based on the continuous flux 
approximation (dashed line). This to be expected from the analyses in \cite{Brodie:2015kza,Marsh:2015zoa}, where 
it was shown that the statistics of generic no-scale vacua (with unconstrained $N_A^0$) can not be described with 
the continuous flux approximation in the strict LCS limit, and that actually in a small neighborhood of the LCS point there are no no-scale 
solutions with $N_A^0\neq0$. Such behavior was also anticipated in \cite{Denef:2004cf}, where the authors 
argued that the techniques presented there could fail to describe vacua statistics restricted to small regions of the moduli 
space. In appendix \ref{app:xiMin} we derive an estimate for the region of validity of the continuous flux approximation (eq. \eqref{eq:xiMinTextBound})
which, in the present ensemble, leads to the additional constraint on the LCS parameter
\be
\xi \ge \xi_{\text{min}}= 0.025\,.
\label{eq:minXi}
\ee

In the following section we will discuss the statistics of the mass spectrum in the regime where the EFT is under control 
and, in addition, where the continuous flux approximation is a good characterisation of the flux ensemble. That is, in the 
moduli region determined by the bounds \eqref{eq:xiBoundInst} and \eqref{eq:minXi}.

Before we end this subsection let us comment briefly on the distribution of the string coupling  constant $g_s$. Both the analytical result  in eq. \eqref{eq:densityFields}, and the numerical  histogram displayed  in figure \ref{fig:fieldSpaceDist}(b), indicate that the probability density
for $\Im (\tau)$ has the form $\rho(\Im \tau) \propto 1/(\Im \tau)^{2}$. Therefore, it is straightforward to check that the string coupling $g_s = (\Im \tau)^{-1}$ is uniformly distributed. This conclusion is relevant in the computation in \cite{Broeckel:2020fdz} of the distribution of the supersymmetry breaking scale in the Landscape which relies on $g_s$ having a uniform distribution.

\subsection{Mass distributions at generic no-scale vacua}

With the joint probability distribution \eqref{eq:vacuaDensity0} at hand it is now straightforward to compute probability 
distributions for the masses, both the fermions and the scalar modes, at the no-scale vacua in our flux ensemble. 

In particular, at a given vacuum with LCS parameter $\xi$ and angular parameter $\theta_W$, the fermion mass spectrum 
normalised by the gravitino mass $m_{\lambda}/m_{3/2}$ is given by
\be
x_\lambda \equiv m_{\lambda}/m_{3/2}=
\left\{
\begin{array}{l c l }
\zeta \,\hat m(\xi)&\qquad \qquad & \lambda=0\\
\zeta \, \hat m(\xi)^{-1}&\qquad \qquad& \lambda=1\\
\zeta(1+\xi)/\sqrt{3(1-2\xi)} &\qquad\qquad & \lambda=2,\ldots, h^{2,1}
\end{array}
\right.\,,
\label{eq:gralFermionSpectrum32}
\ee
where we have used \eqref{eq:gralFermionSpectrum} in combination with \eqref{eq:simpleTadpole}, with $\zeta \equiv \tan^2 \theta_W\in [0,\infty]$.
Then, in order to find the distribution for these masses, we need the joint distribution for $\{\xi,\zeta\}$. This distribution can be obtained from \eqref{eq:vacuaDensity0} and \eqref{eq:detHessian} by
integrating over the phases $\arg(Z_0)$ and $\arg(Z_1)$, the total $D3$-charge induced by fluxes $|Z_A|^2 = N_{\text{flux}}$, and 
the field space directions $\tau$ and $\Im z$. Using the $\zeta$ and $\xi$ variables, this yields
\be
d\mu(\zeta,\xi) = \cN \cdot \frac{(1-2 \xi)}{ \xi^{2/3} (1+ \xi)^2 (1+\zeta)^4} \left|\zeta - \hat m(\xi)^2\right| \left|\zeta - \hat m(\xi)^{-2}\right| \cdot d \zeta \, d \xi\,.
\label{eq:zetaXiPDF}
\ee 
From \eqref{eq:gralFermionSpectrum32} we can see that given a fixed value of $\xi$, we can establish is a 
one-to-one correspondence between $\zeta$ and each of the rescaled fermion masses $x_\lambda$.
Therefore, by performing a change of random variables $\{\zeta,\xi\} \to \{x_\lambda,\xi\}$ in
\eqref{eq:zetaXiPDF}, we can derive three separate distribution functions, each involving a different scaled 
mass $x_\lambda$. After integrating over the LCS parameter on the interval $\xi\in \[\xi_{\text{min}},\xi_{\text{max}}\]$ 
given by \eqref{eq:xiBoundInst} and \eqref{eq:minXi}, the resulting marginal distributions for the fermion masses in the reduced theory read
\be
\rho_0^f(x_0)= \cN \left|x_0^2 -1\right| x_0 \int_{\xi_\text{min}}^{\xi_\text{max}} d\xi \frac{(1-2 \xi)\, \hat m(\xi)^{2}}{\xi^{2/3}(1+ \xi)^2 (\hat m(\xi)^{2}+x_0^2)^4} \left|x_0^2 - \hat m(\xi)^4\right|,
\label{eq:Fmass32_0}
\ee 
and 
\be
\rho_1^f(x_1) = \cN \left|x_1^2 -1\right| x_1 \int_{\xi_{\text{min}}}^{\xi_\text{max}}d\xi \frac{(1-2 \xi)\, \hat m(\xi)^{2}}{ \xi^{2/3}(1+ \xi)^2 (1+x_1^2 \, \hat m(\xi)^{2} )^4} \left| x_1^2\, \hat m(\xi)^4- 1\right|\,.
\label{eq:Fmass32_1}
\ee 
Without further computations, we can already see that the probability of finding vacua with $m_{\lambda=0,1}=m_{3/2}$ 
(equivalently $x_{\lambda=0,1}=1$) is suppressed, i.e., $\rho^f_{\lambda=0,1}(1)=0$. This is a direct consequence of the generalized 
Kac-Rice formula \eqref{eq:vacuaDensity0}. To see this, note that density of vacua is proportional to the square root of the 
Hessian determinant $|\det \cH|^{1/2}$. This means that the probability of finding no-scale solutions with massless scalar 
modes in the flux ensemble should vanish. Since according to \eqref{susyMAsses} massless scalar modes occur precisely 
whenever one fermion mass equals that of the gravitino, we conclude that no-scale vacua with $m_\lambda=m_{3/2}$ are 
quite rare in the Landscape. It is important to emphasize that this does not preclude vacua with $m_{\lambda=0,1}=m_{3/2}$ 
from existing, it just means they represent a very small fraction of the total number of vacua. Actually, the suppression of 
critical points with of zero eigenvalues on the Hessian, which also leads to an apparent repulsion between critical points, 
is a generic feature of random functions (see, e.g., \cite{adler2009random}), and has already been observed in other 
characerisations of the Landscape \cite{Aazami:2005jf,Mehta:2015nva}. 

\begin{figure}[t]
    \centering
    \subfloat[]{
        \includegraphics[width=0.47\textwidth]{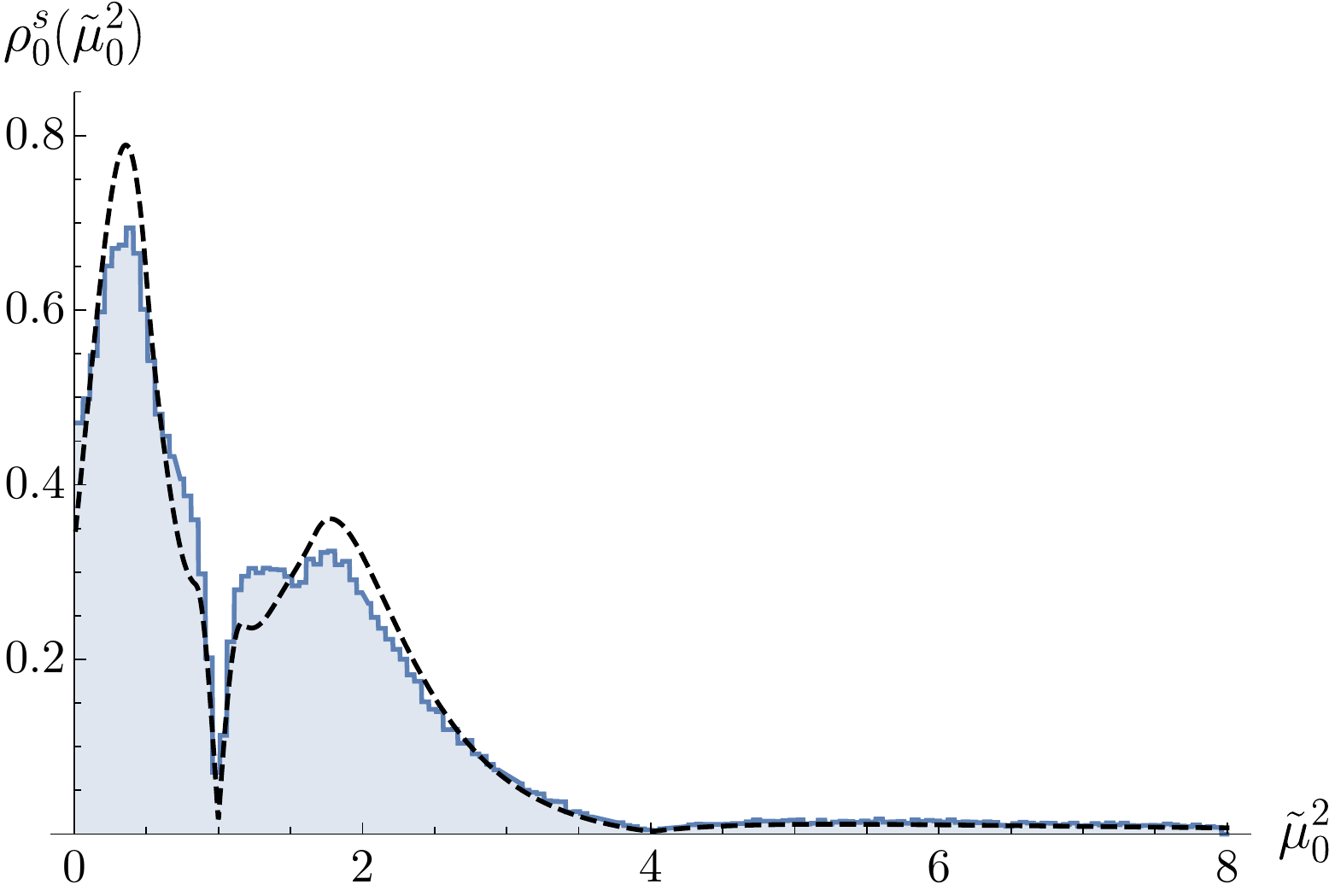}
    }
    \hspace{0.cm}
    \subfloat[]{
        \includegraphics[width=0.47\textwidth]{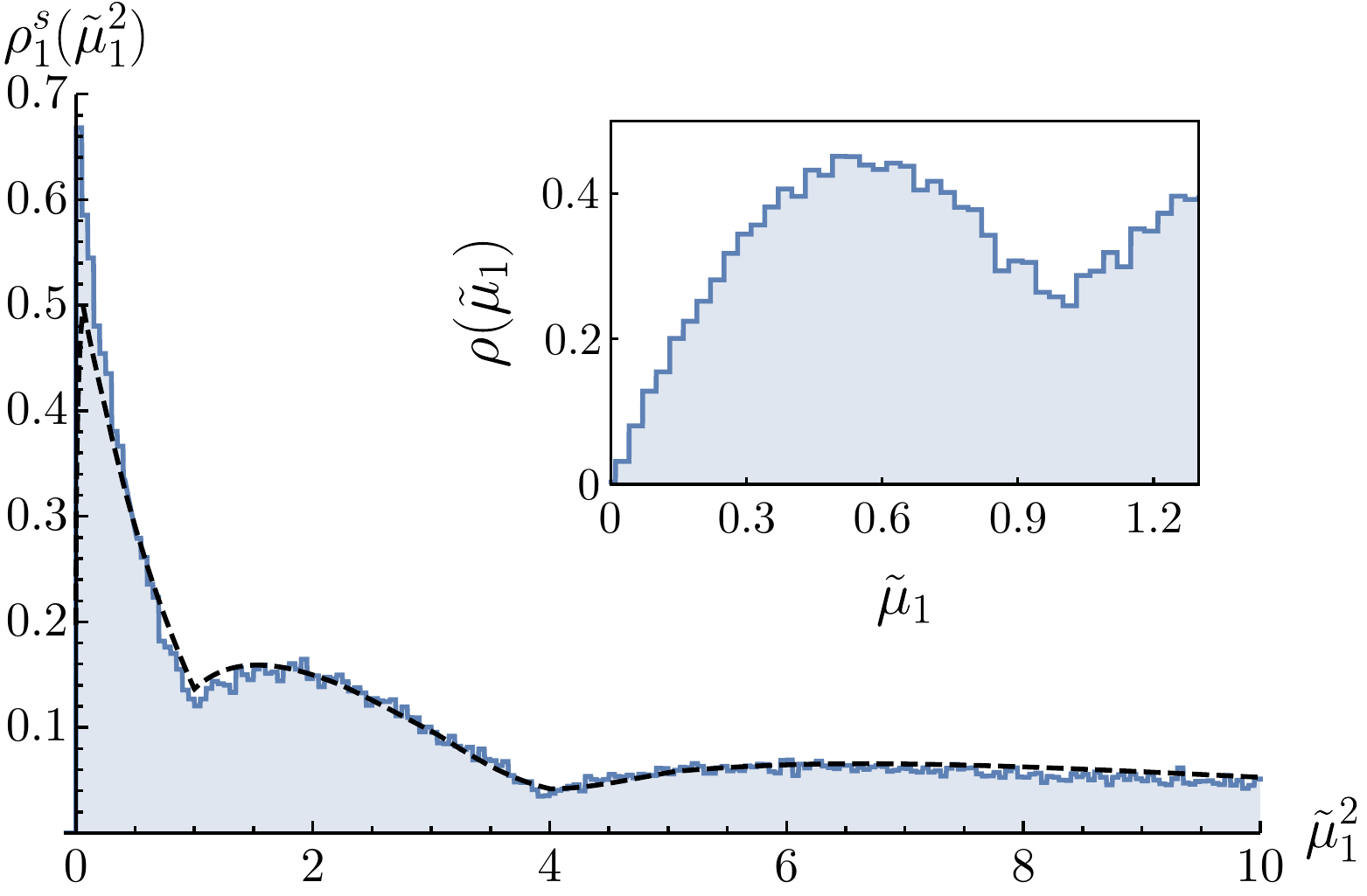}
    }
    \caption{Distribution of the squared scalar masses normalised by the gravitino mass, $\tilde{\mu}^2_\lambda\equiv\mu^2_\lambda/m_{3/2}^2$, in the ensemble with unconstrained fluxes. The dashed lines in (a) and (b) correspond to the probability distribution \eqref{eq:scalar_mass_spectrum} (evaluated with \eqref{eq:Fmass32_0} and \eqref{eq:Fmass32_1}, resp.) for the masses of the scalars in the reduced theory. We also show the mass histograms obtained numerically from the flux ensemble of the $\mathbb{WP}^4_{[1,1,1,1,4]}$ model (blue). The inset in (b) shows the obtained distribution  $\rho(\tilde{\mu}_1) d\tilde \mu_1$ near the origin, which presents a suppression for the massless mode.}
    \label{fig:genericMassSpectra1}
\end{figure} 

Regarding the rescaled mass of the truncated fermions $x_{\lambda'} = m_{\lambda'}/m_{3/2}$, we find the distribution
\be
\rho_{\lambda'}^f(x_{\lambda'}) = \cN x_{\lambda'}\, \int_{\xi_\text{min}}^{\xi_\text{max}} d\xi \frac{(1-2 \xi) f(\xi)^{-2} \left|x_{\lambda'}^2 f(\xi)^{-2} - \hat m(\xi)^2\right| \left|x_{\lambda'}^2 f(\xi)^{-2} - \hat m(\xi)^{-2}\right| }{ \xi^{2/3}(1+ \xi)^2 (1+x_{\lambda'}^2 f(\xi)^{-2})^4}\,,
\label{eq:Fmass32Prime}
\ee
where $f(\xi) \equiv \frac{1+\xi}{\sqrt{3(1-2\xi)}}$.
Note that for generic values of $\xi$ there appears to be no suppression on the probability of finding 
$x_{\lambda'} =1$ in the truncated sector; in other words, $\rho_{\lambda'}^f(1)\neq0$. This in turn shows that massless 
scalar modes on the truncated sector are \emph{not suppressed}. This is a surprising result that is at odds with what 
would be expected from the analysis of generic random functions. This reflects the important role that 
symmetries of the EFT play in shaping the flux Landscape. In particular, this observation is of importance 
for the construction of dS vacua proposed in \cite{Marsh:2014nla,Gallego:2017dvd}, which relies on the 
existence of no-scale solutions with massless modes at tree-level. 

A common feature to the three distributions \eqref{eq:Fmass32_0}, \eqref{eq:Fmass32_1} and \eqref{eq:Fmass32Prime}, is 
that they all vanish for $m_\lambda=0$ (equivalently $x_\lambda=0$). That is, the probability of finding 
no-scale solutions with massless fermions also appears to be suppressed in the ensemble of flux vacua. This 
result can be traced back to the structure of the fermion mass matrix $\cM$ given in \eqref{eq:Hessian}, whose 
eigenvalues come in pairs $\pm m_\lambda$, and the well known \emph{eigenvalue repulsion} 
effect \cite{1929PhyZ...30..467V}, which is characteristic of random matrix ensembles\footnote{The collection of 
mass matrices $\cM$ associated with the ensemble no-scale vacua can be regarded as an statistical ensemble 
of matrices with random entries \cite{Denef:2004cf,Marsh:2011aa,Bachlechner:2012at,Sousa:2014qza,Achucarro:2015kja}.} (see \cite{Guhr:1997ve} for a review).

\begin{figure}[t]
    \centering
    \includegraphics[width=0.75\textwidth]{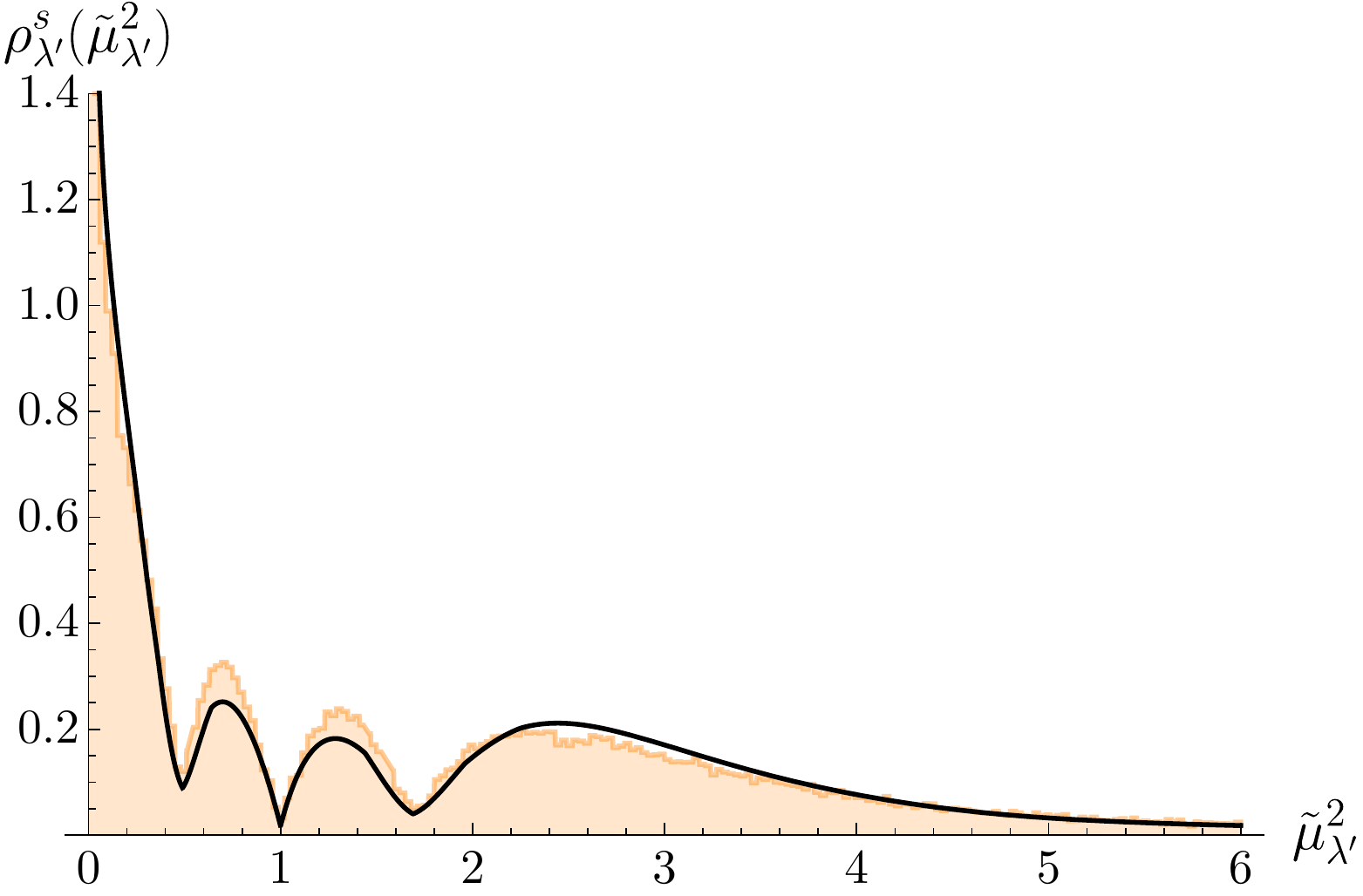}
    \caption{The theoretical prediction for the probability distribution of the normalised squared  masses $\tilde \mu_{\lambda'}^2 = \mu_{\lambda'}^2/m_{3/2}^2$ of the truncated scalar fields (solid line), eqs. \eqref{eq:scalar_mass_spectrum} and \eqref{eq:Fmass32Prime}. For comparison we also show in orange a histogram of scalar masses generated with \eqref{eq:gralSpectrum}.} 
\label{fig:genericMassSpectra2}
\end{figure} 

With the above distributions for $m_\lambda / m_{3/2}$ at hand, the probability density functions for the 
scalar masses can be easily obtained with a simple change of variables. Indeed, for each $\lambda$ the 
probability distribution for the combined two branches of scalar masses $\mu_{\pm \lambda}^2$ reads 
\be
\rho_\lambda^s(\tilde \mu^2_\lambda)  d\tilde \mu^2_\lambda= \cN\cdot \tilde \mu^{-1}_\lambda \, \left[ \rho^f_\lambda(1+\tilde \mu_\lambda) +\rho^f_\lambda(|1-\tilde \mu_\lambda|) \right] d\tilde \mu^2_\lambda,
\label{eq:scalar_mass_spectrum}
\ee
where $\tilde \mu^2_\lambda \equiv \mu^2_\lambda/m_{3/2}^2$.
These theoretical distributions are plotted in figures \ref{fig:genericMassSpectra1} and \ref{fig:genericMassSpectra2}, 
along with the scalar mass histograms obtained from the numerical scan. As described in section \ref{sec:example}, the 
masses of the scalar modes in the sector surviving the truncation, $\mu_{\lambda=0,1}^2$ are obtained via the 
diagonalisation of the fermion mass matrix \eqref{eq:sqFMM} at each vacuum together with the formula \eqref{susyMAsses}, 
while those associated to the truncated modes $\mu_{\lambda'}^2$ are computed from the formula \eqref{eq:gralSpectrum}. The 
plots show a remarkable agreement of the numerical results and analytical predictions in the regime where both the low energy 
EFT and the continuous flux approximation are expected to provide a good description of the theory\footnote{The histograms in figs.~\ref{fig:genericMassSpectra1} and \ref{fig:genericMassSpectra2} show slight  deviations with respect to the theoretical distributions. As argued in  \cite{Denef:2004ze}, the discrepancies might be due to restricting the vacua to lie in a bounded region of moduli space, i.e., within the limits  \eqref{eq:xiBoundInst} and \eqref{eq:minXi}.}.  

It is interesting to note that the spectra in the surviving sector of figure \ref{fig:genericMassSpectra1} show a suppressed 
probability of no-scale solutions with scalar masses\footnote{Note that the $\mu_{1}^2/m_{3/2}^2 = 0$ suppression is not evident in the main plot of figure~\ref{fig:genericMassSpectra1}(b).  This is due to the
factor   $1/\tilde \mu_1$ in \eqref{eq:scalar_mass_spectrum}, which makes it difficult to resolve the suppression in the numerical histogram.  The inset of this figure shows the histogram for $\tilde{\mu}_1$, which does present clearly the suppressed probability  of the massless modes.} $\mu_{\lambda}^2/m_{3/2}^2 = 0,1,4$.
This is consistent with our discussion above, as the first and third cases correspond to vacua with one or more fermion masses
equal to $m_{3/2}$ (see relation \eqref{susyMAsses}), and the second case to vacua with massless 
fermions. Therefore, making contact with our discussion in section \ref{sec:masslessFields}, we can see that the branches of solutions corresponding to $\theta_W=\{\theta_0^W,\theta^W_1\}$ have a low probability to occur, since they are respectively the vacua where the masses $\mu_{-0}$ and $\mu_{-1}$ vanish.

By contrast, in the spectra for the truncated sector, figure \ref{fig:genericMassSpectra2}, we see a suppression 
of vacua with masses $\mu_{\lambda}^2/m_{3/2}^2 =1$ (i.e., with massless fermions), while the distribution 
function \emph{diverges} in the limit $\mu_\lambda^2/m_{3/2}^2\to 0$, that is, for the branch of solutions with  $\theta_W = \theta_2^W$ discussed in  section \ref{sec:masslessFields}. In other words, for a large fraction of 
no-scale solutions, half of the scalar modes in the truncated sector have masses much lower than the gravitino.  It is  precisely these vacua which are in danger of developing tachyonic instabilities upon including quantum 
effects ($\alpha'$ corrections or instanton effects). In particular we observe that approximately $11\%$
 of the vacua contain modes with masses satisfying $\mu^2_{\lambda'} \lesssim 10^{-2} \, m_{3/2}^2$, and about $3\%$  with masses $\mu^2_{\lambda'} \lesssim 10^{-3} \, m_{3/2}^2$. Moreover, a closer  examination of the mass spectra in the ensemble reveals the presence of vacua with large mass hierarchies, with modes  as light as $\mu^2_{\lambda'} \sim 10 ^{-10} \, m_{3/2}^2$.

 Note also that the suppressions seen in figure \ref{fig:genericMassSpectra2} around $\mu_{\lambda'}^2 \approx 0.5\,  m_{3/2}^2$ and $\mu_{\lambda'}^2 \approx1.7\, m_{3/2}^2$ are due to modes of the reduced theory becoming massless, i.e., the branches of vacua with $\theta_W = \{\theta^W_0,\theta^W_1\}$. Indeed, recall that due to the form of the spectrum \eqref{eq:gralSpectrum} the mass distributions for all the modes arise from the same probability density function \eqref{eq:zetaXiPDF}, and thus the suppression of any particular branch of vacua can also be observed in the statistics of all the other masses.
 
As a final remark, let us point out that
 equation \eqref{eq:vacuaDensity0} and \eqref{eq:detHessian} imply that the masses \eqref{eq:gralSpectrum} and $g_s$ are statistically independent from each other. As a consequence, although the ensemble discussed here involves vacua with a marginally small string coupling $g_s\le 1$, the statistical properties of the spectrum would not be affected by restricting the analysis  to vacua with very small string coupling $g_s \ll1$. As a consistency check,  we also computed the numerical histograms represented in figures  \ref{fig:genericMassSpectra1} and \ref{fig:genericMassSpectra2}  for the subset of vacua in our ensemble with $g_s \le 0.1$ ($\sim 5000$ vacua), but no significant changes where observed, and thus, we will not present them here.

\subsection{Statistical properties of the constrained ensemble}

We will now turn to the statistics of the constrained ensemble of vacua, where $N_A^0=0$. As we show in 
appendix \ref{app:DDdistribution} the statistical methods in \cite{Denef:2004ze} can easily be adapted to 
describe this ensemble. In particular, the density of flux vacua with is found to be 
\be
d\mu_{\text{vac}}(z,\tau)\big|_{N_A^0=0} = \cN \cdot
\frac{(1+\xi)\xi^{2/3}}{(2-\xi)^2(\Im \tau)^2 } d^2 z d^2\tau\,,
\label{eq:densityFieldsNA0}
\ee
where $\xi$ should be understood here as a function of $\Re z$. It is worth noting that, provided we 
consider only the weak coupling regime $\Im \tau>1$, the density of no-scale vacua is normalisable within the 
whole  moduli space, even near the conifold point $\xi\to\xi_\text{cnf}$ where the EFT is known to become 
inaccurate. Indeed, contrary to the generic case, the density \eqref{eq:densityFieldsNA0} is not enhanced 
(and remains finite) as we approach the conifold point, and as a result the distribution is well defined in the 
whole range of $z$ (i.e., in $\xi\in[0,1/2]$). Furthermore, as we show in appendix \ref{app:DDdistribution}, 
for this sub-ensemble, flux quantisation and a finite tadpole do not lead to the breakdown of the statistical 
description near the LCS point.\footnote{This observation relies on the fact that when $N_A^0=0$, the flux 
$N_B^0$ is not bounded by the flux tadpole. However, in practice the flux integers are extracted from a 
uniform distribution in $[-50,50]$, which results in deviations from the continuous flux approximation in the 
range $\xi\lesssim \xi_\text{min}=5\cdot10^{-5}$.\label{ft:NA0minXi}} As a consequence, in contrast with the 
case of generic no-scale solutions, the continuous flux approximation provides an excellent characterisation 
of the ensemble in the strict  LCS regime. As we shall see next, these features of the model will lead to an almost perfect 
agreement between the statistical description and the results of the numerical scan in the octic model.

\begin{figure}[t]
    \centering\hspace{-.0cm}
    \includegraphics[width=0.75\textwidth]{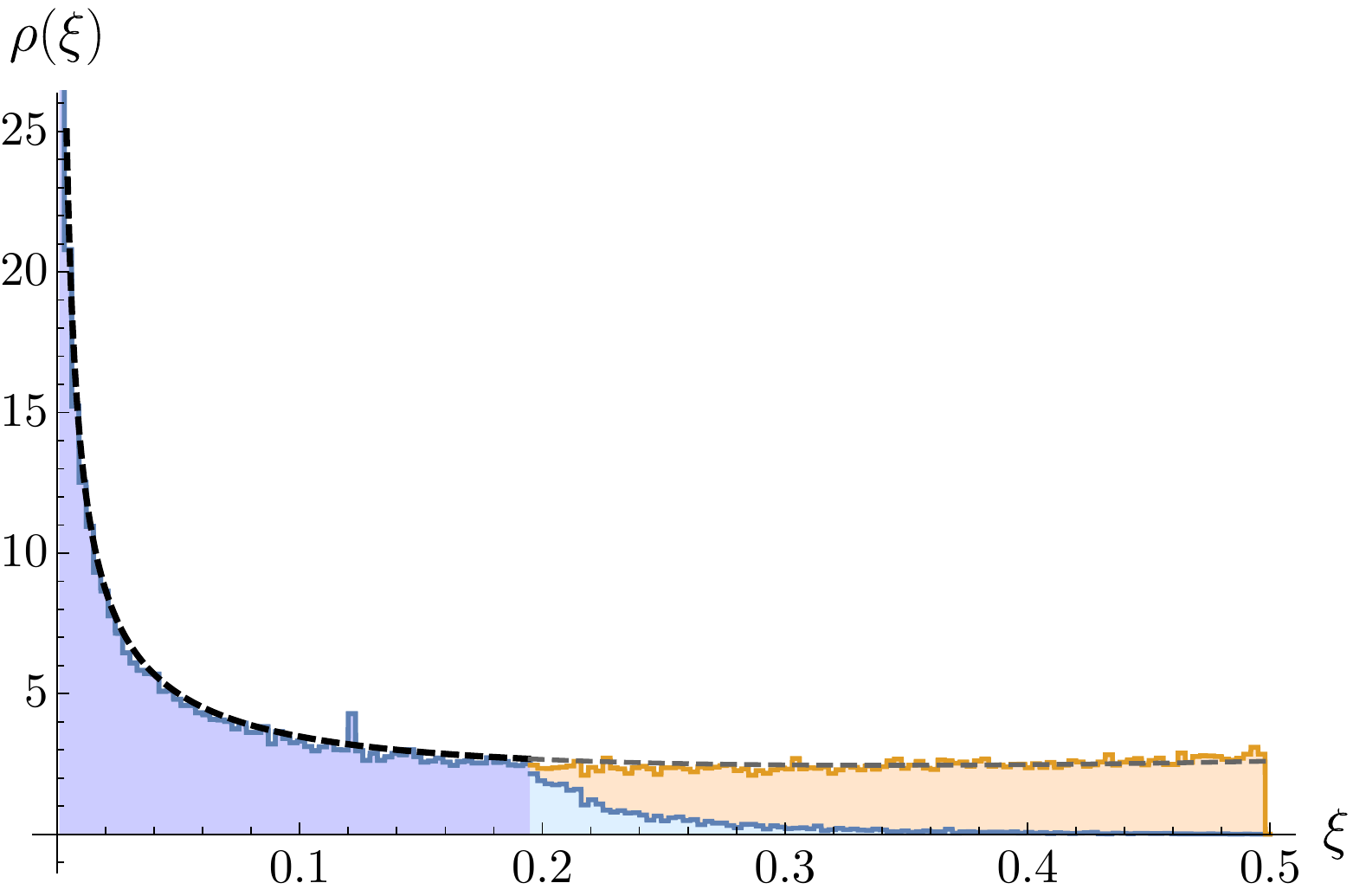}
    \caption{Distribution for the LCS parameter for the constrained flux ensemble. The dashed line represents the theoretical distribution \eqref{fig:xiDistNA0} normalized for data in the range $5.10^{-5} \leq \xi \leq 0.185$ (see footnote \ref{ft:NA0minXi}). We also show the histogram of $\xi$ at no-scale vacua, with colours the same as in figure \ref{fig:fieldSpaceDist}.} 
\label{fig:xiDistNA0}
\end{figure} 

As in the previous section, we begin by computing the probability distribution for the LCS parameter $\xi$, which takes the simple form
\be
\rho(\xi) d\xi = \frac{2^{1/3}(1+\xi)}{(2-\xi)^2\xi^{2/3}} d\xi\,.
\label{eq:xiDistNA0}
\ee
This distribution, together with the histogram obtained from the numerical scan, is plotted in 
figure \ref{fig:xiDistNA0}. As is evident, the analytic formula perfectly matches the histogram over the whole 
range of $\xi$. The histogram includes all of the  no-scale solutions at points where the moduli space metric is well defined, $\xi \in [0,1/2]$, however 
only those shaded in light and dark blue correspond to vacua with small instanton corrections. Excluding solutions with 
sizeable corrections (orange) leads to the fall-off (light blue) observed around $\xi \approx 0.2$. The statistical description 
does not incorporate the effects of truncating the ensemble, and therefore it can only provide a good 
description in the region of $\xi$ where few vacua (or none) are excluded from the ensemble. This region 
of $\xi$, which we shaded in dark blue, represents the set of vacua we will use next to characterise the 
statistics of the mass spectra, both numerically and using the continuous flux approximation.

\begin{figure}[t]
    \centering
    \subfloat[]{
        \includegraphics[width=0.47\textwidth]{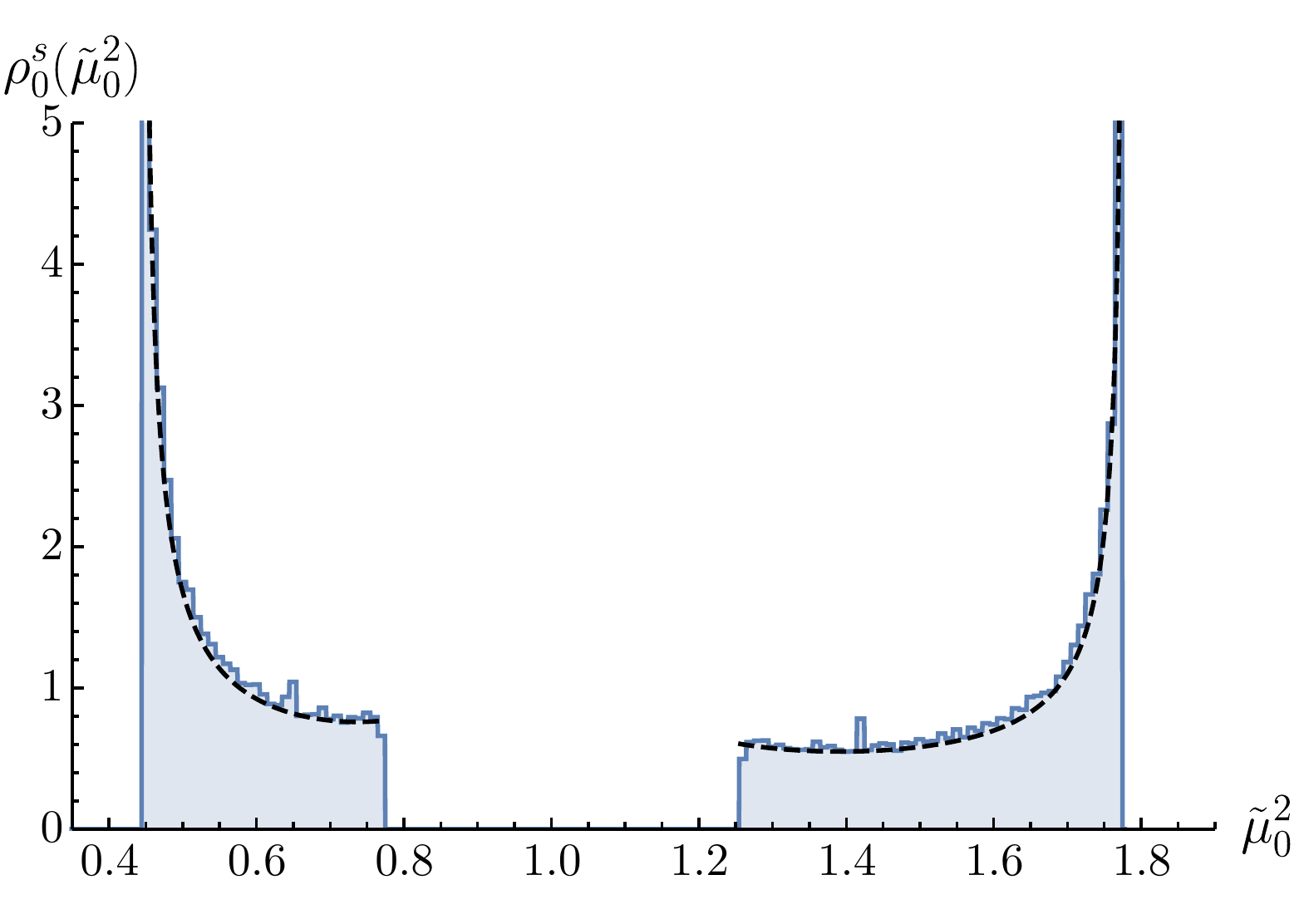}
    }
    \hfill  
    \subfloat[]{
        \includegraphics[width=0.47\textwidth]{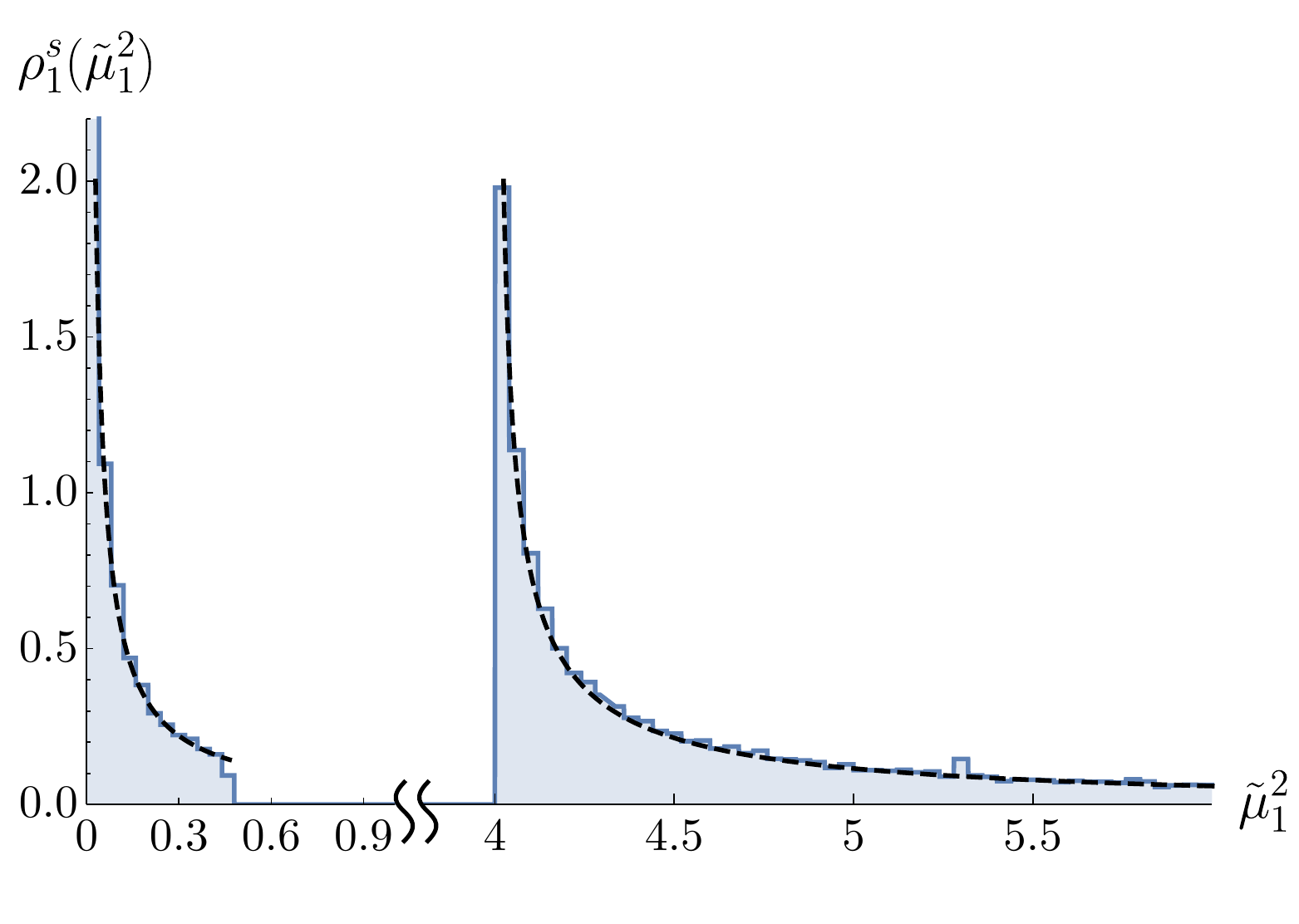}
    }
    \caption{Distribution for the squared scalar masses with constrained fluxes $N_A^0= 0$ normalised by the gravitino mass $\tilde \mu_\lambda^2 = \mu_\lambda^2/m_{3/2}^2$, with $\lambda=0,1$ in (a) and (b), respectively. The dashed lines correspond to the theoretical mass distributions of fields in the reduced theory, \eqref{eq:scalar_mass_spectrum} evaluated with \eqref{eq:fermionSpectrumNA0}. We compare with the histograms obtained numerically from the flux ensemble of the $\mathbb{WP}^4_{[1,1,1,1,4]}$ model.}    
     \label{fig:massSpecrumRedNA0}
\end{figure}

\begin{figure}[t]
    \centering
  \includegraphics[width=0.7\textwidth]{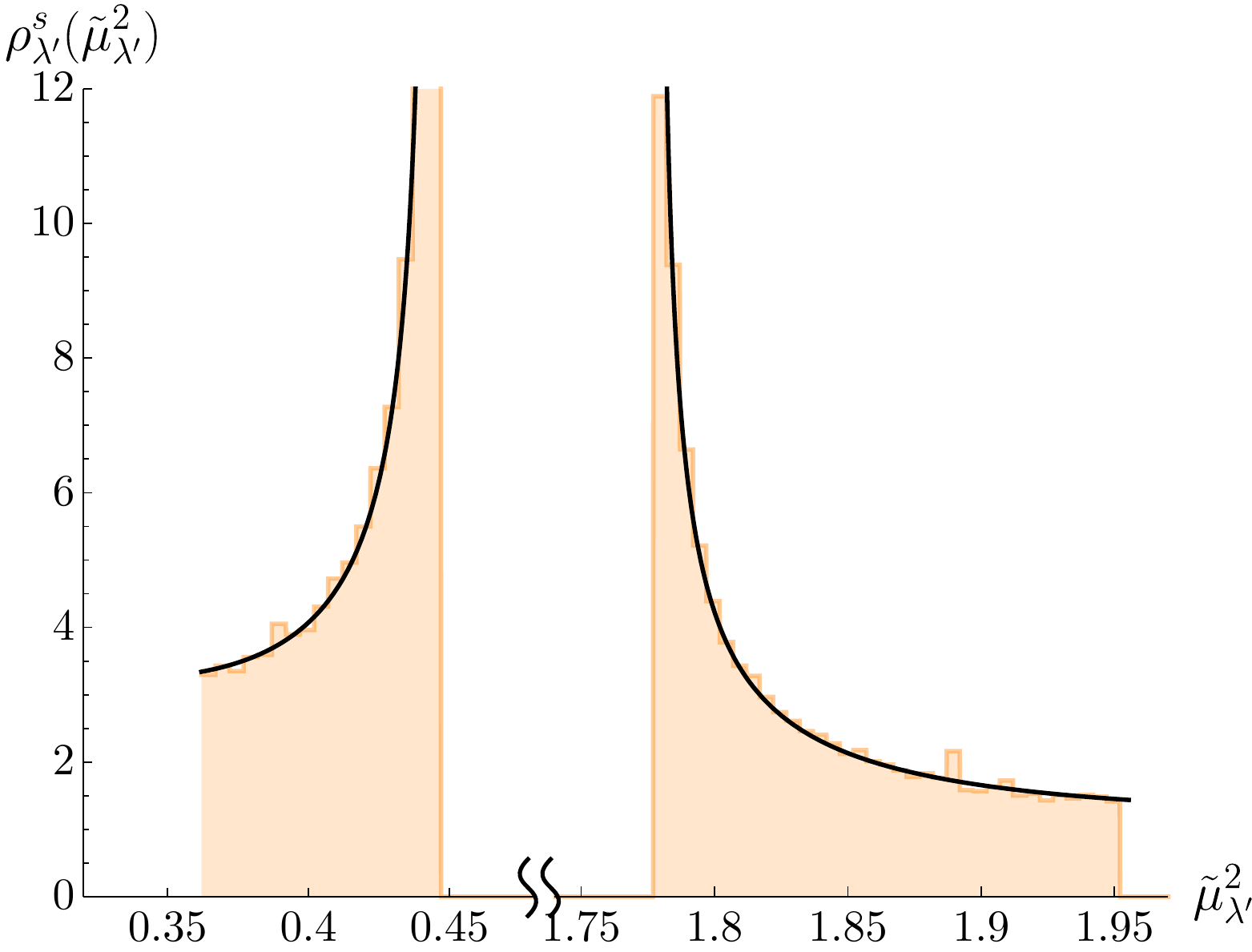}
    \caption{Theoretical prediction (solid line) for the probability distribution of the normalised squared  masses of the truncated scalar fields $\tilde \mu_{\lambda'}^2 = \mu_{\lambda'}^2/m_{3/2}^2$, eqs. \eqref{eq:scalar_mass_spectrum} and \eqref{eq:fermionSpectrumNA0}, in the constrained ensemble. For comparison we display the histogram of values obtained by applying \eqref{eq:gralSpectrum} to the vacua ensemble.}
\label{fig:massSpectrumTruncNA0} 
\end{figure}

As a curiosity, it is worth mentioning the small enhancement\footnote{This spike in the histogram of $\xi$ induces similar enhancements in the mass distributions displayed in figures \ref{fig:massSpecrumRedNA0} and \ref{fig:massSpectrumTruncNA0}, as they all depend on the former.} on the number of vacua with $\xi\approx 0.12$. An 
examination of these solutions reveals that they all correspond to flux configurations satisfying the relation 
$N_A^1=N_B^0$ and $\Im z=0$. Although we have not made further inquiries regarding the origin of the 
enhancement, it seems plausible that this particular choice of fluxes leads to a new symmetry in the 
EFT (exact or approximate), which is known to produce accumulations of no-scale solutions at special 
points of the moduli space \cite{DeWolfe:2004ns}. 

In complete analogy with the previous section, the density function \eqref{eq:xiDistNA0} can be used to derive the distributions for the rescaled fermion masses $ m_\lambda/m_{3/2}$. Performing a change of variables from $\xi$ to each of the normalised masses, and using \eqref{eq:gralFermionSpectrum32} with $\tan \theta_W = \sqrt{(1-2 \xi)/3}$ (see section \ref{fluxAlinged}), we obtain 
\be
\rho^f_\lambda(x_\lambda) dx_\lambda = \frac{2^{1/3}(1+\xi)}{(2-\xi)^2\xi^{2/3} \, (dx_\lambda(\xi)/d\xi)}\Big|_{\xi(x_\lambda)} d x_\lambda\,,
\label{eq:fermionSpectrumNA0}
\ee
where we use the shorthand $x_\lambda = m_\lambda/m_{3/2}$ which was introduced above.
The distributions for the squared scalar masses can then be found using \eqref{eq:scalar_mass_spectrum}, which 
are displayed along with the histograms derived from the numerical scan in figures \ref{fig:massSpecrumRedNA0} 
and \ref{fig:massSpectrumTruncNA0}. As in the case of the generic ensemble, the mass histograms of the scalar 
modes in the reduced theory, $\mu_{\pm 0}^2$ and $\mu_{\pm 1}^2$, have been obtained first by computing the 
eigenvalues of the fermion mass matrix \eqref{eq:sqFMM}, and then via \eqref{susyMAsses}. The histogram 
for the masses in the truncated modes $\mu^2_{\pm \lambda'}$ were found using 
\eqref{eq:N0Spectrum} instead. Here again we can observe the excellent agreement between the 
analytical predictions and the direct numerical computation of the masses in the octic. It is important to mention that, as in the case of the generic ensemble described above, it can be checked that the string coupling $g_s$ is statistically independent from the masses in \eqref{eq:N0Spectrum}. Thus, while the numerical histograms presented here correspond to an ensemble of marginally weakly coupled vacua $g_s\le1$, our conclusions remain valid also for very weakly coupled vacua with\footnote{We checked explicitly that this is indeed the case by computing the numerical mass histograms for the subset of solutions in the constrained ensemble with $g_s \le 0.1$ ($\sim 2500$ vacua). } $g_s\ll1$. 

The most important feature of these distributions is the divergence of the probability density for the masses $\mu_{\pm 1}^2 \ll m_{3/2}^2$ (see figure \ref{fig:massSpecrumRedNA0}  (b)). This implies that in this ensemble the branch of solutions with $\theta_W = \theta^W_1$, defined in \eqref{eq:criticalTheta}, occurs with relatively high frequency.  This contrasts with the results obtained for the generic ensemble, where the same branch  was shown to have a suppressed probability to appear.

\begin{figure}
    \centering
    \includegraphics[width=\textwidth]{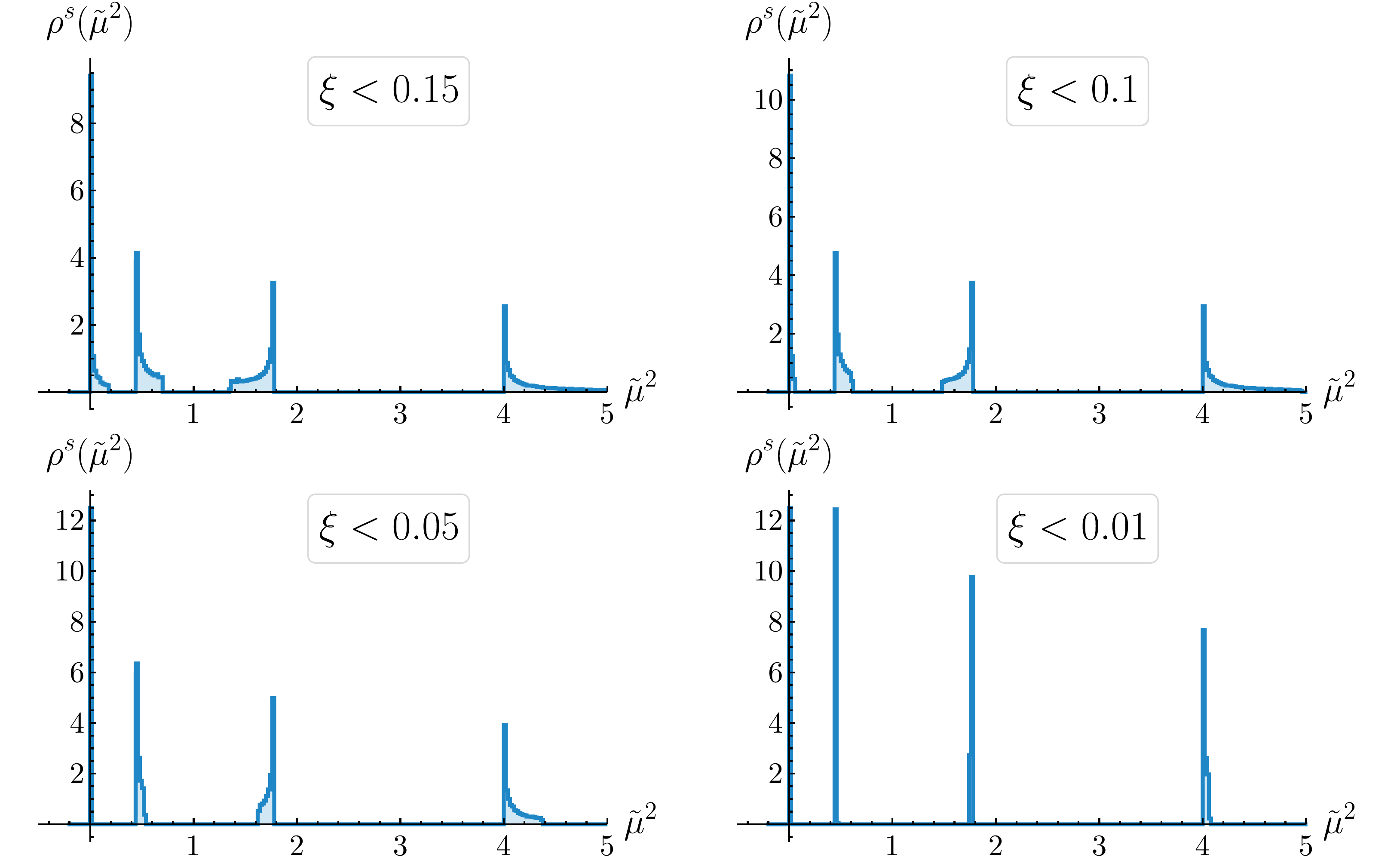}
    \caption{ 
    Histograms for the normalized squared masses $\tilde \mu_{\lambda=0,1}^2= \mu_{\lambda=0,1}^2/m_{3/2}^2$ of the scalars in the reduced theory. The plots represent vacua  in the constrained ensemble, with varying upper bounds on the LCS parameter $\xi$. Note that as the upper bound on $\xi$ decreases the distributions become increasingly deterministic, peaking at the limiting $\xi\to 0$ values given in \eqref{eq:na0_mu_LCS}. The plots also show  the presence of a light mode $\mu_{-1} \ll m_{3/2}$ in all vacua, whose mass  becomes zero $\mu_{-1}\to 0$ in the limit $\xi\to0$. }
    \label{fig:mass_evo}
\end{figure}

Finally, for completeness we have also studied the dependence of the mass spectrum on the distance of vacua from the LCS point. For this purpose, we obtained the mass histograms for subsets of no-scale solutions restricted to be in  neighbourhoods of the LCS point of varying sizes. The results are displayed in figure \ref{fig:mass_evo}, where we have plotted the histograms for four sets of vacua with $\xi \le \xi_\text{max}$, where $\xi_\text{max}= \{0.15,\, 0.1, \, 0.05, 0.01\}$. As it can be seen in the plots, the closer the solutions are to the LCS point, the more deterministic the mass distributions become. Note also that in the case $\xi \le 0.01$ the spectrum is already very peaked at the values given in \eqref{eq:na0_mu_LCS}, which correspond to the strict limit $\xi \to 0$. Interestingly, in this regime the spectrum always contains a (nearly) massless field, $\mu_{-1}^2\approx0$, which belongs to the reduced moduli space (i.e., $\theta_W \approx \theta^W_1$).

\subsection{Comparison with Random Matrix Theory}

Before we conclude this section let us comment on one further simplification, proposed in \cite{Denef:2004cf} to study the 
statistical properties of the flux ensemble. The statistical method we used above to analytically compute the probability distributions 
of the masses relies solely on the continuous flux approximation, but at the theory level it still requires  the knowledge of the couplings of 
the (reduced) EFT. In other words,  the prepotential \eqref{eq:F} needs to be computed. As we mentioned in the 
introduction, in typical Calabi-Yau compactifications the number of moduli fields in the complex structure sector can be 
of the order of hundreds, and in the absence of symmetries to simplify the analysis as done here, the computation 
of the prepotential can be prohibitively difficult. 

To address this problem, in \cite{Denef:2004cf} Denef and Douglas argued that for \emph{sufficiently complex compactifications}
with a \emph{large number of moduli fields}, the mass spectrum of perturbations could be well described in the framework of the 
4d EFT using methods of Random Matrix Theory (RMT). 
Since the universality theorems of RMT ensure that the statistical properties of the mass 
spectrum are independent of the distributions of couplings in the EFT, such a characterisation would avoid the challenges posed by a 
detailed computation. These ideas where developed in \cite{Marsh:2011aa,Bachlechner:2012at}, and RMT models designed 
specifically to describe no-scale vacua  of type-IIB compactifications were presented in \cite{Sousa:2014qza,Achucarro:2015kja}. 

Let us consider for definiteness the normalised fermion mass matrix $\cM/m_{3/2}$. The main prediction of RMT is the 
expectation value of the eigenvalue spectrum, which can be given in terms of the collection of $n=h^{2,1}+1$ rescaled fermion 
masses,
$x_\lambda \equiv m_\lambda/m_{3/2}$, sorted as $x_0\le x_2\le \ldots \le x_{n-2} \le x_{n-1}$. More specifically, the spectrum of random matrices is usually 
characterised by the empirical eigenvalue density function $\rho(x)$, or its alternative (non-standard) definition $\sigma(x)$, which are defined as
\be
\rho(x) \equiv \frac{1}{n}\Big\la \sum_{\lambda=0}^{n-1} \delta(x-x_\lambda)\Big\ra\,, \qquad \sigma(x) \equiv \frac{1}{n}\sum_{\lambda=0}^{n-1} \delta(x-\la x_\lambda\ra)\,,
\label{eq:RMTdef}
\ee
with the average taken over a finite sample of vacua randomly drawn from the ensemble. The alternative definition will be useful below.

The $\rho(x)$ and $\sigma(x)$, which are randomly distributed in the ensemble of flux vacua, become increasingly deterministic as the 
number of fields grows (as $n\to \infty$).
The RMT 
models \cite{Bachlechner:2012at,Sousa:2014qza,Achucarro:2015kja} describing the axio-dilaton/complex structure sector of type-IIB compactifications 
predict the fermion mass spectrum at no-scale vacua to be
\be
\lim_{n\to\infty} \rho(x) =\lim_{n\to\infty} \sigma(x)= \frac{4}{\pi x_h^2} \sqrt{x_h^2-x^2}\,, \qquad x\le x_h\,,
\label{eq:RMTprediction}
\ee
where $x_h$ is a free (model dependent) parameter which determines the typical ratio of the fermion masses to the gravitino mass.\footnote{The parameter $x_h$ is closely related to $\tan \theta_W$, or equivalently to the typical value of $W_0$ \cite{Sousa:2014qza,Achucarro:2015kja}.}

\begin{figure}[t]
    \centering\hspace{-.0cm}
        \includegraphics[width=0.5\textwidth]{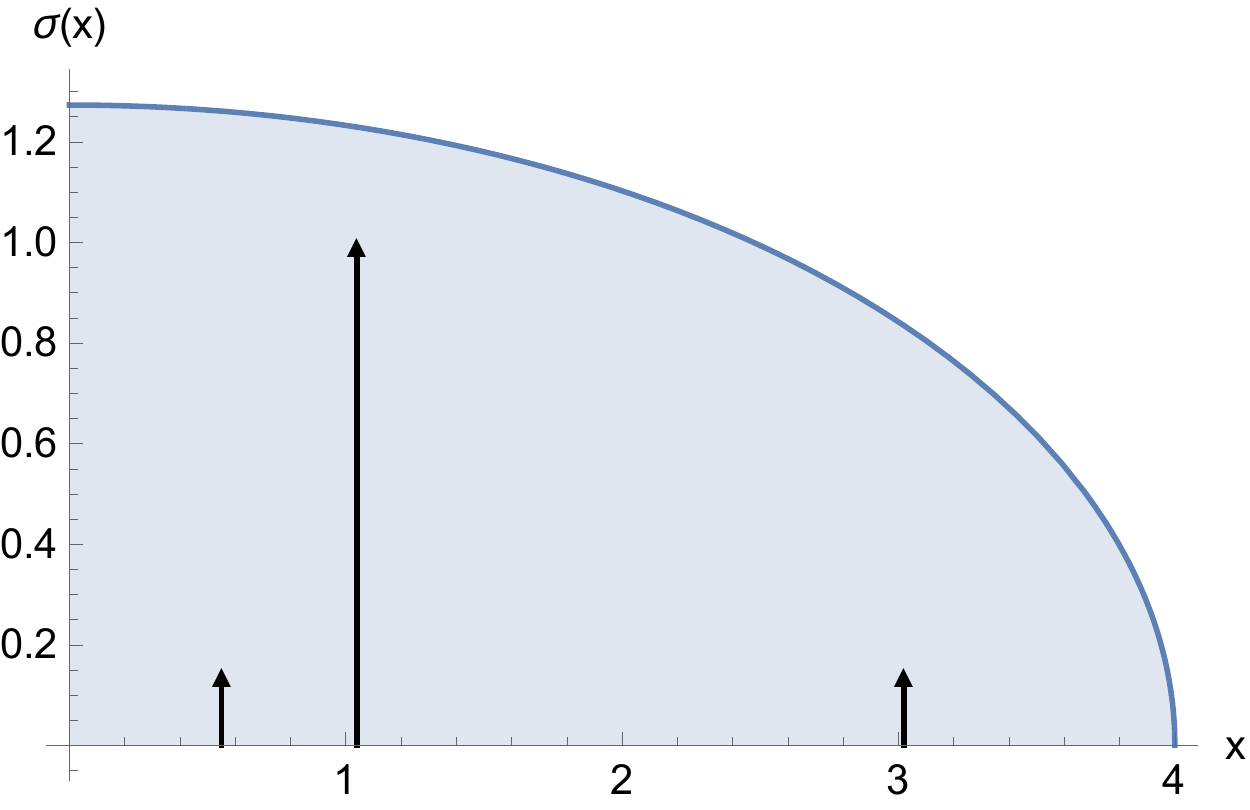}
    \caption{Eigenvalue density function $\sigma(x)$ for the rescaled fermion mass matrix $\cM/m_{3/2}$, with $x=m/m_{3/2}$ . The blue shaded area represents the density function $\sigma(x)$ predicted by Random Matrix Theory \eqref{eq:RMTprediction} ($x_h=4$). The arrows represent the averaged spectral density obtained numerically for the octic model, which is composed of three Dirac-delta contributions at 
    $\la x_\lambda\ra \approx\{0.55,\, 1.04,\, 3.02\}$
    with weights $\{\frac{1}{150}{\scriptstyle (\times20)},\, \frac{148}{150},\, \frac{1}{150}{\scriptstyle (\times20)}\}$.} 
\label{fig:RMT}
\end{figure}

In figure \ref{fig:RMT} we have displayed the RMT prediction together with result of evaluating
$\sigma(x)$ on the numerical ensemble of generic no-scale vacua constructed above for the octic model. The plot 
clearly shows that RMT is not an appropriate choice to represent the fermion mass spectrum in the model at hand. In particular, 
while RMT predicts a spectrum continuously distributed on its support $x \in [0,x_h]$, the expectation value of density 
function \eqref{eq:RMTdef} on the octic flux ensemble consists of three Dirac deltas centered at 
$\la x_\lambda\ra \approx\{0.55,\, 1.04,\, 3.02\}$
with weights given by $\{\frac{1}{150},\, \frac{148}{150},\, \frac{1}{150}\}$ respectively. The values $\la x_\lambda\ra$ are just the averages of the three distinct masses of the spectrum \eqref{eq:gralFermionSpectrum32}. Similar discrepancies were previously 
reported in \cite{Brodie:2015kza,Marsh:2015zoa}.

This negative result is in stark contrast with the successful characterisation provided by the continuous flux approximation 
observed in the previous section. The reason underlying the failure of RMT is that the first assumption on which it is based 
is not applicable here: The large degree of symmetry of the model yields a great simplification of the EFT, which is at odds with
the requirement of complexity\footnote{Note, e.g., that in our numerical example all  the truncated 148 fermions have the exact same mass.}. The 
very same property that allowed us to perform a fully analytical characterisation of the octic 
model prevents RMT from accurately describing the eigenvalue density. 
 
In view of this result, it would be therefore interesting to analyse the statistics of more complicated models, and check if the 
RMT tools become more useful there. By contrast, the present work has already shown that the continuous flux approximation 
works remarkably well, even in simple models with only a moderately large number of active fluxes (only 8 flux integers in the octic). 

\section{More general compactifications}
\label{sec:otherModels}

The main focus of this paper is the study of compactifications invariant under a group of symmetries which effectively reduce 
the complex structure moduli sector to a one-parameter space. In the present section we will discuss the possibility of finding 
the universal spectrum \eqref{eq:gralSpectrum} derived above in more general models. This could include cases where the reduced theory 
involves more than one complex structure field, or where there are no symmetries to reduce the number of moduli. 

The fundamental property that has allowed us to derive the spectrum \eqref{eq:gralSpectrum} for the compactifications 
discussed above is the possibility of truncating all complex structure fields except one. As a direct consequence of this, 
the vector of derivatives of the superpotential $Z_{0a}$ points along the no-scale direction $e_1^a$. This fact, together
with some generic properties of type-IIB compactifications at large complex structure, led us to the full mass 
spectrum. Thus, in order to generalise this result to more general compactifications we just have to find out 
under which circumstances it is possible to find vacua where $Z_{0a}$ aligns with the no-scale direction.
 
We will now show that, provided we neglect the quantisation of fluxes, it is possible to find no-scale vacua with the 
spectrum \eqref{eq:gralSpectrum} in the LCS regime of any compactification of type-IIB superstrings. More 
specifically, \emph{at any point of the moduli space} $\{\tau_0,z^i_0\}$ parametrised by the axio-dilaton and the 
$h^{2,1}$ complex structure moduli, it is possible to find a two complex dimensional family of fluxes $\{N_A^I,N^B_I\}$ 
such that the point $\{\tau_0,z^i_0\}$ is a no-scale vacuum, where the corresponding set of $2(h^{2,1}+1)$ tree-level 
masses is given by \eqref{eq:gralSpectrum}. 
 
To see this we can begin with the Hodge decomposition of the flux vector $N$ (see appendix \ref{app:HodgeDecomp}), 
which allows us to express this vector as a linear combination of the period vector and its K\"ahler covariant derivatives. 
This decomposition holds at any generic point of the moduli space $\{\tau,z^i\}$, excluding certain singular field 
configurations known as $D$-limits \cite{Ashok:2003gk}. In particular, at no-scale vacua where $ D_{ 0 }  W =D_a W=0$ 
the decomposition has the simpler form given in \eqref{eq:hodgeN1}. Conversely, at any given point of the moduli 
space $\{\tau_0,z_0^a\}$, any flux configuration which can be expressed as a linear combination
\be
N = \bar a^0 \bar \Pi + a^a \, D_a \Pi\,,
\label{eq:susyHodge}
\ee
is guaranteed to satisfy the no-scale vacuum equations \eqref{eq:susyeqs} at that point. In other 
words, given an arbitrary point in the moduli space, we can always choose a flux vector to make that point a no scale vacuum. The 
possible choices of flux in that case are in one-to-one correspondence with a set of $h^{2,1}+1$ complex parameters $\{a^0,a^a\}$ which can be freely chosen. 
In particular, we can use this freedom to make the complex vector $a^a$ point along the no scale direction $e_1^a$. 
Then, comparing \eqref{eq:susyHodge} with equation \eqref{eq:hodgeN1} we see that the vector $Z_{0a}$ will also 
point along $e_1^a$, which is precisely the condition that guarantees that total the tree-level spectrum is given by \eqref{eq:gralSpectrum}. 

To find such a choice of flux explicitly we can use \eqref{eq:noscaleDirection}, which implies that the 
vector $K^a$ points along the no-scale direction. Choosing the direction of $a^a$ accordingly, and using the 
definitions of $\theta_W$ and $m_{3/2}$ we obtain the following parametrisation for fluxes consistent with the no-scale 
spectrum \eqref{eq:gralSpectrum}
\be
N = m_{3/2}\, \left( \rmi \rme^{\rmi \alpha_W} \bar \Pi -\frac{1}{\sqrt{3}} \rme^{-\rmi \alpha_K} \tan \theta_W \sqrt{(1- 2 \xi)} K^i D_i \Pi \right) \frac{ \sqrt{\pi}\cV \,\rme^{K_{cs}/2} }{ g_s}\,,
\label{eq:gralVacua}
\ee
where $\alpha_W = \arg(W)$ and $\alpha_K = \arg(D_0 D_1 W)$. Note that in this formula the Calabi-Yau volume $\cV$, 
the string coupling $g_s$, the expectation value of the K\"ahler potential and the LCS parameter are all determined by 
the configuration of moduli fields. Nevertheless, we are still free to select the gravitino mass scale $m_{3/2}$, the 
angular parameter $\theta_W$ and the phases $\alpha_W$ and $\alpha_K$. Thus, as anticipated above, for a given 
Calabi-Yau geometry the set of fluxes compatible with the spectrum \eqref{eq:gralSpectrum} has complex dimension two. 

It is important to emphasize that the vector $N$ obtained by this method will not be in general compatible with the 
quantisation of fluxes. Our ability to tune the parameters $\{a^0, a^a\}$ in \eqref{eq:susyHodge} will be limited by 
the maximum value of the flux $D3$-charge, which is set by the tadpole constraint. Indeed, since the fluxes can only 
be changed in integral steps, and the maximum value they can attain is of the order of $\sqrt{N_{\text{flux}}}\le \sqrt{L}$, we expect 
that the maximum accuracy that may be achieved in aligning $Z_{0a}$ to the no scale direction, i.e., making $Z_{0a'}$ small, 
is given by 
\be
|Z_{0a'}|/|Z_{01}| \sim \cO(1/\sqrt{N_{\text{flux}}}) \gtrsim \cO(1/\sqrt{L}) \sim \cO(10^{-1}-10^{-2}).
\ee
Therefore, in generic compactifications we will only be able to find no-scale vacua with quantised fluxes that satisfy 
\eqref{eq:gralVacua} approximately. Still, with a large flux tadpole,
and provided the moduli space is non-singular at the point of interest so that the Yukawa couplings $\mathring \kappa_{abc}$ 
have a finite value, it might be possible to find vacua where the corrections to the spectrum \eqref{eq:gralSpectrum} 
are small, $\delta m_{\lambda'} \ll m_{\lambda'}^{(0)}$.
To estimate the magnitude of these corrections we can consider the diagonalisation of the fermion mass matrix 
\be
(Z^\dag Z)_{A\bar B} =(Z^\dag Z)_{A\bar B}^{(0)} + \delta (Z^\dag Z)_{A\bar B}+ \ldots
\ee
to first order in perturbation theory, where $(Z^\dag Z)^{(0)}$ is the matrix obtained setting $Z_{0a}^{(0)} = \delta_a^1 \, Z_{01}$, 
and we regard the remaining components $Z_{0a'}$ as small deformation. The perturbed eigenvalues are then given 
by $\delta  m_\lambda^2 = \delta (Z^\dag Z)_{\lambda \bar \lambda}$. A straightforward computation shows that 
the first-order corrections to the rescaled fermion masses $\tilde m_{\lambda} = m_{\lambda}/m_{\text{susy}}$ are 
\be
\delta \tilde m_0 =\delta \tilde m_1 =0\,, \qquad \delta \tilde m_{\lambda'} = - \mathring \kappa_{\lambda' \lambda' a'} \Re\left(\frac{\delta Z_{0a'}}{Z_{01}}\right)\,,
\ee 
where we made use of \eqref{eq:Yukawas}, and for simplicity we define $m_{\text{susy}} = |Z_{01}|$. Without further 
computations we can already see that generically, these corrections will lift the degeneracy of the fermion masses in the 
truncated sector.

To understand under which circumstances these deformations can be regarded as small corrections, it is fundamental to 
have some estimate of the size of the Yukawa couplings. We will further investigate these issues in future publications.
 
\section{Conclusions}
\label{sec:conclusions}

No-scale vacua of type-IIB flux compactifications are an essential stepping stone in the construction of dS vacua and inflationary 
models in KKLT and Large Volume Scenarios. Guaranteeing the validity of these constructions requires a good understanding 
of the perturbative spectrum of the no-scale solutions. Indeed, while the no-scale property ensures the absence of tachyons 
in the axio-dilaton/complex structure sector at tree-level, this does not prevent the existence of arbitrarily light fields, which 
may turn tachyonic upon including quantum corrections, uplifting terms, or the effect of matter fields 
\cite{Conlon:2006tq,Achucarro:2015kja}. These light modes might also lead to difficulties when implementing viable 
inflationary models in these scenarios, as the backreaction effects caused by the inflaton might also result in their destabilisation.

In generic situations, the absence of such light fields can be argued using scaling arguments in the KKLT and LVS settings, as done 
in \cite{Gallego:2009px,Gallego:2008qi,Abe:2006xi} and \cite{Balasubramanian:2005zx,Conlon:2005ki}, respectively. However,
the most interesting vacua for phenomenological applications are often those in the neighbourhood of special (and thus non-generic) points 
of the moduli space, such as the conifold or the LCS points, where the presence of light fields 
may become unavoidable \cite{Bena:2018fqc,Demirtas:2019sip}. Moreover, as argued in \cite{Achucarro:2015kja}, in 
the specific case of LVS vacua with an only moderately large compactification volume 
\cite{Rummel:2014raa,Rummel:2013yta,Maharana:2015saa,Cicoli:2015wja}, typical vacua might still contain a sizeable 
fraction of light modes susceptible to becoming tachyonic. 

Despite its importance, a complete analytic understanding of the perturbative spectrum at no-scale vacua has remained 
elusive, primarily due the complexity of the corresponding EFTs and the large number of fields involved. In the present 
paper we have considered a particular class of Calabi-Yau compactifications with an arbitrary number of moduli fields, and 
computed \emph{analytically} the complete mass spectrum of the axio-dilaton/complex-structure sector at no-scale vacua 
in the LCS regime (see \eqref{eq:gralSpectrum}). The Calabi-Yau geometries we considered here are invariant 
under large discrete isometry groups, which allows for a consistent reduction of the complex structure sector to a single field. An 
important feature of this class of models is that the Calabi-Yau symmetries make the computation of an EFT for the 
unique complex structure modulus surviving the truncation feasible~\cite{Candelas:1990pi,Candelas:1990rm,Klemm:1992tx,Font:1992uk}. Then, 
using only symmetry arguments, together with certain universal properties satisfied by the EFT couplings at LCS, we derived the 
mass spectrum of the full axio-dilaton/complex structure sector, including the truncated fields. Remarkably, the full spectrum can 
be expressed solely in terms of the couplings of the reduced EFT theory, which can be determined. This result applies to plenty 
of interesting compactifications such as: the family of quintic hypersurfaces in $\mathbb{WP}_{[1,1,1,1,1]}$ \cite{Candelas:1990rm} admitting 
the discrete symmetry groups discussed in \cite{Doran:2007jw}; the close relatives to the quintic (i.e., the sextic, octic and dectic) with 
analogous symmetric configurations \cite{Klemm:1992tx,Font:1992uk} and quotients thereof \cite{Klemm:1992tx,Candelas:2017ive}; or the Complete Intersection Calabi-Yau
described in \cite{Braun:2011hd} and its quotients (see Table \ref{table:ModelSelection}). Moreover, we can also use these results to describe the LCS regime of the 
hundreds of one-parameter models listed in \cite{Batyrev:2008rp}. We should remark that  these discrete global symmetries are nevertheless expected to be broken upon including all sub-leading $\alpha'$ and quantum effects \cite{Banks:1988yz,Kallosh:1995hi,Banks:2010zn}, what will induce small corrections in the spectrum \eqref{eq:gralSpectrum},  lifting, in particular, the large degeneracy of the truncated sector. 

A crucial step in the derivation of the spectrum is the computation of the Yukawa couplings which determine the fermion masses 
at LCS. While the universal behaviour of the canonically normalised Yukawa couplings in the strict LCS limit has been well known 
for a long time \cite{Cremmer:1984hj}, here we have extended those results to the complete LCS regime. That is, we have computed 
the relevant subset of these couplings in the whole region of moduli space where the instanton corrections to the complex structure K\"ahler potential and 
flux superpotential can be safely neglected\footnote{This regime is sometimes referred to as the \emph{nilpotent orbit} approximation, 
and the strict-LCS as the $sl(2)$-\emph{orbit} approximation \cite{Grimm:2019ixq}.}. As in the strict LCS limit, we found that the
Yukawa couplings exhibit a universal behaviour, independent of the details of the compactification or the number of complex structure 
fields (see \eqref{eq:Yukawas}). It is important to emphasize that this result applies to any Calabi-Yau compactification, and 
does not require the invariance of the manifold under a large discrete isometry group.

In the class of models that we consider here, the strict LCS/weak-coupling limit is of particular interest, as it is the region of moduli 
space where one has the best perturbative control of the EFT. In \cite{Brodie:2015kza,Marsh:2015zoa}, the authors considered 
compactifications of type-IIB and F-theory at the strict LCS limit where the superpotential was dominated by its cubic or quartic terms, and 
proved the absence of vacua (AdS, dS or Minkowski) in this region of moduli space. More specifically, the no-scale potential was 
shown to satisfy the relation $|\del V| = (\sqrt{7}/2) \, V > 0$, consistent with the de Sitter conjecture in \cite{Ooguri:2018wrx}, which 
forbids the existence of de Sitter vacua when approaching an infinite distance limit in moduli space (see \cite{Garg:2018reu} for the necessary second-derivative conditions).

These conclusions can nevertheless be avoided by setting to zero the flux associated with the period which grows without bound in this 
limit ($N_A^0\equiv f_A^0-\tau h_A^0=0$ in \eqref{fluxW}). Indeed, in this case the higher order terms in the flux superpotential 
are identically zero, and thus the above no-go theorem does not apply. In section \ref{fluxAlinged} we computed the mass spectrum 
at this class of no-scale vacua for the models described above, and proved it to have the universal form in the strict LCS limit. Namely, 
the spectrum of squared masses in the axio-dilaton/complex structure sector is given by
\be
\text{Spectrum at the \underline{strict LCS limit}:} \qquad \mu^2_\lambda =\Big\{0, \, \frac{4}{9} m_{3/2}^2,\, \frac{16}{9} m_{3/2}^2,\, 4 m_{3/2}^2\Big\},
\label{eq:detSpectrum}
\ee
with the first and last masses appearing with multiplicity $1$, and each of the other two with multiplicity $h^{2,1}+1$. In particular it 
can be observed that the spectrum always contains exactly one massless field, while the rest of the moduli have masses of the 
order of the gravitino mass $m_{3/2}$. It is also worth mentioning that closely related classes of vacua surviving in the strict LCS limit were 
also discussed in \cite{Magda2,Danielsson:2006xw}, and in particular those of \cite{Demirtas:2019sip} also present a massless field 
in the no-scale spectrum, which is nevertheless lifted by instanton corrections.

The previous results are consistent with \cite{Junghans:2018gdb,Grimm:2019ixq}, where it was argued that
obtaining vacua parametrically close to the LCS point requires turning on \emph{unbounded fluxes}, that is, fluxes not contributing 
to the total $D3$-charge and therefore unconstrained by the tadpole condition. Furthermore, as discussed in \cite{Grimm:2019ixq}, the contribution to the flux potential due to the unbounded fluxes must also be asymptotically vanishing in the strict LCS limit. Interestingly, the class of no-scale solutions (and thus Minkowski vacua) described above satisfies both of these 
conditions, and is therefore consistent with the no-go theorems derived in \cite{Grimm:2019ixq} (see section \ref{fluxAlinged}). On the 
one hand, setting to zero the flux $N_A^0\equiv f_A^0- \tau h_A^0$ on the diverging period implies that the flux on the dual 
$B$-cycle, i.e., $N^B_0\equiv f_0^B-\tau h_0^B$, does not contribute to the tadpole. On the other hand, as the term in the flux superpotential associated to   $N_0^B$  is just a constant, it also follows that its contribution to the no-scale potential is  asymptotically vanishing  at the LCS point. Note that the 
analyses in \cite{Junghans:2018gdb,Grimm:2019ixq} only refer to the strict LCS limit, while the results presented here also allow one to
characterise the properties of the no-scale potential away from the LCS point, i.e., over a region of moduli space not captured in those works.

For generic flux vacua, not necessarily close to the LCS point, the mass spectrum will not have the deterministic form 
of \eqref{eq:detSpectrum}, and thus will in general be dependent on the choice of fluxes. Therefore, in order to obtain a characterisation 
of the spectrum independent of the choice of flux we have studied the statistical properties of the moduli masses in the ensemble of flux 
vacua. More specifically, using the continuous flux approximation, we computed analytically the probability distributions for the density of 
vacua and the masses, both for the generic ensemble of vacua and for the constrained ensemble with vanishing flux $N_A^0$. Moreover, we 
verified the validity of the obtained distributions by comparing them with the result of a numerical scan on the octic model 
$\mathbb{WP}^4_{[1,1,1,1,4]}$. As can be seen in figures~\ref{fig:fieldSpaceDist}--\ref{fig:massSpectrumTruncNA0}, the analytical and 
empirical distributions show an excellent agreement in the expected regime of applicability of the continuous flux approximation.

Regarding the density of vacua,
for the generic ensemble the result of the numerical scan in the octic model shows a suppression on the density of vacua close to 
the LCS point with respect to the statistical predictions (see figure~\ref{fig:genericXi}). This discrepancy with the theoretical distributions 
was nevertheless already anticipated in \cite{Brodie:2015kza,Marsh:2015zoa} (see also \cite{Denef:2004ze}), due to a breakdown 
of the continuous flux approximation. In the generic ensemble, vacua with $N_A^0=0$ represent only $0.08\%$ of 
total vacua and, as we mentioned above, the results of \cite{Brodie:2015kza,Marsh:2015zoa} show that only vacua with vanishing 
flux $N_A ^0$ may be found parametrically close to the LCS point. By contrast, as can be seen in figure~\ref{fig:xiDistNA0}, the constrained
ensemble exhibits no suppression near the LCS point, indicating that this subclass of solutions will dominate in 
this region of the moduli space. 

Away from the LCS point, the computed probability distributions based on the continuous flux approximation describe very accurately 
the result of the numerical scan. Therefore, we can use these results to have a precise analytic understanding of the features displayed 
by the mass spectra observed in the flux ensemble. In the case of the generic ensemble, when considering only the reduced 
theory, we observe that vacua with modes much lighter than the gravitino appear with very low frequency (figure \ref{fig:genericMassSpectra1}),  
$\text{Prob}(\mu^2_{\text{red}} < 0.01 \,m_{3/2}^2) \approx 2\%$. This result is actually a well-known consequence of the generalized 
Kac-Rice formula, which characterises the density of critical points in random fields (see \cite{adler2009random,Mehta:2015nva}). However, 
when considering the truncated sector, the situation changes: The mass distribution of the truncated fields diverges 
in the limit $\mu^2\ll m_{3/2} ^2$ (see figure~\ref{fig:genericMassSpectra2}), indicating that a sizeable fraction of vacua contain light 
fields in this sector, $\text{Prob}(\mu^2_{\text{trunc}}<0.01 \, m_{3/2}^2) \approx 11\%$, with masses as low as $\mu^2_{\text{trunc}} \sim 10^{-10} \, m_{3/2}^2$. 
The reason for this is that, in the models we consider, only the critical points of the reduced scalar potential 
can be described as extrema of random fields, and thus appropriately characterised by the Kac-Rice formula. By contrast, the expectation 
values of the fields in the truncated sector are fully determined by the action of the Calabi-Yau symmetry group, and thus the Kac-Rice formula 
cannot be used to obtain the distribution of extrema for these moduli, or their mass spectra.

Concerning the statistics of vacua in the constrained ensemble, our results show that the mass spectra change significantly 
due to the condition imposed on the fluxes. In particular, contrary to the generic ensemble, in this class of vacua the lightest field is always 
in the reduced moduli space (see figure~\ref{fig:massSpecrumRedNA0}). In order to understand the dependence of the spectra on the distance 
to the LCS point, we considered subsets of vacua constrained to be in neighbourhoods of this point with varying sizes. This analysis 
showed that the smaller the neighbourhood around the LCS point, the more deterministic the mass spectrum becomes, recovering the 
limiting form \eqref{eq:detSpectrum} in the strict LCS limit. In other words, for the dominant class of vacua near the LCS point, 
the spectrum was always observed to contain a very light (and asymptotically massless) field in the reduced moduli space (see figure \ref{fig:mass_evo}).

Finally, let us comment on the applicability of our results. As we mentioned above, our results can be used to describe no-scale vacua in 
compactifications invariant under a large discrete isometry group. Therefore, it would be desirable to understand if the class of vacua 
with an analytic spectrum discussed here can be embedded in more general compactifications.
As we discussed in section \ref{sec:otherModels}, as long as we neglect the quantisation of the fluxes,  in 
\emph{any Calabi-Yau compactification}, and \emph{for every point of the moduli space at the LCS regime}, it is possible to find an 4-real-dimensional 
family of ISD fluxes (i.e., satisfying the no-scale condition), such that the mass spectrum in the axio-dilaton/complex structure sector is 
given by \eqref{eq:gralSpectrum}. This already suggests that these vacua might be encountered in compactifications where the $D3$-charge 
tadpole is large, as in type-IIB compactifications arising as the orientifold limit of F-theory compactified on a fourfold. Embedding this class 
of vacua in generic compactifications, while retaining the flux quantisation condition, is an interesting problem that we will address in future 
work. To conclude we will briefly comment on the possible extension of our results to other regimes away from the LCS limit. In general, such an analysis would require a specific treatment which is beyond the reach of the present analysis, as our derivations depend crucially on the  universal properties satisfied by the couplings of the EFT at LCS. However, in the specific case of conifold limits of the moduli space, it might be possible to make some progress following an analogous procedure to the one described in \cite{Demirtas:2020ffz,Blumenhagen:2020ire} (see also \cite{Crino:2020qwk}). In those works, it was explicitly demonstrated that one can stabilise a subset of the complex structure moduli near a conifold point, while fixing the rest near the  LCS point, i.e., at a \emph{conifold-LCS regime}. As shown in  \cite{Demirtas:2020ffz,Blumenhagen:2020ire}, provided the moduli at LCS are sufficiently massive, the stabilisation of this sector can be treated independently, ignoring consistently the presence of moduli near the conifold limit to leading order.  Therefore, another interesting future direction would be to study the application of our results  to characterise the spectrum of those complex structure at LCS for compactifications in a conifold-LCS regime, such as those described in \cite{Demirtas:2020ffz,Blumenhagen:2020ire}.

\section*{Acknowledgments}

We are grateful to Igor Bandos, I\~naki Garc\'ia-Etxebarria and Irene Valenzuela for useful suggestions and discussions. We also thank to Igor Broeckel and  Savdeep S. Sethi for  comments on the preprint. This work is supported in part 
by the Spanish Ministry MCIU/AEI/FEDER grant (PGC2018-094626-B-C21), the Basque Government grant (IT-979-16) 
and the Basque Foundation for Science (IKERBASQUE). KS is supported by the Czech science foundation 
GA\v CR grant (19-01850S). MAU is also supported by the University of the Basque Country grant (PIF17/74). 
The numerical work carried out in this paper has been possible thanks to the computing infrastructure of the 
ARINA cluster at the University of the Basque Country, (UPV/EHU).

\appendix

\section{Hodge decomposition of the flux vector}
\label{app:HodgeDecomp}

In this appendix we review the Hodge decomposition of the flux vector $N$ \cite{Denef:2004cf}. This decomposition
was used in sections \ref{fluxAlinged} and \ref{sec:otherModels} of the main text, and is also the starting point for 
the derivations of the probability distributions of the type-IIB flux ensemble. 

The flux vector $N$ has complex dimension $2h^{2,1} +2$ and transforms non-trivially under the symplectic group 
$\mathrm{Sp}(2h^{2,1} +2,\mathbb{Z})$, i.e., it is a \emph{symplectic section}. As we reviewed in section \ref{typeIIB}, the 
set of $2 h^{2,1}+2$ vectors $\cB = \{\Pi,\bar \Pi, D_a \Pi, D_{\bar a} \bar \Pi\}$ evaluated at any given point $\{\tau,z^a\}$ is 
also composed of symplectic sections, which can be shown to be linearly independent. In other words, the set $\cB$ forms a 
basis in the space of sections. To prove the linear independence of the elements of $\cB$ we introduce the symplectic 
product $\la A,B \ra$ of two sections $A$ and $B$,
\be
\la A ,B \ra = A^T \cdot \Sigma \cdot B, 
\ee
where $\Sigma$ is the symplectic invariant matrix \eqref{symInvMat}. Then, it can be checked from the definition of $\Pi$ that the elements of $\cB$ satisfy the orthogonality relations
\bea
\la\Pi, \bar \Pi\ra &=& \rmi \rme^{-K_{cs}}\,, \nonumber \\
\la\bar \Pi, \bar \Pi\ra &=&0\,, \nonumber \\
\la \Pi, D_a \Pi\ra &=&0\,, \nonumber \\
\la \Pi, D_{\bar a} \bar \Pi\ra &=&0\,, \nonumber \\
\la D_a\Pi, D_{b} \Pi\ra &=&0\,, \nonumber \\
\la D_a\Pi, D_{\bar b} \bar \Pi\ra &=&-\rmi\rme^{-K_{cs}}\delta_{a\bar b}\,,
\label{eq:hodge_prod}
\eea
from which the linear independence of the set $\cB$ follows. In this setting the Hodge decomposition of the flux vector can be obtained as the decomposition in the basis of sections $\cB$,
\be
N =\sqrt{4 \pi}( a_0 \Pi + \bar b_0 \overline \Pi + a ^a D_a \Pi + \bar b^{a} D_{\bar a}\overline \Pi\,).
\label{eq:hodgeNgen}
\ee
Using these orthogonality relations it is straightforward to find that the coefficients $\{a_0,a^a,b_0,b^a\}$ are determined by the values of the superpotential and its derivatives at the point $\{\tau,z^a\}$
\be
a^0 = - \rme^{K_{cs}} D_{\bar 0 } \bar W, \qquad \bar b^0 = \rmi \rme^{K_{cs}} W, \qquad a^a =\rme^{K_{cs}} D_{\bar 0} D_{\bar a} \bar W, \qquad \bar b^a = - \rmi \rme^{K_{cs}} D_a W. 
\ee
Therefore, the basis elements $\Pi$, $D_a \overline \Pi$, $D_{\bar a} \Pi$ and $\overline \Pi$ correspond to the $(3,0)$, $(1,2)$, $(2,1)$ and $(0,3)$ components of the flux $G_3=F_3- \tau H_3$, respectively. In particular, at no-scale vacua \eqref{eq:susyeqs}, which is the moduli space locus where the parts $(3,0)$ and $(1,2)$ of $G_3$ vanish (i.e., $G_3$ is ``imaginary self-dual''), the Hodge decomposition reduces to 
\be
N = \sqrt{4 \pi} \rme^{K_{cs}} (\rmi W \, \bar \Pi + D_{\bar 0} D_{\bar a} \bar W \, D_a \Pi)\,,
\label{eq:hodgeN}
\ee
Substituting this expression into \eqref{eq:tadpole_def} we can obtain an expression for the $D3$-charge induced by imaginary self-dual fluxes:
\bea
N_{\text{flux}} &=&-\rmi 4 \pi \rme^{2 K_{\text{cs}}+K_{\text{d}}} \left( \left| W \right|^2 \left\langle \Pi , \bar\Pi \right\rangle + \left| D_0 D_1 W \right|^2 
\left\langle D_{\bar{1}} \bar{\Pi} , D_1 \Pi \right\rangle \right. \nonumber \\[5pt]
&&\qquad \qquad \quad\; \; \left. +2 \Im \left[ \bar{W} D_{\bar{0}} D_{\bar{1}} \bar{W} \left\langle \Pi , D_1 \Pi \right\rangle \right] \right) \nonumber \\[5pt]
&=&4 \pi \rme^{K_{\text{cs}}+K_\text{d}} \left( \left| W \right|^2 + \left| D_0 D_1 W \right|^2 \right),
\eea
where, in the last step, we have 
applied the identities \eqref{eq:hodge_prod}. Finally, using the definitions of the gravitino mass and $m_\text{susy}$, we find $N_\text{flux}$ to be positive semidefinite, and given by
\be
0 \le N_{\text{flux}} =4 \pi \cV^2 \, \left(m_{3/2}^2 + m_{\text{susy}}^2 \right) \le L\,.
\ee

\section{Numerical method: Paramotopy}
\label{app:paramotopy}

In this appendix we describe in detail the numerical method used in this work to obtain the ensemble of no-scale solutions 
for the $\mathbb{WP}^4_{[1,1,1,1,4]}$ model, also known as the octic.
As discussed in section \ref{sec:example}, this model features a single complex structure modulus and an axio-dilaton, which we seek to stabilize at
no-scale configurations \eqref{eq:susyeqs}. 

It can easily be checked, using the machinery described in section~\ref{typeIIB}, that the no-scale conditions
can be expressed as a system of non-linear polynomial equations near the LCS point, where the instanton contributions to the 
prepotential (\ref{eq:octic_prep}) (or, more generally, (\ref{eq:F})) can be neglected \cite{MartinezPedrera:2012rs}. In the following 
we will denote this polynomial form of the no-scale conditions \eqref{eq:susyeqs} by 
\begin{equation}
P_i (z,\bar{z},\tau,\bar{\tau};f,h)=0, \quad i=\lbrace 1,2 \rbrace ,
\label{eq:param_poly}
\end{equation}
where $f$ and $h$ are the \emph{quantized} flux vectors defined in (\ref{eq:flux_defs}), and which are subject to the \emph{tadpole condition} (\ref{eq:tadpole1}). 
The main numerical difficulty of this problem lies in solving the polynomial system of equations (\ref{eq:param_poly}) for the huge number of allowed choices of $f$ and $h$.

\subsection{Polynomial homotopy continuation and Paramotopy}

In recent years, one of the most outstanding algorithms to solve systems of the form of (\ref{eq:param_poly}) has been that of \emph{polynomial homotopy continuation} (PHC), coined within the field of numerical algebraic geometry \cite{SW96,NSSP}. Schematically, this method works as follows:
\begin{enumerate}
    \item Given the set of polynomial equations to be solved, $P(x)=0$, the first step is to construct an auxiliary system of 
    equations, $Q(x)=0$, which is easily solvable and that has the same maximal number of solutions. We then define
    \begin{equation}
    H(x,t) = \gamma (1-t) Q(x) + t P(x),
    \end{equation}
    which is known as the \emph{homotopy} function, where $t \in [0,1]$ and $\gamma$ is a random complex number.\footnote{This parameter ensures no singularities will occur during the deformation of $Q(x)$ into $P(x)$. For more detail, see \cite{NSSP}.}
    
    Note that the roots of $H(x,0)$ correspond to those of $Q(x)$, while the roots of $H(x,1)$ are those of $P(x)$, the ones we are interested in. 
    
    \item Once the roots of $Q(x) \propto H(x,0)$ are determined, it can be shown that, as $t$ increases, the roots of $H(x,t)$ will be continuously deformed from their original values \cite{NSSP}. Thus, we can easily track the path each solution takes as we vary $t$ up to $t=1$, where the solutions correspond to the solutions to our problem. 
\end{enumerate}

Many different implementations of the PHC method can be found in the literature, such as \texttt{phcpy}~\cite{DBLP:journals/corr/abs-1907-00096}, \texttt{StringVacua}~\cite{Gray_2009}, and \texttt{Bertini}~\cite{BHSW06}. In this work, we have used \texttt{Paramotopy}\footnote{Software available at \url{www.paramotopy.com}}~\cite{bates2018paramotopy}, a highly efficient PHC-based algorithm specially suited for polynomial systems like (\ref{eq:param_poly}) which depend on parameter tuples. 

In short, \texttt{Paramotopy} works in two steps. Given a certain parametrically-dependent polynomial system $P(x;p)$, it first performs the above PHC algorithm for a random $p_0$. Once that has been solved, it performs the same algorithm for the homotopy
\begin{equation}
H (x,t;p_0,p)= \gamma (1-t) P(x;p_0) + t P(x;p)
\end{equation}
where $p$ corresponds to one of the parameter choices we are interested in. However, in this second run, the number of paths that have to be tracked is qualitatively smaller than in the first step, as only those paths that led to proper solutions of $P(x;p_0)$ have to be followed. In cases where the number of well-behaved paths is orders of magnitude lower than the maximum number of solutions, this second step proves to be crucial for an efficient solution \cite{bates2018paramotopy}. 

\subsection{Construction of the flux ensemble and search for no-scale solutions}

In order to perform a consistent exploration of the moduli space vacua of the octic, we took random integer flux values from a 
uniform distribution, that the components of the flux vectors satisfy $f, h \in [-f_{max},f_{max}]$. Only those flux tuples satisfying the tadpole condition 
\begin{equation}
0 < h \cdot \Sigma \cdot f \leq L
\end{equation}
were kept where, for our purposes, we took $L=972$, and chose $f_{max}=50$ to be sufficiently large for the distribution of the $D3$-charge $N_{\text{flux}}$ \eqref{eq:tadpole_def} to converge to a flat distribution. This way, we avoid artificially induced boundary effects associated to having set a finite value for $f_{max}$.

For the case of generic flux vacua, we generated $10^7$ flux tuples consistent with the tadpole condition. On the other hand, for the constrained case, we generated $10^6$ consistent flux tuples\footnote{Random fluxes are more prone to high corrections as opposed to those with $N_A^0=0$, mostly due to the difference in the number of solutions near the LCS point. Thus, to keep a considerable amount of solutions in the former case, we generated more flux tuples.} following the algorithm above, while manually keeping $f_A^0 = h_A^0 = 0$.

With these parameter choices in hand, we employed \texttt{Paramotopy} to solve the system of no-scale conditions given by \eqref{eq:susyeqs}. Note 
that \eqref{eq:susyeqs} involves both $\tau,z$ and $\bar{\tau},\bar{z}$. 
One possibility to deal with this would be to solve for the real and imaginary parts of each variable. However, here we solved for barred and unbarred 
variables separately, and then only kept those solutions which actually satisfied the conjugation relation between the variables. We found that this second 
choice was easier to solve by the software, and kept the equations simple. The whole process took around 5 hours in a 46-core machine for the constrained ensemble
($10^6$ tuples) and 50 hours for the generic ensemble ($10^7$ tuples). 

From the resulting ensemble of no-scale solutions, we only considered (in section~\ref{sec:statisticalAnalysis}) those with moderately small instanton corrections. 
When performing this cut in our ensemble data, we made sure that the EFT we used to describe the octic model is indeed reliable.
The sizes of these corrections were considered \emph{a posteriori}, once the tree-level equations had been solved. More specifically, at each 
of the obtained solutions, we computed the K\"ahler potential and K\"ahler metric, both neglecting entirely the instanton contributions to the 
prepotential and considering the leading correction \eqref{eq:instCorrectDom}. First, we  selected only those solutions for which the K\"ahler metric was still well defined  
 after including the corrections, i.e., where it was non-degenerate and positive. Then, with these solutions, we computed the gravitino mass $m_{3/2}$, rescaled Yukawa coupling $\mathring \kappa$, and K\"ahler metric (with and without considering the leading instanton), and selected those vacua where the relative corrections were $<20\%$. The resulting ensemble of solutions is represented in blue in the histograms of figures~\ref{fig:fieldSpaceDist}, \ref{fig:genericXi} (generic ensemble), and \ref{fig:xiDistNA0} (constrained ensemble).

\subsection{Redundancies of the EFT and solution duplicates}

As discussed in sections \ref{typeIIB} and \ref{sec:example}, the low energy supergravity description of type-IIB string compactifications has two inherent redundancies: one associated with the modular $\mathrm{SL}(2,\mathbb{Z})$ transformations \eqref{eq:SL2Z}, and one associated with the symplectic transformations acting as in \eqref{eq:symplectitTrans} and \eqref{eq:fluxTransform}. Thus, no-scale solutions which can be related to each other by any combination of these transformations should be regarded as equivalent.

To avoid double-counting no-scale solutions related by 
the $\mathrm{SL}(2,\mathbb{Z})$ symmetry,
we transported each solution to the fundamental domain of the modular group, given by the complex upper half-plane with $|\tau|>1$ and $|\text{Re} (\tau)|<1/2$. This operation can be easily performed by successively applying the generators of the group, given by
\begin{equation}
\begin{array}{c}
T_b =
\begin{pmatrix}
1 & b \\ 0 & 1
\end{pmatrix}
\\[12pt]
\tau \rightarrow \tau + b
\end{array}
\quad , \quad
\begin{array}{c}
R =
\begin{pmatrix}
0 & -1 \\ 1 & 0
\end{pmatrix}
\\[12pt]
\tau \rightarrow -1/\tau
\end{array}\,.
\label{eq:FDtrans1}
\end{equation}
Note that while these transformations change the value of $f$ and $h$, it can be shown that the quantity $N_{\text{flux}}$ remains invariant, so the transported solutions will nevertheless satisfy the tadpole condition. 

As for the symplectic transformations, at large complex structure the corresponding source of redundancy comes from the monodromy around the LCS point, which acts on the complex structure field $z$ and the fluxes as~\cite{Klemm:1992tx,Font:1992uk,Conlon:2004ds,DeWolfe:2004ns}
\begin{align}
z \rightarrow z - i \ n \qquad \equiv \qquad 
\left\lbrace
\begin{array}{l}
N \rightarrow A^n \cdot N \\
\Pi \rightarrow A^n \cdot \Pi
\end{array}
\right.\,,
\label{eq:FDtrans2a}
\end{align}
where $n \in \mathbb{Z}$ and
\begin{align}
A=
\begin{pmatrix}
1 & 0 & 0 & 0 \\
1 & 1 & 0 & 0 \\
4 & -2 & 1 & -1 \\
-4 & -2 & 0 & 1
\end{pmatrix}
\label{eq:FDtrans2b}
\end{align}
for the octic\footnote{See \cite{Klemm:1992tx} for more detail on this and other one-parameter models.}. This symmetry allows us to define a fundamental domain on the $z$ plane, which we chose to lie at $\left| \Im (z) \right| \leq 1/2 $. 

Both sets of transformations, \eqref{eq:FDtrans1} together with \eqref{eq:FDtrans2a} and \eqref{eq:FDtrans2b} can be used to transport $\tau$ and $z$ to their respective fundamental domains. Vacua with the same flux and moduli values (up to $10^{-8}$, corresponding to the error estimate of \texttt{Paramotopy}) are then removed to avoid double-counting solutions in the numerical scan.

\section{Density distribution of no-scale flux vacua}
\label{app:DDdistribution}

In this appendix we present a derivation of the theoretical probability distributions for the density of no-scale 
vacua \eqref{eq:vacuaDensity0} and \eqref{eq:densityFieldsNA0}, which describe the generic 
ensemble and the one constrained by the condition $N_A^0=0$, respectively 

The proof below relies only on the continuous flux approximation, and 
closely follows the one presented in \cite{Denef:2004ze}. We begin by deriving the probability distribution for the variables 
\eqref{eq:fluxModulusVariables} at no-scale vacua using the Hodge decomposition \eqref{eq:hodgeNgen}. We then combine 
the result with a generalised version of the Kac-Rice formula to derive the density of flux vacua. In each of these two steps, we 
will present the argument first for the generic ensemble, as obtained in \cite{Denef:2004ze}, and then we will adapt the it to the case 
the constrained ensemble.

\subsection{Derivation of the Denef-Douglas distribution}

Following \cite{Denef:2004ze}, our starting point is a flat distribution for the $4m=4(h^{2,1}+1)$ integer flux parameters 
\be
\{f_A^I, h_A^I, f^B_I,h^B_I\}.
\ee
Note that this distribution also matches the numerical procedure we followed to obtain the ensemble in the 
$\mathbb{WP}^4_{[1,1,1,1,4]}$ model, where flux realisations are drawn from a flat distribution.

In addition, we will also consider the situation when the tadpole constraint is large, $L\gg1$. In this case the typical values of these flux parameters are also large, and can be regarded as a continuous random variables. The corresponding probability distribution is therefore 
\be
d\mu_{\text{flux}}(f,h) = \cN \, (d f d h)^{4m}\,.
\ee
Begin by switching variables to the complex flux parameters $N = f - \tau h$ and their conjugates $\bar N = f - \bar \tau h$. The associated Jacobian is
\be
J= \frac{\pd(N_A, \bar N_A,N^B, \bar N^B )}{\pd (f_A,h_A,f^B,h^B)}= \begin{pmatrix}
    \frac{\pd(N_A, \bar N_A)}{\pd(f_A,h_A)} & 0 \\
    0 & \frac{\pd(N^B, \bar N^B)}{\pd(f^B,h^B)}
\end{pmatrix}\,,
\ee
where
\be
\frac{\pd(N_A, \bar N_A)}{\pd(f_A,h_A)} =\frac{\pd(N^B, \bar N^B)}{\pd(f^B,h^B)} = \begin{pmatrix}
    \unity & \unity \\
    -\tau \unity & - \bar \tau \unity
\end{pmatrix}\,.
\ee
Then, 
\be
\det(J) = \det \begin{pmatrix}
    1 & 1 \\
    -\tau & \bar \tau
\end{pmatrix}^{2 m} = (-2 \rmi \Im \tau)^{2 m}\,,
\ee
implying that the resulting probability distribution for $\{N,\bar N\}$ reads
\be
d\mu_{\text{flux}}(N,\bar N) =\cN\, (d N d \bar N)^{4m} \rme^{2m K_d}\,.
\label{eq:firstChange}
\ee
Next, we consider the change of variables between $\{N,\bar N\}$ and $\{Z_0, F_a, F_0,Z_a,\, c.c.\}$, defined by\footnote{In the following derivations, in order to simplify the notation, we will ignore the overall $1/\sqrt{4 \pi}$ factor in the definition of $W$  \eqref{fluxW}, since it plays no role in the final result. }
\bea
F_0 &\equiv& \rme^{K/2} D_0 W= -\rmi \rme^{K/2} N^\dag \cdot \Sigma\cdot \Pi\,, \nonumber \\
F_a&\equiv& \rme^{K/2}D_a W= \rme^{K/2} N^T \cdot \Sigma\cdot D_a \Pi \,,\nonumber \\
Z_0&\equiv& \rme^{K/2}W = \rme^{K/2} N^T \cdot \Sigma\cdot \Pi\,, \nonumber \\
Z_a &\equiv& \rme^{K/2}D_{0} D_a W = -\rmi \rme^{K/2} N^\dag \cdot \Sigma\cdot D_a \Pi\,,
\label{eq:fluxModulusVar_app}
\eea
where the components of the vectors have been expressed in a canonically normalised basis. Note that these definitions 
coincide with those in the main text, \eqref{eq:fluxModulusVariables}, up to a volume factor $\cV$. Since the volume is independent 
of the complex structure moduli or the dilaton, the effect of the rescaling necessary to make contact with \eqref{eq:fluxModulusVariables} amounts 
to a redefinition of the normalisation constant, and thus we will ignore the volume prefactor in the following. The Jacobian 
$J =\pd(Z_0, F_a, F_0,Z_a,c.c.) /\pd (N,\bar N)$ of the transformation above reads
\be
J= \rme^{K/2} \begin{pmatrix}\Sigma \cdot \Pi \; & \Sigma\cdot D_a \Pi &0&0&0&0 &\rmi \Sigma \cdot \bar \Pi\; & \rmi \Sigma \cdot D_{\bar a} \bar \Pi \\
    0&0 & -\rmi \Sigma \cdot \Pi\; &- \rmi \Sigma \cdot D_a \Pi & \Sigma \cdot \bar \Pi& \Sigma\cdot D_{\bar a} \bar \Pi&0&0
\end{pmatrix}.
\ee
Then,
\be
\det(J) = \rme^{2 m K}|\det M|^2\,.
\ee
where
\be
M \equiv \begin{pmatrix}\Sigma \cdot \Pi, \; & \Sigma\cdot D_a \Pi, \; & \rmi \Sigma \cdot \bar \Pi,\; & \rmi \Sigma \cdot D_{\bar a} \bar \Pi \,.
\end{pmatrix}
\ee
To compute the determinant, we use the trick
\be
|\det M|^2 = |M^\dag M|= |M^\dag\cdot \Sigma \cdot M|\,,
\ee
since $|\Sigma| = 1$. Then, using the orthogonality properties of the basis $\{\Pi, \bar \Pi, D_a \Pi , D_{\bar a} \bar \Pi\}$ for the space of symplectic sections under the product defined by $\Sigma$, we obtain~\cite{Denef:2004ze}
\be
M^\dag\cdot \Sigma \cdot M = \begin{pmatrix}
    \rme^{-K_{cs}} & 0& 0& 0\\
    0& -\rmi \rme^{-K_{cs}}\delta_{a\bar b} &0&0\\
    0&0&\rme^{-K_{cs}} &0\\
    0&0&0& \rmi \rme^{-K_{cs}}\delta_{a\bar b} 
\end{pmatrix}\,.
\ee
The determinant of this matrix is then $\det (M^\dag\cdot \Sigma \cdot M)= \rme^{-2mK_{cs}}$. Using this result we find that the determinant of the Jacobian of the change of variables is 
\be
\det(J) = \rme^{2m(K_d+K_k)}\,,
\ee
and therefore, noting that the factor $\rme^{2m K_d}$ cancels with that of \eqref{eq:firstChange}, we find that the probability distribution on the variables $\{F_0,F_a, Z_0,Z_a\}$ is flat 
\be
d\mu_{\text{flux}}(F_A,Z_A,\bar F_A, \bar Z_A) = \cN\; (d F_A d\bar F_A d Z_A d\bar Z_A)^{4m} \rme^{-2m K_k}. 
\label{eq:gralDDdist}
\ee
In these variables no-scale vacua correspond to those configurations with $F_A=0$, and the tadpole constraint requires 
\be
N_{\text{flux}} = Z_A \bar Z_A \equiv |Z|^2 \le L.
\label{eq:tadpoleSphere}
\ee
The no-scale conditions can be imposed by introducing a delta function $\delta^{2m}(F_A,\bar F_A)$ in \eqref{eq:gralDDdist}, which is equivalent to considering simply the distribution
\be
d\mu_{\text{flux}}(Z_A,\bar Z_A)|_{\text{no-scale}} = \cN\; (d Z_A d\bar Z_A)^{2m}\,.
\label{eq:DDdistribZZ}
\ee
In other words, the variables $Z_A = \{\rme^{K/2} W, \rme^{K/2} D_{0} D_a W\}$ and their complex conjugates form a set of $2(h^{2,1}+1)$ independent complex variables uniformly distributed on the sphere \eqref{eq:tadpoleSphere} defined by the tadpole constraint. We refer to the previous probability density function as the \emph{Denef-Douglas distribution}.

\subsection{Constrained flux distribution}

We will now repeat the above computation for the constrained ensemble of vacua. We begin by noting that this constraint can be implemented in the continuous flux approximation with Dirac delta functions as
\be
d \mu_{\text{flux}}( f, h) = \cN \, (d f d h)^{4m} \, \delta(f_A^0) \delta(h_A^0)\,. 
\ee
After changing to complex flux coordinates, we obtain
\be
d\mu_{\text{flux}}(N,\bar N) =\cN\, (d N d \bar N)^{4m} \rme^{2m K_d} \delta(N_A^0) \delta(\bar N_A^0) |J_0|\,,
\ee
where 
\be
|J_0| =\det \frac{\pd(N_A^0,\bar N_A^0)}{\pd(f_A^0, h_A^0)} = \det \begin{pmatrix}
    1 & 1\\
    -\tau & -\bar \tau
\end{pmatrix} = -2 \rmi \Im \tau\,,
\ee
and so 
\be
d\mu_{flux}(N,\bar N) =\cN\, \rmi (d N d \bar N)^{4m} \rme^{(2m-1) K_d} \delta(N_A^0) \delta(\bar N_A^0)\,.
\ee
Using the Hodge decomposition of the flux vector
\be
N = \rme^{\frac{1}{2} (-K_k-K_d +K_{cs})}(-\bar F_0 \Pi + \rmi Z_0 \bar \Pi + \bar Z_{a} D_a \Pi -\rmi F_a D_{\bar a} \bar \Pi)\,,
\ee
we can obtain the form of the constraint in terms of the variables $\{Z_A,F_A,\bar Z_A, \bar F_A\}$. It is given by
\be
N_A^0 = \rme^{\frac{1}{2} (-K_k-K_d +K_{cs})}(-\bar F_0+ \rmi Z_0 -\sqrt{3/(1- 2 \xi)} \bar Z_{1} +\sqrt{3/(1- 2 \xi)} \rmi F_1) =0\,.
\ee
We can implement this condition as a constraint on the variables 
\be
Z_0 = Z_0^*(Z_a,F_A)\,,\qquad \bar Z_0 = \bar Z_0^*(Z_a,F_A)
\label{eq:ZFNA0const}
\ee
with the aid of Dirac delta functions, so that the final density function, written in terms of $\{Z_A,F_A,\bar Z_A, \bar F_A\}$, reads
\be
d\mu_{\text{flux}} = \cN\; \rmi (d F_A d\bar F_A d Z_A d\bar Z_A)^{4m} \rme^{-K_{cs}} \delta(Z_0-Z_0^*) \delta(\bar Z_0 -\bar Z_0^*)\,. 
\label{eq:constrainedMeasure}
\ee
Here we have absorbed the Calabi-Yau volume factor, which is independent of the axio-dilaton and complex structure fields, in the normalisation constant. 

\subsection{Density of generic no-scale vacua}

We now turn to the computation of the density of flux vacua in the generic ensemble. The number of no-scale vacua $\cC_{\text{vac}}(N)$ for a given choice of flux $N$ can be obtained using the generalised Kac-Rice formula \cite{kac1943,rice1944mathematical}
\be
\cC_{\text{vac}}(N) = \int d^{2m} u\; \delta^{2m}(DW) \; |\det D^2 W|\,,
\ee 
with $u^A = \{\tau, z^i\}$ and 
\be
\delta^{2m}(DW) = \delta^m(D_A W) \; \delta^m(D_{\bar A} \bar W), \qquad 
D^2 W = \begin{pmatrix}
    D_{A} D_{\bar B} W & D_A D_B W \\
    D_{\bar A} D_{\bar B} \bar W & D_{\bar A} D_{B} \bar W
\end{pmatrix}\,.
\ee
Let us now consider the total number of no-scale vacua in the ensemble of fluxes satisfying the tadpole constraint $N_{\text{flux}} \le L$, which is given by
\be
\cC_{\text{vac}}(N_{\text{flux}}\le L) = \sum_{N} \Theta(N_{\text{flux}}-L)\int d^{2m} u \;\delta^{2m}(DW) |\det D^2 W|\,.
\label{eq:TotalNumVacua}
\ee
Note that here we have chosen to count all choices of flux with equal weight, consistent with our initial assumption that fluxes are drawn from an underlying uniform distribution. 

As discussed in \cite{Denef:2004ze}, using the integral representation of the Heaviside theta function we can rewrite the previous formula as
\be
\cC_{vac}(N_{\text{flux}}\le L) = \frac{1}{2 \pi \rmi}\int_C\frac{d\alpha}{\alpha} \rme^{\alpha L^*} \cC(\alpha)
\ee
with 
\bea
\cC(\alpha) &\equiv& \sum_{N} \int d^{2m} u \; \rme^{-\alpha N_{\text{flux}}}\;\delta^{2m}(DW) |\det D^2 W| \nonumber \\
&\approx& \int (d N d \bar N)^{4m} \int d^{2m} u \; \rme^{-\alpha N_{\text{flux}}}\;\delta^{2m}(DW) |\det D^2 W| \rme^{2m K_d}\,,
\eea
where in the last line we have approximated the sum over the integer fluxes by an integral with measure given by \eqref{eq:firstChange}. 

Expressing the gradient, $DW$, and the Hessian of the superpotential, $D^2 W$, in a canonically normalised basis, we find
\bea
\cC(\alpha)
&=& \int (d N d \bar N)^{4m} \int d^{2m} u |\det g|\; \rme^{-\alpha N_{\text{flux}}}\;\delta^{2m}(D_aW) |\det \cH|^{1/2} \rme^{2m K_d}\nonumber \\
&=& \int (dZ d F)^{4m} \int d^{2m} u |\det g|\; \rme^{-\alpha \cV^2 |Z|^2}\; \delta^m(F_A) \delta^m(\bar F_A) |\det \cH|^{1/2} \rme^{-2m K_k}\nonumber \\
&=& \cV^{4m}\int (dZ d \bar Z)^{2m} \int d^{2m} u |\det g|\; \rme^{-\alpha \cV^2 |Z|^2}\; |\det \cH|^{1/2}\,.
\label{eq:NvacChangeVars}
\eea
In the second line, the term of the form $\rme^{mK}$ arises as a result of the change of variables \eqref{eq:fluxModulusVar_app} in the argument of 
the delta function, which then is reabsorbed when expressing $\det D^2 W$ in terms of $\cH$.\footnote{Recall that in canonically normalized coordinates 
we have the relation $\cH =\rme^K (D^2 W)^2 = (m_{3/2}\unity +\cM)^2$, from \eqref{eq:FMM}.} Rescaling 
$Z_A \to Z_A/(\cV\sqrt{\alpha})$, it is possible to see that $\cC(\alpha) = \cC(1,\cV=1) \alpha^{-2m}$ and the overall volume factor disappears, so 
we can explicitly perform the integral in $\alpha$ to give
\be
\cC_{vac}(N_{\text{flux}}\le L) = \Lambda(L,m) \cdot \int d^{2m} u |\det g| \int (dZ d \bar Z)^{2m}\; \rme^{- |Z|^2}\; |\det \cH|^{1/2}\,.
\ee
where $\Lambda(L,m)$ is a constant depending on the tadpole parameter $L$ and the moduli space dimension $m$.
Consistent with the previous equation, the density of flux vacua is then defined by
\be
d\mu_{\text{vac}}(z,\tau) = \cN \cdot d^{2m} u \, |\det g| \, \Big[\int (dZ d \bar Z)^{2m}\; \rme^{- |Z|^2}\; |\det \cH|^{1/2}\Big]\,.
\ee 

\subsection{Density of constrained vacua}

For the ensemble with constrained fluxes the argument proceeds as before, but when changing to the variables \eqref{eq:fluxModulusVar_app} in \eqref{eq:NvacChangeVars}, we should use the measure \eqref{eq:constrainedMeasure} rather than \eqref{eq:DDdistribZZ}. As a result, the density of no-scale vacua is given by 
\be
d\mu_{\text{vac}}(z,\tau) = \cN \cdot d^{2m} u \, |\det g| \, \Big[\int (dZ d \bar Z)^{2(m-1)}\; \rme^{- |Z|^2}\; |\det \cH|^{1/2} \rme^{-K_{cs}}\Big]\,,
\ee 
where the integral is over the subspace of $\{Z_A,\bar Z_A\}$ defined by the constraints \eqref{eq:ZFNA0const} with $F_A = \bar F_A=0$ (no-scale conditions).
In the case when the complex structure moduli space is one dimensional ($m=2$), the integral appearing in the previous equation reduces to 
\be
\int dZ_1 d \bar Z_1 \rme^{- \frac{2(2-\xi)}{1-2 \xi}|Z_1|^2 }\; |Z_1|^4 \Big|1+ \frac{9}{(1-2 \xi)^2} - \frac{3 (2+ \mathring \kappa^2)}{(1-2 \xi)} \Big| \frac{\xi +1}{\xi}\,,
\ee 
where we used the relation $|Z_0| = \sqrt{3/(1-2 \xi)} |Z_1|$ (given in \eqref{correlation}) and the definition of the LCS parameter $\rme^{-K_{cs}} = |2\Im \kappa_0| (\xi+1)/\xi$. Integrating the previous expression over the complex variable $Z_1$, and over the directions $\Im z$ and $\Re \tau$ we find
\be
d\mu_{\text{vac}}(z,\tau) = \cN \cdot \frac{(1+\xi)}{(2-\xi)^2} \frac{1}{r^2 s^2} dr ds\,,
\ee 
expressed in terms of the variables \eqref{eq:srDefs}, which agrees with \eqref{eq:densityFieldsNA0}.

\section{Bounds on the LCS parameter}
\label{sec:bounds}

The theoretical distributions \eqref{eq:vacuaDensity0} and \eqref{eq:densityFieldsNA0} were derived in order 
to have an analytical description of the ensemble of no-scale vacua, constructed as described in section \ref{sec:example} 
and appendix \ref{app:paramotopy}. However, this characterisation is only a faithful representation of the Landscape in the 
regime of moduli space where both the EFT and the continuous flux approximation are valid.

Indeed, vacua with large instanton corrections should be discarded, since the EFT we used cannot be 
trusted in those cases. As the vacua density functions \eqref{eq:vacuaDensity0} and \eqref{eq:densityFieldsNA0} 
contain no information regarding the size of the instanton contributions, they are bound to be inaccurate in the regime where 
these corrections are large. Furthermore, from \cite{Brodie:2015kza,Marsh:2015zoa}, we know that in the 
generic ensemble the density of vacua should be suppressed with respect to the theoretical distribution \eqref{eq:vacuaDensity0}, due 
to the breakdown of the continuous flux approximation. In this appendix we discuss the parameter space where the
statistical description can be applied, providing an analytic estimate for this region in terms of the LCS parameter $\xi$.

\subsection{No-scale equations near the LCS point}

We begin by rewriting the no-scale equations \eqref{eq:susyeqs} in the LCS limit in a more convenient way for the derivations below.

In this regime, $\xi\to0$, the K\"ahler potential of the complex structure sector can be expressed as
\begin{equation}\label{eq:Y}
Y \equiv e^{-K_{cs}} \approx \frac{1}{6} \kappa_{ijk} (z^i + \bar{z}^i) (z^j + \bar{z}^j) (z^k + \bar{z}^k)\,,
\end{equation}
and the canonically normalized Yukawa couplings satisfy~\cite{Brodie:2015kza,Marsh:2015zoa}
\begin{align}
\kappa_{111} = \frac{2}{\sqrt{3}} Y\,, \qquad 
\kappa_{11\tilde a} = 0\,, \quad
\kappa_{1\tilde a\tilde b} = - \frac{1}{\sqrt{3}} Y \delta_{\tilde a\tilde b} \,,
\label{eq:lcs_simp}
\end{align}
where the direction ``1'' corresponds to the no-scale direction \eqref{eq:noscaleDirection} and $\tilde a, \tilde b=2,\ldots h^{2,1}$. 
We also have
\begin{align}
z^a = \frac{\sqrt{3}}{2} \delta_1^a + i \lambda^a\,, \qquad
K_a = - \sqrt{3} \delta_1^a\,,
\end{align}
where $z^a$ (with $a=1,\ldots,h^{2,1}$) are the canonically normalised fields at the vacuum, and $\lambda^a = \Im (z^a)$. After some algebra, it can be shown that the superpotential has the form
\begin{align}
W =&-\frac{i}{6} \kappa_{abc} N_A^0 z^a z^b z^c + \frac{1}{2} \kappa_{abc} N_A^a z^b z^c + i \left( \kappa_a N_A^0 - \kappa_{ab} N_A^b -N_a^B \right) z^a 
\nonumber\\[5pt]
&+ \kappa_0 N_A^0 + \kappa_a N_A^a - N_0^B\,,
\end{align}
while the no-scale conditions \eqref{eq:susyeqs} read
\begin{align}
D_0 W = &-\frac{1}{6} \kappa_{abc} \bar{N}_A^0 z^a z^b z^c - \frac{i}{2} \kappa_{abc} \bar{N}_A^a z^b z^c + \left( \kappa_a \bar{N}_A^0 - \kappa_{ab} \bar{N}_A^b -\bar{N}_a^B \right) z^a 
\nonumber\\
&- i \kappa_0 \bar{N}_A^0 - i \kappa_a \bar{N}_A^a + i \bar{N}_0^B = 0\,, 
\nonumber\\[7pt]
D_a W = &- \frac{i}{2} \kappa_{abc} N_A^0 z^b z^c + \kappa_{abc} N_A^b z^c + i \kappa_a N_A^0 - i \kappa_{ab} N_A^b - i N_a^B - \sqrt{3} W \delta_a^1 = 0\,.
\end{align}
To analyse these equations it is convenient to introduce the flux parameter redefinitions
\begin{align}
\hat{N}_0^B \equiv &- \frac{1}{6} \kappa_{abc} N_A^0 \lambda^a \lambda^b \lambda^c - \frac{1}{2} \kappa_{abc} N_A^a \lambda^b \lambda^c - \left( \kappa_a N_A^0 - \kappa_{ab} N^b_A - N_a^B \right) \lambda^a 
\nonumber\\
&+ \kappa_0 N_A^0 + \kappa_a N_A^a - N_0^B\,, \label{eq:N0Bhat}
\\[7pt]
\hat{N}_a^B \equiv &- \frac{1}{2} \kappa_{abc} N_A^0 \lambda^b \lambda^c - \kappa_{abc} N^b_A \lambda^c - \kappa_a N_A^0 + \kappa_{ab} N_A^b + N_a^B\,. \label{eq:NABhat}
\end{align}
Note that the flux parameters $\hat f^B_I$ and $\hat h^B_I$ in $\hat N = \hat f -\tau \hat h$ are still real, but are not integers in general. 
Using these new parameters, the superpotential now reads
\begin{equation}
W = \hat{N}_0^B - i \frac{\sqrt{3}}{2} \hat{N}_1^B + \left[ \left( \frac{\sqrt{3}}{4} \lambda^1 - \frac{i}{8} \right) N_A^0 + \frac{\sqrt{3}}{4} N_A^1 \right] Y
\end{equation}
and the no-scale conditions are given by
\begin{align}
\left[ \left( \frac{1}{2} \lambda^1 + \frac{i}{4\sqrt{3}} \right) N_A^0 + \frac{1}{2} N_A^1 \right] Y + i \hat{N}_1^B &= - \frac{2}{\sqrt{3}} \hat{N}_0^B\,, \\[5pt]
\left[ \left( \frac{1}{2} \lambda^1 - \frac{i\sqrt{3}}{4} \right) N_A^0 + \frac{1}{2} N_A^1 \right] Y + i \hat{N}_1^B &= 2\sqrt{3} \hat{N}_0^B\,, \\[5pt]
\left[ \frac{1}{2} \lambda^{a'} N_A^0 + \frac{1}{2} N_A^{a'} \right] Y + i \hat{N}_{a'}^B &= 0\,.
\end{align}
These expressions can be rewritten in a more compact way as
\begin{align}
\boxed{
    \hat{N}_0^B = - \frac{i}{8} Y N_A^0, \qquad \hat{N}_a^B = \frac{i}{2} Y \left( N_A^0 \lambda^a + N_A^a \right)\,.
}
\label{eq:LCS_result}
\end{align}
The second equation can be equivalently written, after contracting it with the vielbein $e^a_i$, as
\be
\hat{N}_i^B = \frac{i}{2} Y \left( N_A^0 \lambda^j + N_A^j \right) g_{ji}\,,
\ee
where we have used that $\hat{N}_a^B e^a_i = \hat{N}_i^B$, and $N_A^a e^a_i = N_A^j e_j^a e^a_i= N_A^j g_{ji}$, and similarly for the terms involving $\lambda^a$. The redefined fluxes can also be rewritten as
\begin{align}
\hat{N}_0^B \equiv &- \frac{1}{6} \kappa_{ijk} N_A^0 \lambda^i \lambda^j \lambda^k - \frac{1}{2} \kappa_{ijk} N_A^i \lambda^j \lambda^k - \left( \kappa_i N_A^0 - \kappa_{ij} N^j_A - N_i^B \right) \lambda^i 
\nonumber\\
&+ \kappa_0 N_A^0 + \kappa_i N_A^i - N_0^B\,,
\\[7pt]
\hat{N}_i^B \equiv &- \frac{1}{2} \kappa_{ijk} N_A^0 \lambda^j \lambda^k - \kappa_{ijk} N^j_A \lambda^k - \kappa_i N_A^0 + \kappa_{ij} N_A^j + N_i^B\,.
\end{align}

\subsection{Lower bound on the LCS parameter}
\label{app:xiMin}

We will now determine the regime of applicability of the continuous flux approximation near the LCS point on the generic flux ensemble, 
expressed as a lower bound for the LCS parameter $\xi$. 

Let us first discuss the equation in \eqref{eq:LCS_result} for $\hat{N}_0^B$ in the $\xi \to 0$ limit when $N_A^0\neq0$. From the definition of the LCS parameter $\xi$ (\ref{eq:LCS_param}), we can rewrite $Y$ as
\begin{align}
Y = 2 \left| \Im \kappa_0 \right| \frac{1+\xi}{\xi} \approx 2 \left| \Im \kappa_0 \right| \xi^{-1}
\end{align}
in the LCS limit, that is, $\xi \rightarrow 0$. On the other hand, from the definition of $N_A^0$, we can obtain the following lower bound:
\begin{align}
\left| N_A^0 \right|^2 = (f_A^0 - \Re (\tau) h^0_A)^2 + \Im^2 (\tau) (h_A^0)^2 \geq 1\,,
\label{eq:N_cond}
\end{align}
where in the last step we used $\Im (\tau) > 1$, as required for the vacua to be in the weak string coupling regime. This implies that at no-scale vacua near the LCS limit, we must have
\be
|\hat{N}_0^B | \approx \frac{\left| \Im \kappa_0 \right| \left| N_A^0 \right|}{4\xi}\,.
\ee
Note that, since by assumption $|N_A^0|\neq0$, the right hand side blows up when $\xi\to0$, and so $ |\hat{N}_0^B |\gtrsim \cO(\xi^{-1})$, which will require some contributions in \eqref{eq:NABhat} to become large. From \eqref{eq:N_cond} we can see that the previous condition will be the least restrictive when $|N_A^0|$ and $\Im \tau$ are both $\cO(1)$, leading to 
\be
|\hat{N}_0^B | \gtrsim \frac{\left| \Im \kappa_0 \right|}{4\xi}\,.
\ee
In order to solve this condition, one could try to tune the parameters $\lambda^i =\Im z^i$ to be large, $\lambda^i=\cO(\xi^{-1/3})$; however, in that case the cubic terms in 
\eqref{eq:NABhat} would dominate, and the first equation in \eqref{eq:LCS_result} would become 
\begin{align}
\frac{1}{6} \kappa_{ijk} \lambda^i\lambda^j \lambda^k \overset{!}{\approx} \rmi\frac{\left| \Im \kappa_0 \right|}{4\xi}\,,
\end{align}
which cannot be solved, as the left hand side is real and the right hand side purely imaginary. This conclusion is in agreement with the results of \cite{Brodie:2015kza,Marsh:2015zoa}, where it was shown that if the superpotential is dominated by its cubic term, the no-scale equations \eqref{eq:susyeqs} cannot admit solutions in a neighbourhood of the LCS point.

As a consequence, we need terms with different powers of $\lambda^i$ in \eqref{eq:NABhat} to be comparable and, assuming the constant coefficients of the prepotential to be $\cO(1)$, due to the tadpole condition (\ref{eq:tadpole_def}) we will typically have $|\hat N_0^B|=\cO(\sqrt{N_{\text{flux}}})$. Combining all of this, we arrive at the bound we were looking for:
\be
\xi \gtrsim\xi_{\text{min}}\big|_{N_A^0\neq0} \equiv \frac{\left| \Im \kappa_0 \right|}{4\sqrt{N_{\text{flux}}}}\,.
\ee
As an example, for the $\mathbb{WP}^4_{[1,1,1,1,4]}$ model, we have $N_\text{flux} = 972$ and $2 |\Im \kappa_0| \approx 2.9$, which implies
\be
\xi_\text{min} =\cO(10^{-2})\,.
\label{eq:appXiBound}
\ee

In the case of the constrained ensemble, the previous argument does not apply.
In the limit $\xi \to 0$ with $f_A^0=h_A^0=h_A^z=0$ and 
\begin{align}
f_0^B = f_A^z \ k_z + \frac{h_0^B}{2 (h_z^B)^2} \left[ 2 h_z^B (f_z^B + f_A^z \ k_{zz}) - f_A^z \ h_0^B \ k_{zzz} \right]\,,
\end{align}
the LCS parameter $\xi$ can be used to parametrize no-scale solutions. This leads to a flat direction along $\Im \tau \approx\left(\xi_0/\xi\right)^{1/3} |f_A^z/h_z^B|$, where $\xi_0 \equiv (3/2^4) |\Im \kappa_0|k_{zzz}^2$, which allows solutions arbitrarily close to the LCS/weak-coupling limit.

\subsection{Maximum value consistent with small instanton corrections}
\label{sec:instantonEstimate}

In this section, we give an upper bound $\xi_{\text{max}}$, independent of the flux vector $N$, by requiring the instanton corrections to the metric of the complex structure moduli space to remain small. As we mentioned above, the numerical analysis shows that the field space metric is typically the object where these  corrections have the largest effect, and thus it is particularly suitable for estimating the regime of validity of the EFT. For convenience we will split the prepotential as
\begin{equation}
\cF(z) = \hat{\cF} (z) + \cF_* (z)+ \ldots\,,
\end{equation}
where $\hat \cF$ is the perturbative part of the prepotential, and $\cF_*$ denotes the leading term of the instanton contributions $\cF_{\text{inst}}$.
In the case of the octic model, near the LCS point the prepotential is given by \eqref{eq:octic_prep} and \eqref{eq:instCorrectDom}, which have the form 
\be
\hat{\cF} (z) = \frac{i}{6} \kappa_3 z^3 + \frac{1}{2} \kappa_2 z^2 + i \kappa_1 z + \frac{1}{2} \kappa_0\,, \qquad 
\cF_* (z) = - \frac{i n_1}{(2\pi)^3} e^{-2\pi z}\,,
\ee
the latter being perturbatively small when compared to $\hat{\cF}$ in the LCS regime. The period vector \eqref{eq:Period} is then split as
\begin{align}
\Pi = 
\begin{pmatrix}
1 \\ i z \\ 2 \cF - z \pd_z \cF \\ -i \pd_z \cF
\end{pmatrix}
=
\begin{pmatrix}
1 \\ i z \\ 2 \hat{\cF} - z \pd_z \hat{\cF} \\ -i \pd_z \hat{\cF}
\end{pmatrix}
+
\begin{pmatrix}
0 \\ 0 \\ 2 \cF_* - z \pd_z \cF_* \\ -i \pd_z \cF_* 
\end{pmatrix}
\equiv \hat{\Pi} + \Pi_*\,.
\end{align}
Keeping only to leading instanton contribution, the K\"ahler potential \eqref{eq:CSkahler} is then
\begin{align}
e^{-K_{cs}} &= i \ \Pi^\dag \cdot \Sigma \cdot \Pi \approx i \ \hat{\Pi}^\dag \cdot \Sigma \cdot \hat{\Pi} + i \ \hat{\Pi}^\dag \cdot \Sigma \cdot \Pi_* + i \ \Pi_*^\dag \cdot \Sigma \cdot \hat{\Pi} \nonumber \\
&= e^{-\hat{K}} - 2 \ \text{Im} (\hat{\Pi}^\dag \cdot \Sigma \cdot \Pi_*)\,,
\end{align}
Defining $\cI \equiv \text{Im} (\hat{\Pi}^\dag \cdot \Sigma \cdot \Pi_*)$ we obtain to leading order
\begin{align}
K_{cs} \approx \hat{K} + 2 \cI e^{\hat{K}}.
\end{align}
Denoting the zeroth order metric and its leading instanton correction as
\be
G_{z \bar z} \equiv \pd_z \pd_{\bar z} \hat{K}\,, \qquad 
g_{z \bar z} \equiv \pd_z \pd_{\bar z} ( 2 \cI e^{\hat{K}} )\,,
\ee
we find that the field space metric is given by
\begin{align}
\pd_z \pd_{\bar{z}} K_{cs} &\approx G_{z\bar{z}} + g_{z\bar{z}} \nonumber \\
&= G_{z\bar{z}} + 2 e^{\hat{K}} \left( G_{z\bar{z}} \cI + \hat{K}_z \hat{K}_{\bar{z}} \ \cI + \hat{K}_{\bar{z}} \cI_z + \hat{K}_z \cI_{\bar{z}} + \cI_{z\bar{z}} \right), 
\end{align}
where subindices denote partial derivatives. 
Letting $e^1_z$ be a \emph{real vielbein} with respect to the zero-order metric $G_{z \bar z}$ (so that $G_{1\bar{1}} = 1$), we then find
\begin{align}
g_{1\bar 1} =2 e^{\hat{K}} \left( \left( 1 + (\hat{K}_1)^2 \right) \cI + \hat{K}_1 \left( \cI_1 + \cI_{\bar{1}} \right) + \cI_{1\bar 1} \right)\,,
\label{eq:g11}
\end{align}
where we used $\hat{K}_1 = e^z_1 \hat{K}_z = e^{\bar{z}}_{\bar{1}} \hat{K}_{\bar{z}} = \hat{K}_{\bar{1}}$.
The value of $\cI$ and its partial derivatives are found by direct computation. Defining $\theta = 2\pi \text{Im}(z)$ and using \eqref{eq:vielbein}, we get
\begin{align}
\cI &= - \frac{n_1}{4\pi^3} e^{-2\pi \text{Re}(z)} (1+2\pi \text{Re}(z) ) \cos (\theta) \label{eq:first}\,,\\[7pt]
\cI_1 + \cI_{\bar{1}} &= \frac{2 n_1}{\pi} e^{-2\pi \text{Re}(z)} \frac{\text{Re}^2(z)}{x} \cos (\theta)\,, \\[7pt]
\cI_{1\bar{1}} &= \frac{2 n_1}{\pi} e^{-2\pi \text{Re}(z)} \frac{\text{Re}^2(z)}{x^2} \cos (\theta)\,.
\end{align}
From the analysis in section~\ref{sec:totalSpectrum}, we know that the relations
\begin{align}
e^{\hat{K}} &= \frac{1}{2|\text{Im} \ \kappa_0|} \frac{\xi}{1+\xi}\,, \quad 
\hat{K}_1 = - \sqrt{\frac{3}{1- 2\xi}}\,,
\\
x^2&=\frac{3(1-2 \xi)}{(1+\xi)^2}\,, \quad
\text{Re} \ z = \left( \frac{3 |\text{Im} \ \kappa_0|}{2 \kappa_3} \right)^{1/3} \frac{1}{\xi^{1/3}}
\label{eq:last}
\end{align}
hold. Thus, substituting \eqref{eq:first} through \eqref{eq:last} into \eqref{eq:g11}, and defining $\alpha \equiv \left( \frac{3 |\text{Im} k_0|}{2 k_3} \right)^{1/3}$, we get
\begin{align}
g_{1\bar 1} = - \frac{n_1 \cos \theta }{6 \pi^3 |\text{Im} \ \kappa_0|} \frac{(2-\xi) \xi^{1/3}}{(1+\xi)(1-2\xi)} e^{-\frac{2\pi \alpha}{\xi^{1/3}}} \Big( (2\pi\alpha)^2 (1+\xi) + 3 \xi^{2/3} +6\pi \alpha \xi^{1/3} \Big)\,. 
\nonumber
\end{align}
Requiring the relative corrections to the canonically normalised metric to remain below $20\%$ (or equivalently $|g_{1\bar 1}| \leq 0.2$, since $G_{1\bar 1}=1$), we find
\begin{align}
\xi \le \xi_{\text{max}} \approx 0.2\,.
\label{eq:appXiBound2}
\end{align}

\subsection{Accuracy of the statistical description}
\label{sec:corrections}

\begin{figure}[t]
    \centering
    \subfloat[]{
        \includegraphics[width=0.45\textwidth]{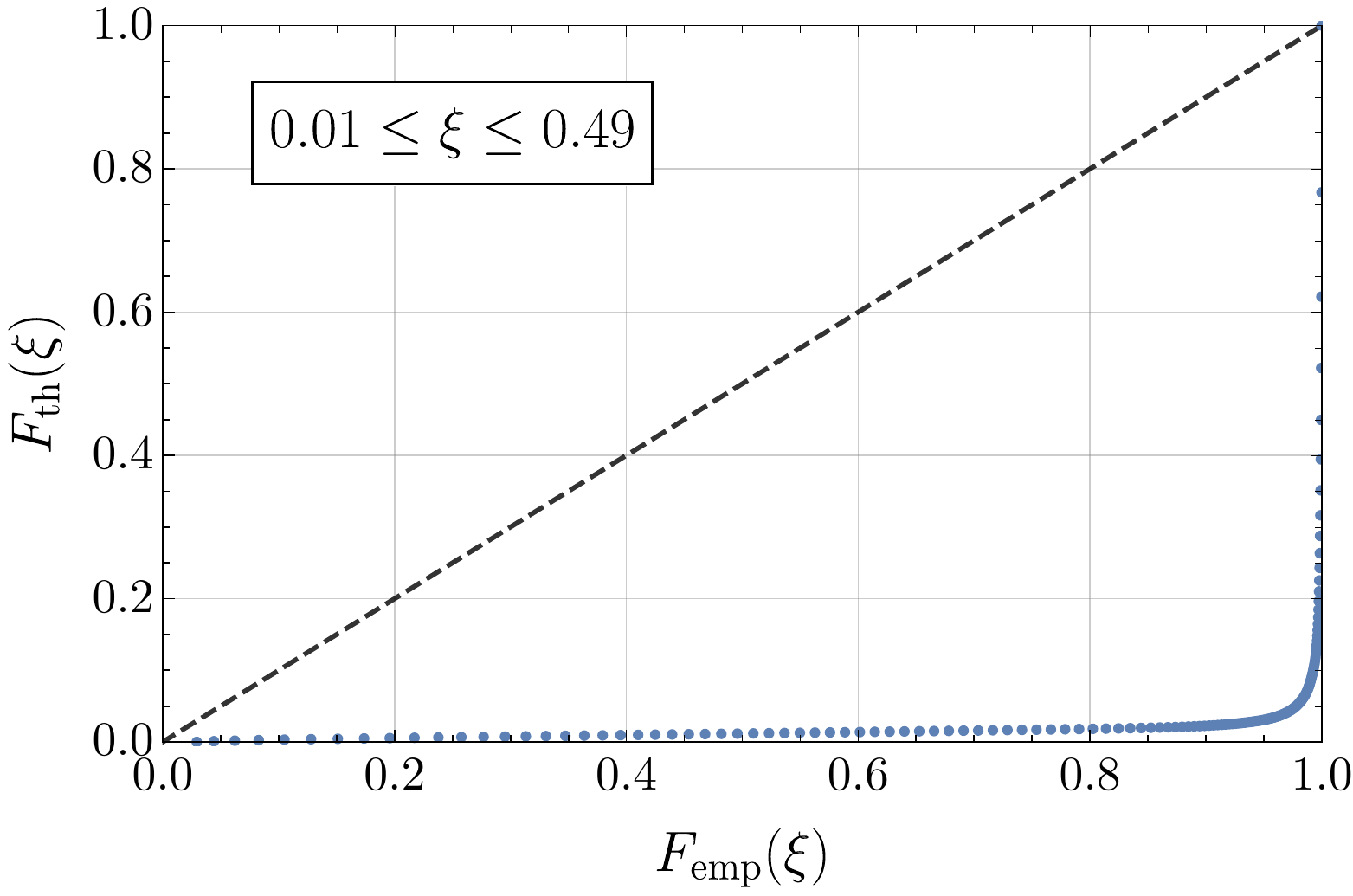}
    }
    \hfill  
    \subfloat[]{
        \includegraphics[width=0.45\textwidth]{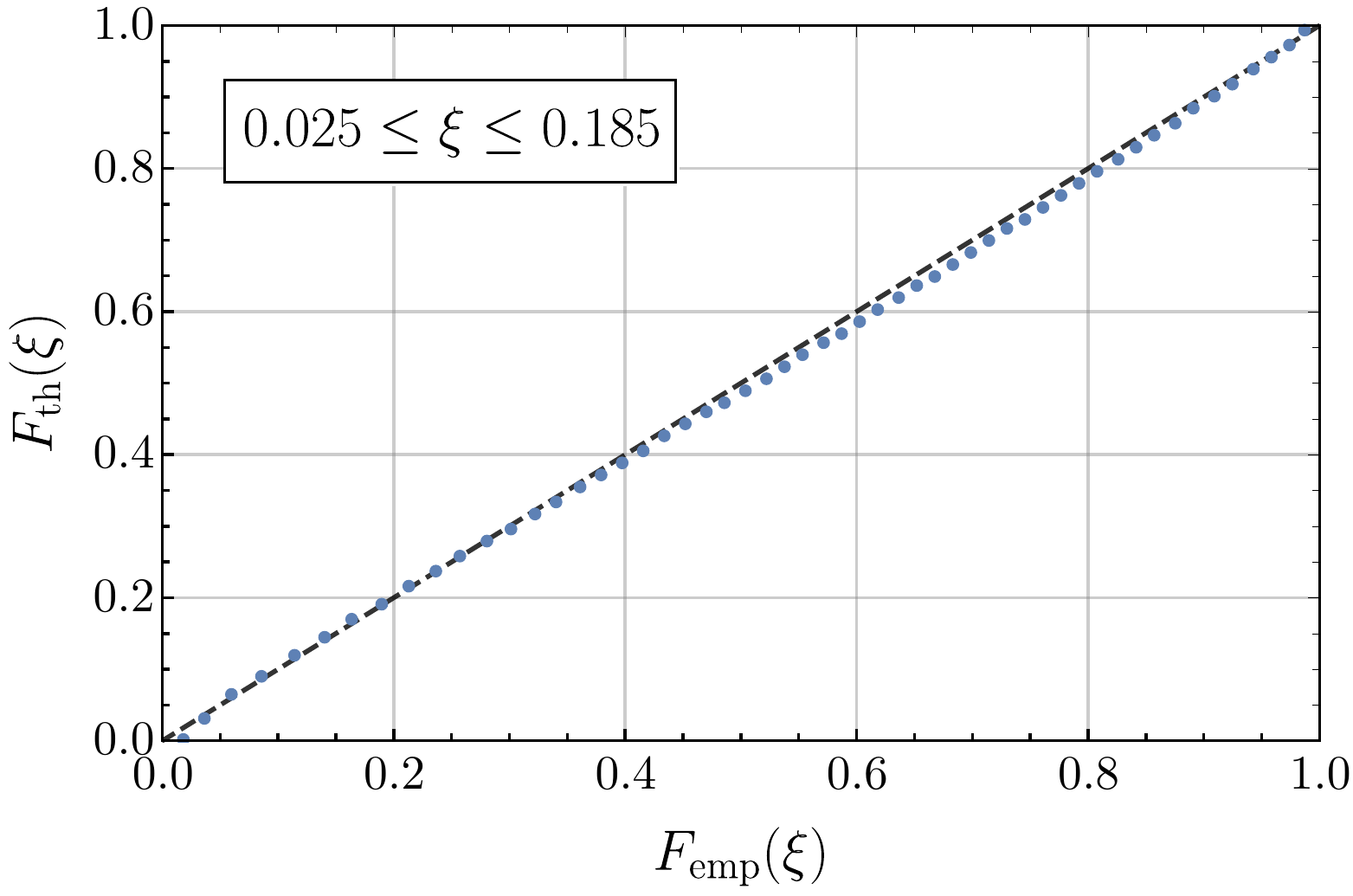}
    }
    \caption{P-P plots of the theoretical distribution of $\xi$, \eqref{eq:distributionXi}, versus the data obtained within the generic ensemble for (a) $0.01\leq \xi \leq 0.49$ and (b) $0.025\leq\xi\leq 0.185$. For reference, we plot the curve $F_{\text{th}}(\xi)=F_{\text{emp}}(\xi)$ corresponding to perfect agreement with a dashed line. We find that the distribution only fits the data within the range plotted in (b).}
    \label{fig:pp_plots}
\end{figure}

The statistical description of the flux ensemble presented in section \ref{sec:statisticalAnalysis} relies on two assumptions: first that the prepotential defining the EFT \eqref{eq:octic_prep} contains only polynomial terms (we neglect instanton contributions, $\cF_{\text{inst}} \equiv 0$), and second that the flux vector $N$ can be regarded as a \emph{continuous} random variable. These two conditions were imposed to make the problem analytically tractable, and also to simplify the numerical analysis so that the homotopy continuation methods could be applied. Thus, it is expected that, in regimes of parameter space where the vacua fail to be consistent with these assumptions, we should observe discrepancies between the theoretical probability distributions and the histograms obtained from the numerical scan. Here, we present the method we used to identify these deviations in the data in a systematic way. For definiteness we will focus the discussion on the distribution of the LCS parameter $\xi$, whose theoretical distributions for the generic and constrained ensembles are given by \eqref{eq:distributionXi} and \eqref{eq:xiDistNA0}, respectively.

As we discussed above, the predicted distribution cannot be applied in the whole domain of $\xi$. Vacua near the conifold point $\xi_{\text{cnf}}\approx0.39$ are prone to big instanton corrections while, for the generic ensemble, those in the LCS limit $\xi \to 0$ are expected to be suppressed due to the breakdown of the continuous flux approximation. Thus, we will now establish empirical bounds where the theoretical distributions for $\xi$ are applicable.
A widely used graphical method to verify how an empirical data distribution performs against a reference distribution is that of \emph{P-P} plots (see, e.g., \cite{PPplot}). 
The method consists in plotting the empirical and theoretical cumulative distributions functions, $F_\text{em}(\xi)$ and $F_{\text{th}}(\xi)$, one against the other. If the  agreement is perfect  then  the resulting plot is a straight line at 45 degrees.

As shown in figure~\ref{fig:pp_plots}, in the generic ensemble case, the best agreement with the Denef-Douglas distribution was found for 
\begin{equation}
0.025 \leq \xi \leq 0.185\,.
\end{equation}
The large discrepancy observed in figure~\ref{fig:pp_plots} (a) is due to the deficit of vacua near the LCS point in the generic ensemble. Note that these bounds are in very good agreement with the estimates \eqref{eq:appXiBound} and \eqref{eq:appXiBound2} obtained above, which correspond to the limits of applicability of the continuous flux approximation and of the EFT, respectively.

A similar analysis shows that for the constrained ensemble, the theoretical and empirical distributions agree in the interval
\begin{equation}
5\cdot10^{-5} \leq \xi|_{N_A^0 = 0} \leq 0.185\,.
\end{equation}
As can also be seen in the histograms of section~\ref{sec:statisticalAnalysis}, the Denef-Douglas distribution \eqref{eq:vacuaDensity0} provides an accurate description of the LCS parameter and other physical quantities (Yukawa couplings, scalar and fermion masses...) within the limits established above.

\bibliographystyle{JHEP}
\bibliography{refs}

\end{document}